\RequirePackage{silence}                                                                         
\documentclass[bibyear]{aa} 

\usepackage[varg]{txfonts}
\usepackage{amsmath}
\usepackage{amssymb}
\usepackage{graphicx}
\usepackage{hyperref}
\usepackage{float}
\usepackage{multirow}
\usepackage{booktabs}
\usepackage[usenames, dvipsnames]{color}
\usepackage{placeins}
\hypersetup{
        colorlinks = true,  
        urlcolor   = blue,  
        linkcolor  = red,   
        citecolor  = blue   
}
\usepackage{bookmark}

\usepackage{natbib}
\bibpunct{(}{)}{;}{a}{}{,} 
\title{The VLA-COSMOS 3 GHz Large Project: Average radio spectral energy distribution of highly star-forming galaxies}
\titlerunning{Radio SED}
\author{
         K. \Tisanic    \inst{1}    \thanks{\emph{ktisanic@phy.hr}}
    \and V. \Smolcic    \inst{1}    
    \and J. Delhaize    \inst{1, 6}  
    \and M. Novak       \inst{1}
    \and H. Intema      \inst{2}
    \and I. Delvecchio  \inst{1}
    \and E. Schinnerer  \inst{3}
    \and G. Zamorani    \inst{4}
    \and M. Bondi       \inst{4}
    \and E. Vardoulaki  \inst{5}
}
\institute{
    Department of Physics, Faculty of Science, University of Zagreb, Bijeni\v{c}ka cesta 32, 10000  Zagreb, Croatia
\and Leiden Observatory, Leiden University, Niels Bohrweg 2, 2333 CA Leiden, The Netherlands
\and MPI for Astronomy, K\"onigstuhl 17, D-69117, Heidelberg, Germany
\and INAF - Osservatorio di Astrofisica e Scienza dello Spazio, via Gobetti 93/3, 40129, Bologna, Italy
\and Argelander Institute for Astronomy, University of Bonn, Auf dem H\"ugel 71, D-53121 Bonn, Germany
\and Department of Astronomy, University of Cape Town, Private Bag X3, Rondebosch 7701, South Africa
}
\newcommand{\Smolcic}{Smol\v{c}i\'{c} }
\newcommand{\Tisanic}{Tisani\'{c} }

\newcommand{\Hz}{\,\mathrm{Hz}}
\newcommand{\kHz}{\,\mathrm{kHz}}
\newcommand{\MHz}{\,\mathrm{MHz}}
\newcommand{\GHz}{\,\mathrm{GHz}}
\newcommand{\THz}{\,\mathrm{THz}}
\newcommand{\mum}{\,\mathrm{\mu m}}

\newcommand{\km}{\,\mathrm{km}}

\newcommand{\Mpc}{\,\mathrm{Mpc}}

\newcommand{\hr}{\,\mathrm{hours}}
\newcommand{\yr}{\,\mathrm{yr}}

\newcommand{\W}{\,\mathrm{W}}
\newcommand{\Lrad}{L_{1.4\GHz}}
\newcommand{\LTIR}{L_{\mathrm{TIR}}}

\newcommand{\muJy}{\,\mathrm{\mu Jy}}

\newcommand{\Msun}{\,\mathrm{M_{\odot}}}

\newcommand{\SFR}{\,\mathrm{SFR}}

\newcommand{\percent}{\%}

\begin{document}

\date{Received 2 November 1992 Accepted 7 January 1993}         
\abstract{We construct the average radio spectral energy distribution (SED) of highly star-forming galaxies (HSFGs) up to $z\sim 4$. Infrared and radio luminosities are bound by a tight correlation that is defined by the so-called $q$ parameter.
This infrared-radio correlation provides the basis for the use of radio luminosity as a star-formation tracer. Recent stacking and survival analysis studies find $q$ to be decreasing with increasing redshift.
It was pointed out that a possible cause of the redshift trend could be the computation of rest-frame radio luminosity via a single power-law assumption of the star-forming galaxies' (SFGs) SED.
To test this, we constrained the shape of the radio SED of a sample of HSFGs.  To achieve a broad rest-frame frequency range, we combined previously published Very Large Array observations  of the COSMOS field at $1.4\GHz$ and $3\GHz$ with unpublished Giant Meterwave Radio Telescope (GMRT) observations at $325\MHz$ and $610\MHz$ by employing survival analysis to account for non-detections in the GMRT maps. We selected a sample of HSFGs  in a broad redshift range ($z\in[0.3,4]$,$\SFR\geq100\Msun/\yr$) and constructed the average radio SED. By fitting a broken power-law, we find that the spectral index changes from $\alpha_1=0.42\pm0.06$ below a rest-frame frequency of $4.3\GHz$ to  $\alpha_2=0.94\pm0.06$ above $4.3\GHz$. Our results are in line with previous low-redshift studies of HSFGs ($\SFR>10\Msun/\yr$) that show the SED of HSFGs  to differ from the SED found for normal SFGs ($\SFR<10\Msun/\yr$). The difference is mainly in a steeper spectrum around $10\GHz$, which could indicate a smaller fraction of thermal free-free emission. Finally, we also discuss the impact of applying this broken power-law SED in place of a simple power-law in K-corrections of HSFGs and a typical radio SED for normal SFGs drawn from the literature. We find that the shape of the radio SED is unlikely to be the root cause of the $q-z$ trend in SFGs.}
\keywords{Galaxies: evolution, Galaxies: statistics, Radio continuum: galaxies, Galaxies: star formation}

\maketitle      
\makeatother
\section{Introduction}\label{sect:Introduction}
Star formation rate (SFR) measurements  derived from the ultraviolet and optical emission are sensitive to dust \citep{Kennicutt98a}, and infrared-derived SFRs are easy to understand only in the optically thick case \citep{Kennicutt98a, Condon92}.
On the other hand, radio emission in star-forming galaxies (SFGs) below  $\sim30\GHz$  is unbiased by dust \citep{Condon92}. 
Furthermore, it has been found that the infrared and radio luminosities are bound by a tight correlation over many orders of magnitude \citep{Kruit71, Helou85,Condon92}.
This infrared-radio correlation provides the basis for radio luminosity as a star-formation tracer, as infrared luminosity is linked to the SFR \citep{Kennicutt98b}. 
It is  defined by the so-called $q$ parameter, defined as $q=\log L_{TIR}/3.75\THz-\log(\Lrad)$, where $\LTIR$ is the total infrared luminosity ($8-1000\mum$) in units of $\W$ and $\Lrad$ is the luminosity density at $1.4\GHz$ in units of $\W/\Hz$. A frequency of $3.75\THz$ is used to obtain a dimensionless quantity.
In the calorimeter model \citep{Voelk89}, $q$ is expected to be constant over a wide range of magnetic field strengths through electrons that lose their energy because of synchrotron and inverse Compton losses.
This produces a tight correlation, even though individual galaxies may show deviations.
Theoretical considerations \citep{Murphy09} predict $q$ to increase with redshift, mostly because of the synchrotron electrons' inverse Compton scattering off cosmic microwave background photons. 
Recent stacking and survival analysis studies find $q$ to be decreasing with increasing redshift  \citep{Ivison10,Magnelli15, Delhaize17, CalistroRivera17}.
\citet{Delhaize17}  pointed out that the computation of rest-frame radio luminosity via a K-correction using only a simple, single power-law assumption of the star-forming galaxies' spectral energy distribution (SED) might cause such a trend.

The exact shape of the radio SED of star-forming galaxies is usually assumed to be a superposition of the steep synchrotron spectrum, described by a power law, and a $\sim10\percent$ contribution at $1.4\GHz$ of a flat free-free spectrum \citep{Condon92}. These results are supported by observations of  nearby galaxies at $3-30\Mpc$ ($z\approx 0.01$) with an $\SFR<10\Msun/\yr$ \citep{Tabatabaei17}.

For (ultra)luminous infrared galaxies ((U)LIRGs, $\SFR>10 \Msun/\yr$), observations in the local Universe find their radio SEDs to be steep around $10\GHz$, corresponding to a spectrum that shows less flattening above $10\GHz$ than in galaxies with an $\SFR<10\Msun/\yr$  \citep{Clemens08, Leroy11, Galvin17}. 
Both free-free self-absorption and modifications to the spectrum due to the aging of the electron population would also be present. These processes result in deviations from a simple power-law at low  ($\nu_{rest}\lesssim1\GHz$) and high ($\nu_{rest}\gtrsim \mathrm{a\,few}\GHz$) frequencies  that are observationally hard to constrain. 
    
In this paper, we constrain the shape of the average radio spectral energy distribution (SED)\ of a $1.4\GHz$-selected sample ($\SFR>100\Msun/\yr$) of highly star-forming galaxies (HSFGs)  in the $\sim 0.4-10\GHz$ frequency range within the COSMOS field. To obtain a broad frequency range, we use the fluxes from the $1.4\GHz$ Very Large Array (VLA)-COSMOS Joint Project \citep{Schinnerer10} and the $3\GHz$ VLA-COSMOS Large Project \citep{Smolcic:17a,Smolcic:17b} and reduce previously unpublished Giant Meterwave Radio Telescope (GMRT) observations at $325\MHz$ and $610\MHz$. We construct the average {radio} SED by using survival analysis to account for the {lower sensitivity} of the GMRT maps. 

In Sect. \ref{sect:Data} we describe the VLA and GMRT datasets used in this paper, in Sect. \ref{sect:Sample} we describe the galaxy sample of HSFGs, in Sect. \ref{sect:Average radio SED} we construct the average radio SED for HSFGs and test the applicability of our method. In Sect. \ref{sect:Results} we report the average radio spectral indices of HSFGs, while in Sect. \ref{sect:Discussion} we discuss the shape of the radio spectrum of SFGs and the impact on the infrared-radio correlation.
In this paper we use the convention for the spectral index $F_\nu\sim\nu^{-\alpha}$, where $F_{\nu}$ is the flux density and $\alpha$ is the spectral index.

\section{Data}\label{sect:Data}

To assess the radio spectral energy distribution, we combined catalogs of the COSMOS field at four  observer-frame frequencies: $325\MHz$ and $610\MHz$, obtained with the GMRT,  and at $1.4\GHz$ 
\citep{Schinnerer10} $\text{and }3\GHz$ \citep{Smolcic:17a}, obtained by the VLA.
Here we briefly describe these VLA (Sect. \ref{sect:Available VLA data} ) and GMRT (Sect. \ref{sect:GMRT data}) data.

\subsection{Available VLA data}\label{sect:Available VLA data}
The VLA COSMOS $3\GHz$ Large Project map \citep{Smolcic:17a} was constructed from $384\hr$ of observations of the  $2$ deg2 COSMOS field. 
Observations were carried out over $192$ pointings in the S-band with VLA in A and C antenna configurations.
This was a wide-band observation with a total bandwidth of $2084\MHz$ derived from $16$ spectral windows, each $128\MHz$ wide. The final mosaic reached a median root-mean-square (RMS) of $2.3\muJy/\mathrm{beam}$ at a resolution of $0.75\arcsec$.
Considering the large bandwidth and the volume of the data, each pointing was imaged separately using the multi-scale multi-frequency synthesis algorithm \citep[hereafter MSMF,][]{Rau11} and then combined into a single mosaic. 
The MSMF algorithm had been found to be optimal for both resolution and image quality in terms of the RMS noise and sidelobe contamination \citep{Novak15}.
The $3\GHz$ catalog has been derived using \textsc{blobcat} \citep{Hales12} with a $5\sigma$ threshold.
In total, 10830 sources were recovered (67 of which were multi-component sources). For more details, see \citet{Smolcic:17a}.

The $1.4\GHz$ catalog is a joint catalog comprised of the VLA COSMOS $1.4\GHz$ Large \citep{Schinnerer07} and Deep surveys \citep{Schinnerer10}.  The joint catalog was constructed by combining the observations of the central $50\arcmin\times 50\arcmin$ subregion of the COSMOS field \citep[VLA-COSMOS Deep Project][]{Schinnerer10} at a resolution of $2.5\arcsec\times2.5\arcsec$ and previously published observations of the entire $2\deg^2$ COSMOS field at a resolution of $1.5\arcsec\times1.4\arcsec$ \citep[VLA-COSMOS Large Project][]{Schinnerer07}. The average RMS in the resulting map was found to be $12\muJy$.
The large survey consisted of $240\hr$ of observations over $23$ pointings spread out over the entire COSMOS field in the L band centered at $1.4\GHz$ and with a total bandwidth of $37.5\MHz$ in VLA A configuration and $24\hr$ in the C configuration. 
\citet{Schinnerer10} supplemented the $7$ central pointings with an additional $8.25\hr$ of observations per pointing using the A configuration and the same L-band configuration. 
These new measurements were then combined in the uv-plane with the Large Project observations.
The joint catalog was constructed by using the SExtractor package \citep{BertinArnouts96} and the AIPS task SAD, yielding 2865 sources \citep{Schinnerer10}.

\subsection{GMRT data}\label{sect:GMRT data}
The GMRT $325$ and $610\MHz$ data reduction, imaging, cataloging, and testing is described in detail in Appendix \ref{sect:GMRT data reduction}.
Here we give a brief overview of the procedure. 

The overall data reduction procedure is similar for both {observing} frequencies. {The observations were carried out} with $30$ antennas with their longest baseline being $25\km$. The {data} were then reduced by using the source peeling and atmospheric modeling pipeline \citep[SPAM, described in detail in][]{Intema17}. Finally, the source extraction package \textsc{blobcat} was used to extract sources down to $5\sigma$, where $\sigma$ is the local RMS value computed from RMS maps produced by the AIPS task RMSD. The values of $\sigma$ are shown in the visibility functions, presented in top left panels of Figs. \eqref{fig:325AstroBWS} and \eqref{fig:610AstroBWS} for the $325\MHz$ and $610\MHz$ RMS maps, respectively. The fluxes were corrected for bandwidth-smearing, and resolved sources were {identified}.

The $325\MHz$ observations of a single pointing were carried out under the project 07SCB01 (PI: S. Croft) (at a reference frequency of $325\MHz$ and a bandwidth of $32\MHz$).
The observations lasted for $45\hr$ in total, {comprising four observations with a total on-source time of $\sim40\hr$.} They were reduced using the SPAM pipeline and imaged at a resolution of $10.8\times 9.5\,\mathrm{arcsec}^2$.  A primary beam correction was applied to the pointing.
{We {measured} a median RMS of $97\muJy/\mathrm{beam}$ over the $\sim 2 \deg^2$ COSMOS field.}
In total, 633 sources were identified using \textsc{blobcat} down to $5\sigma$. {By employing a peak-over-total flux criterion, we consider 177 of these sources to be resolved.}

The $610\MHz$ observations were carried out at a central frequency of $608\MHz$ using a bandwidth of $32\MHz$.
Observations lasting $86\hr$ ({spread over eight observations with an average time on source per pointing of $\sim 4\hr$}) were conducted under the project 11HRK01 (PI: H. R. Kl\"ockner).
A total of $19$ pointings were observed. The data were reduced using the SPAM pipeline, and imaged at a resolution of $5.6\times 3.9\,\mathrm{arcsec}^2$. Following \citet{Intema17}, primary beam correction and the average pointing error corrections were applied to each pointing prior to mosaicing.
{We {measured} a median RMS of $39\muJy/\mathrm{beam}$ in the final mosaic over the $\sim 2 \deg^2$ COSMOS field. }
\textsc{Blobcat} was used to extract 999 sources down to $5\sigma$, 196 of which we consider to be resolved.

\section{Sample}\label{sect:Sample}
\begin{figure}[t]
\includegraphics[width=\columnwidth]{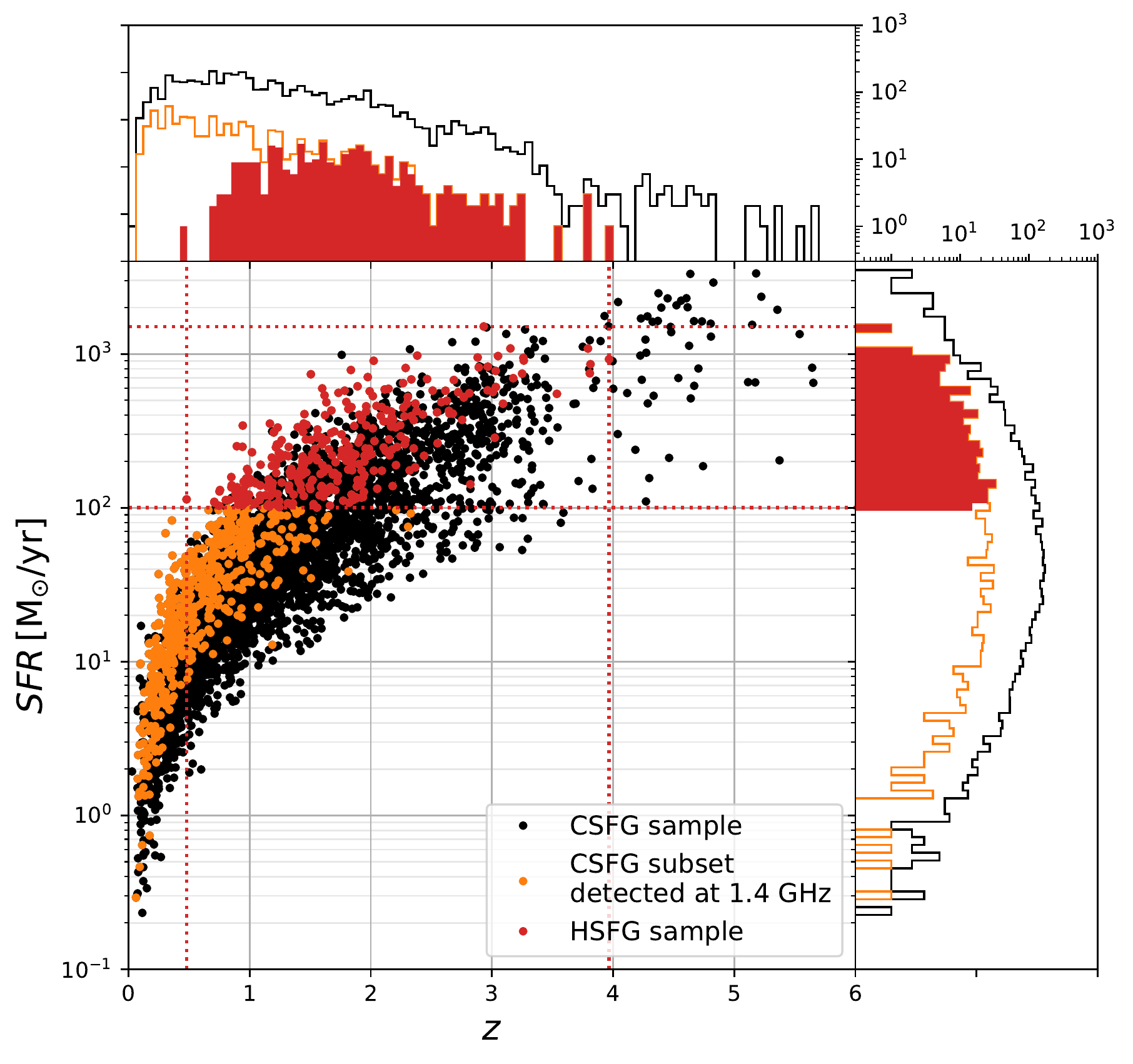}
\caption{Star formation rate vs. redshift for the  ``clean star-forming galaxy sample'' (CSFG, black points), as defined by \citet{Smolcic:17b}, and described in Sect. \ref{sect:Sample}.
Orange points indicate a subset of the CSFG sample that was detected at both $1.4\GHz$ and $3\GHz$.
Red points indicate the HSFG sample used here, i.e., the subset of the CSFG sample with an additional cut of the infrared-derived SFR, $SFR>100\Msun/\yr$. Histograms show the redshift and SFR distributions with the different subsamples colored as in the main figure. For greater clarity, the HSFG sample is marked by filled histograms, and its limits in the main figure are denoted by dotted red lines.}\label{fig:SFRz}
\end{figure}

To construct a complete and pure sample of star-forming galaxies over a wide range in redshift, we used the criteria described in \citet{Smolcic:17b}, which we briefly summarize below.
Starting from the VLA COSMOS Large Project $3\GHz$  Source Catalog, cross-correlated with multi-wavelength counterparts detected in the COSMOS field, we selected the clean star-forming galaxy sample (CSFG), as defined in \citet{Smolcic:17b}, and imposed an additional cut of infrared-derived SFRs of $\SFR>100\Msun \yr^{-1}$.

\citet{Smolcic:17b} cross-correlated the $3\GHz$ Source Catalog with  multi-wavelength  counterparts, drawn from three catalogs: i) the COSMOS2015 catalog \citep{Laigle16},  which contains sources detected on a $\chi^2$ sum of $Y$, $J$, $H$, $K_s$  and $z^{++}$ images, ii) the i-band catalog \citep{Capak07}, which contains sources  detected from a combination of the CFHT $i^*$  and Subaru $i^+$original point spread function (PSF) images, and iii) the IRAC catalog, which contains sources detected in the 3.6~$\mu$m band \citep{Sanders07}. \citet{Smolcic:17b} combined counterparts from the three catalogs to maximize the number of counterparts as the i-band (IRAC) catalog is sensitive to potentially blue (red) sources undetected in the  $Y$, $J$, $H$, $K_s$ , $z^{++}$  stack. 
Over the inner $1.77\,\mathrm{deg}^2$ subarea of the COSMOS field, which excludes all regions affected by saturated or bright sources in the optical to near-infrared bands, and thus contains sources with the best possible photometry, \citet{Smolcic:17b} found 7729 COSMOS2015, 97 i-band, and 209 IRAC counterparts to the 
8696 3 GHz sources within this area. 
The COSMOS2015 and i-band catalogs offer precise photometric redshifts, with a joint precision of $\sigma_{\Delta z/(1+z)}=0.01$.

The CSFG sample was constructed from the full $3\GHz$ galaxy sample with COSMOS2015 or i-band counterparts as the cleanest sample of star-forming galaxies by excluding AGN through a combination of criteria, such as X-ray luminosity, mid-infrared  (color-color and SED-based) criteria, $r^+-$NUV colors, and ($>3\sigma$) excess in the distribution of the ratio of 1.4 GHz radio luminosity and IR-based SFRs as a function of redshift \citep[see Sect. 6.4 and Fig. 10 in][]{Smolcic:17b}. 
To determine the physical properties of the galaxies, a three-component SED-fitting procedure \citep{Delvecchio17} was applied using all  of the available photometry.
The fitting procedure was based on MAGPHYS \citep{daCunha08} and SED3FIT \citep{Berta13} to account for the AGN component of the SED.
SFRs were computed from the 8–1000 $\mathrm{\mu m}$ rest-frame infrared luminosity that was {obtained by using the best-fit galaxy template and was then} converted into SFR through the \citet{Kennicutt98a} conversion factor and scaled to a \citet{Chabrier03} initial mass function.
Star formation rates are shown for the entire $3\GHz$ Large Project catalog and the subset of highly star-forming galaxies ($\SFR\geq 100\Msun/\yr$) in Fig. \eqref{fig:SFRz} as a function of redshift.
The figure shows that a cut above $100\Msun/\yr$ minimizes incompleteness above $z\approx 2$, which strongly influences the sample below $100\Msun/\yr$. To check whether  the remaining incompleteness could influence our results, we repeated our analysis of Sect.  \ref{sect:Results} for the more complete subsample of HSFGs out to $z\sim 2$. We find no significant difference in the resulting  SED and therefore conclude that the $100\Msun/\yr$ cut sufficiently reduces incompleteness out to $z\sim 4$.

\begin{figure}
\includegraphics[width=\columnwidth]{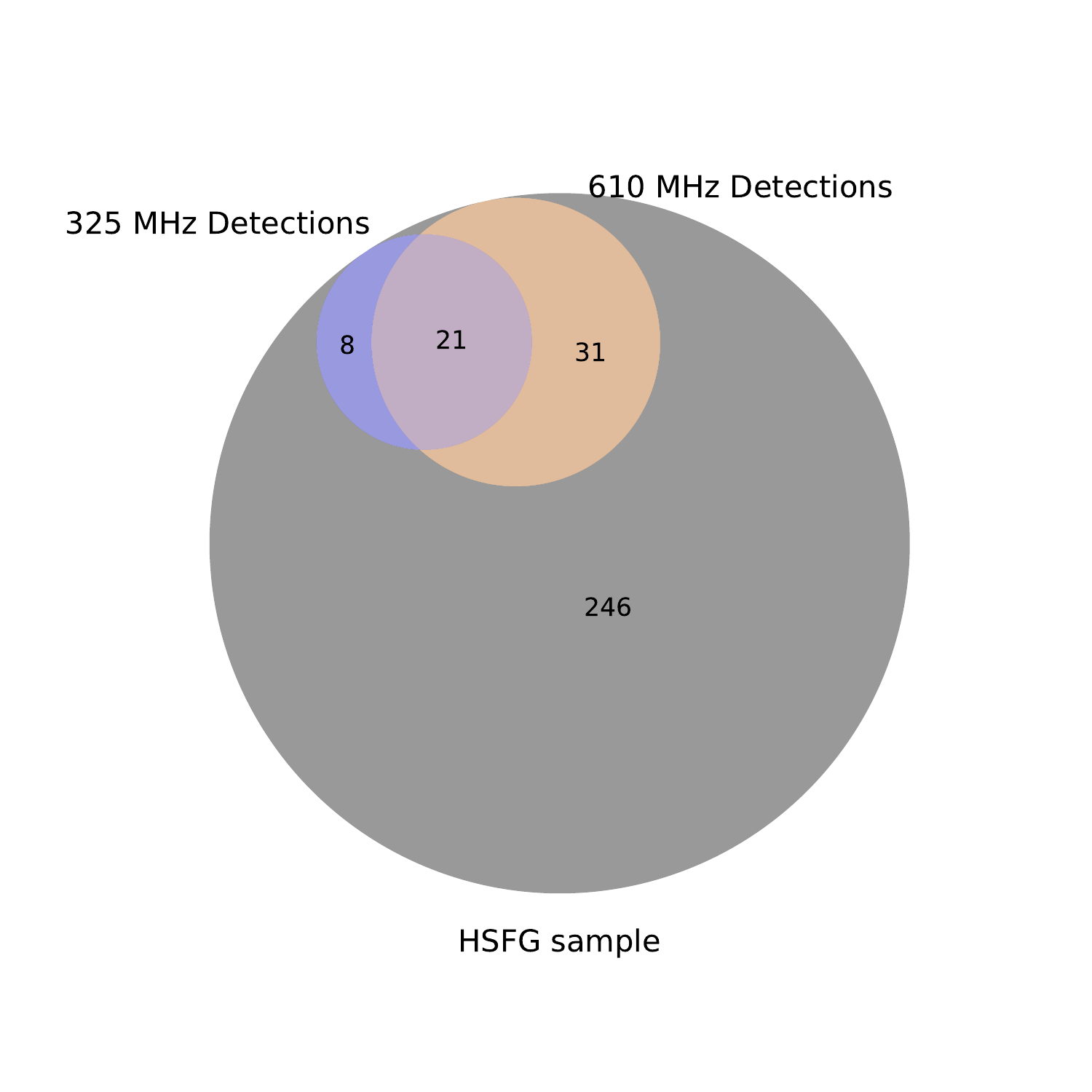}
\caption{Number of detections in GMRT catalogs  of sources in the HSFG samples. The gray area represents the sources that are not detected in the GMRT maps, while blue and orange areas represent sources detected at $325\MHz$ and $610\MHz$, respectively. }
\label{fig:HSFGsummary}
\end{figure}

Finally, we cross-matched the $3\GHz$ catalog with the above described cuts and the $1.4\GHz$ joint catalog, to arrive at a sample of $306$ galaxies that have two reliably determined fluxes in the rest-frame frequency-flux space.
We call the subset of the CSFG sample that is detected at both $1.4$ and $3\GHz$ and has $SFR>100\Msun/\yr$ the highly star-forming galaxy (HSFG) sample.
The $325\MHz$ and $610\MHz$ catalogs were then cross-matched with the HSFG sample catalog {with a matching radius of 1 beam}, leading to 29 and 52 sources in the HSFG sample, respectively. The summary of the number of detections and non-detections in the catalogs used to construct the radio SED of the HSFG sample is {provided} in Fig. \eqref{fig:HSFGsummary}.\footnote{The table of the cross-matched fluxes is available in electronic form
at the CDS via anonymous ftp to \url{cdsarc.u-strasbg.fr} (130.79.128.5)
or via \url{http://cdsweb.u-strasbg.fr/cgi-bin/qcat?J/A+A/}.
}

In summary, the selected sample is effectively a $1.4\GHz$ data-selected HSFG ($\SFR>100\Msun/\yr$). {With this dataset,} we associated the $3\GHz$ data,  with a much lower RMS ($\sim 12\,\mathrm{\mu Jy/beam}$ and $\sim 2\,\mathrm{\mu Jy/beam}$ at $1.4\GHz$ and $3\GHz$, respectively), minimizing potential biases that could arise in the $1.4$ to $3\GHz$ spectral index distribution, given that almost all $1.4\GHz$ sources \citep[~90$\percent$,][]{Smolcic:17a} have counterparts at $3\GHz$. In total, the HSFG sample contains 306 galaxies, out of which $9\percent$ (29) were detected at $325\MHz$ and $17\percent$ (52) at $610\MHz$.

\begin{figure*}[ht]
        \includegraphics[width=\textwidth]
            {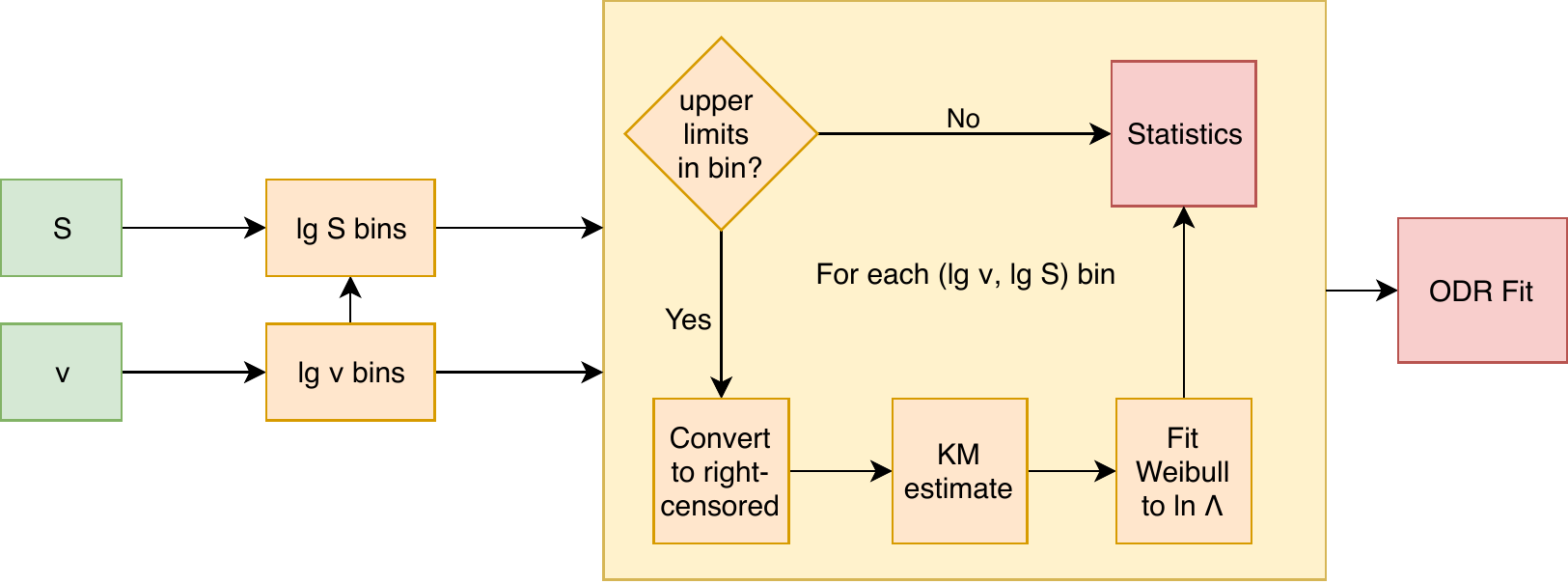}
        \caption{Flowchart of the radio SED fitting procedure described in Sect. \ref{sect:Construction}. The input values are source fluxes, $S,$ and rest-frame frequencies,  $\nu$. The log fluxes are then binned in log-rest-frame frequency bins. If there are no upper  limits in a certain bin, the mean and standard deviation are calculated immediately, and if that is not the case, the log fluxes are converted into a right-censored dataset, and the Kaplan-Meier estimate is calculated. This estimate is then fit to a Weibull distribution, yielding the analytically determined means and standard deviations. The resulting different bins are then treated as data points with errors in both axes and are fit by a particular model by employing the ODR method.}\label{fig:FlowFit}
 \end{figure*}

\begin{figure*}[hbtp]
 
\centering\includegraphics[width=\textwidth]{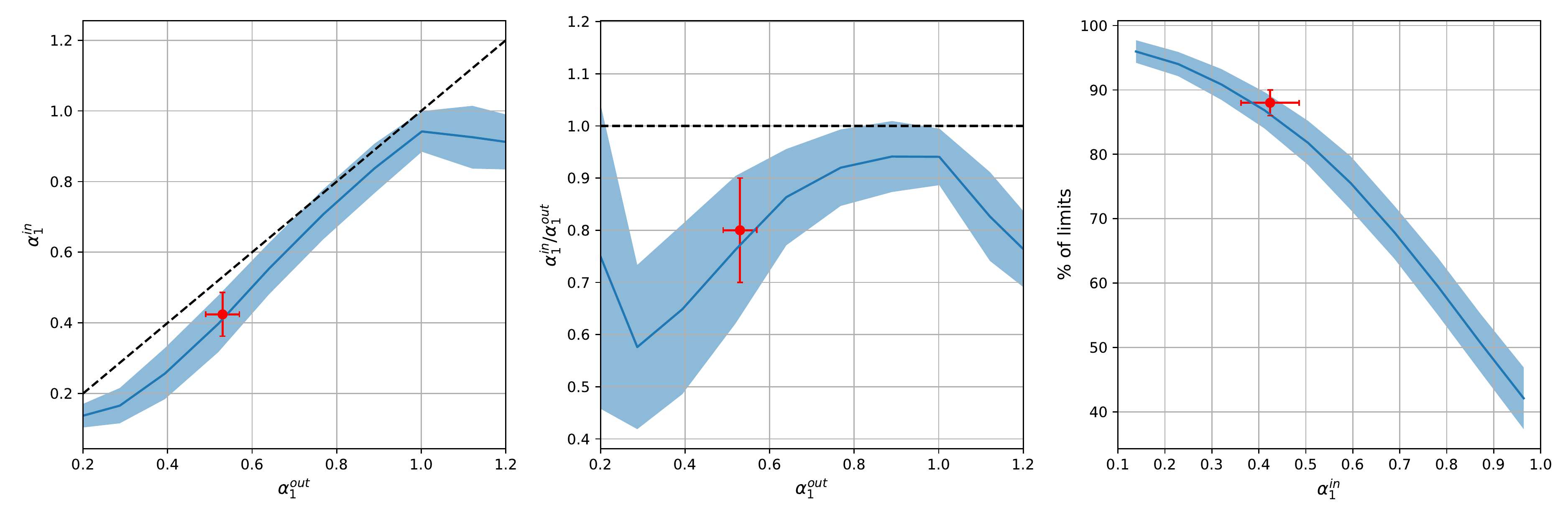}
\caption{Results of simulations produced by varying {the mean value of} the input spectral index $\alpha_1$ (labeled as $\alpha_1^{in}$). The figure shows the interdependence of the following properties: the input spectral index $\alpha_1^{in}$, the derived spectral index $\alpha_1^{out}$ , and the percentage of upper limits below $1\GHz$. The left panel shows the dependence of the input spectral index, $\alpha_1^{in}$,  on $\alpha_1^{out}$; the middle panel shows the dependence of the  $\alpha_1^{in}/\alpha_1^{out}$ ratio on the output spectral index, $\alpha_1^{out}$; and the right panel shows the dependence of the percentage of upper limits below $1\GHz$ on $\alpha_1^{in}$. The solid lines show the means derived for different bins in the x-axes of their respective plot, while the colored intervals show the $1\sigma$ interval for the same bins. The red point shows the values of spectral index $\alpha_1$ before and after correction, and the percentage of upper limits below $1\GHz$, as presented in Sect. \ref{sect:Results}. The dashed lines are the equality lines.} \label{fig:simulations}
\end{figure*}
\section{Average radio SED}\label{sect:Average radio SED}
In this section, we describe the method used to construct the radio SED and the models used to analyze it.
In Sect. \ref{sect:Construction} we describe how we constructed the radio SED of HSFGs from the GMRT and VLA observations of the COSMOS field while accounting for lower sensitivities of the GMRT maps using survival analysis.
In Sects. \ref{sect:SED fitting} and \ref{sect:Method verification via simulations} we describe the models of the radio SED and the Monte Carlo simulations that test the applicability of the fitting procedure, respectively.
\subsection{Construction of the average radio SED of HSFGs}\label{sect:Construction}

To construct a typical SED for HSFGs in the COSMOS field, we normalized the spectra of individual galaxies at a particular rest-frame frequency.
Considering that the redshifts of sources in our sample vary from $0.3$ to $4$, we obtain a well-constrained SED from rest-frame frequencies of $\sim0.5\GHz$ to $\sim 15\GHz$.

A complication to this basic idea is that there are very few detections of HSFGs in the $325$ and $610\MHz$ maps.
For this reason, we normalized the flux densities at all frequencies  based on a linear fit to the $1.4$ and $3\GHz$ spectra. 
{We chose the normalization rest-frame frequency to be the median log-frequency of sources in the HSFG sample at $1.4\GHz$ and $3\GHz$. For the HSFG sample, this frequency is $5.7\GHz$.}
We constrained the $325$ and $610\MHz$ flux densities of non-detections by $\text{five}$ times the local RMS flux density of their respective maps. 
The problem is therefore to constrain the spectral index when a significant portion of the data is left-censored, meaning that many of the data points below a rest-frame frequency of $1\GHz$, corresponding to observer frame $325$ and $610\MHz$ frequencies,  are upper limits. 
To correctly deal with upper limits, we decided to employ the survival analysis technique described in the following paragraph and illustrated in Fig. \eqref{fig:FlowFit}.

We combined into a single dataset the normalized fluxes and normalized upper limits of individual galaxies at different rest-frame frequencies.
As we have four different observer-frame frequencies, the number of our data points is four times larger than the number of galaxies.
To achieve uniform frequency binning, we used 20 bins that were equally separated in log space of the rest-frame frequencies. 
The number of bins was chosen empirically to achieve a satisfactory number of bins while retaining a sufficient number of data points within each bin for the statistical analysis.

For each bin, $i$, if there were no upper limits within the bin, we computed the mean of the log rest-frame frequency, $\nu_i=10^{\langle\log \nu_r \rangle_i}$,  and its standard deviation, $\sigma_{\log \nu_i}$, along with the mean normalized log flux, $\langle\log F_\nu\rangle_i$, and its standard deviation, $\sigma_{\log F_i}$.
If there were upper limits within a particular bin, we constructed the Kaplan-Meier estimate of the survival function of fluxes within the bin using the \emph{lifelines} package\footnote{See \url{https://lifelines.readthedocs.io/}.}. We did this by converting this left-censored dataset into a right-censored one in accordance with \citet{Feigelson85} as the \emph{lifelines} package handles right-censored datasets. 

We then fit the survival function with the Weibull distribution \citep{Lai06}. 
We have found this method to be satisfactory because the Weibull distribution survival function,  $S(x|\lambda, k)=e^{-(x/\lambda)^k}$, where $\lambda$ and $k$ are free parameters of the model, can be easily linearized.
We estimated  $\langle {\log F_\nu}\rangle$ from $k$ and $\lambda$ and calculated $\sigma_{\log F}$ as the standard deviation of the Weibull distribution.

Survival analysis assumes that upper limits have the same underlying distribution as the data, and the Kaplan-Meier estimator further assumes that detections and upper limits are mutually independent \citep{Kaplan58}.
{The latter condition is satisfied by normalizing spectra to one frequency, as the normalization redistributes both the data and the upper limits based on the source's normalization flux.}
Further tests of this procedure are described in Sect. \ref{sect:Method verification via simulations} using Monte Carlo simulations of the properties of the HSFG sample.

\subsection{SED fitting}\label{sect:SED fitting}
Previous studies, listed in Sect. \ref{sect:Introduction}, have shown that the shape of the radio SED deviates from a simple power law. 
As the SED at higher frequencies ($\nu\gtrsim10\GHz$) is expected to have a contribution not only by synchrotron, but also by free-free emission, the simplest modification to a simple power-law would be adding a thermal contribution with a flat spectral index of $0.1$ \citep{Condon92} that would produce a flattening of the SED above $10\GHz$:
\begin{equation}
F\left(\nu_r\middle| \begin{matrix}\alpha_2\\ b\\ f\end{matrix}\right) = 10^b\left[\left(\frac{\nu_r}{\nu_n}\right)^{-\alpha_2} (1-f) + f\left(\frac{\nu_r}{\nu_n}\right)^{-0.1}\right],
\end{equation}
where $\nu_r$ is the rest-frame frequency, and the parameters of the model are $\alpha_2$, the nonthermal spectral index, $b$, the normalization constant, $\nu_n$, the normalization frequency, and $f=f(\nu_n)$, the thermal fraction at $\nu_n$.
In our HSFG sample, however, we observe that the SED flattens below $3-4\GHz$ (see Sect. \ref{sect:Results} and figures therein).
To account for this, we assumed a broken power-law instead of using only the nonthermal spectral index.
Therefore, the full SED would be
\begin{equation}
F\left(\nu_r\middle| \begin{matrix}\alpha_1\\\alpha_2\\ b\\ f\\ \nu_b\end{matrix}\right)=
\begin{cases}
        10^b\left[\left(\frac{\nu_r}{\nu_n}\right)^{-\alpha_2} (1-f) + f\left(\frac{\nu_r}{\nu_n}\right)^{-0.1}\right], & \nu_r>\nu_b\\
    10^b\left[\left(\frac{\nu_r}{\nu_n}\right)^{-\alpha_1} (1-f) \left(\frac{\nu_b}{\nu_n}\right)^{\alpha_1-\alpha_2}+ f\left(\frac{\nu_r}{\nu_n}\right)^{-0.1}\right],&\nu_r<\nu_b
\end{cases},\label{eq:SEDfull}
\end{equation}
where $\nu_n$ is the normalization frequency, $\nu_b$ the break frequency, $\alpha_1$ ($\alpha_2$) the spectral index below (above) $\nu_b$ , and $b$ the normalization constant.

In the HSFG sample, as described in Sect. \ref{sect:The shape of the radio SED}, we do not find $f$ to be significantly greater than 0 {for $\nu_n\sim1\GHz$.}
We therefore simplified Eq. \eqref{eq:SEDfull} to a broken power-law model for the log-normalized flux
\begin{equation}
\log F\left(\nu_r\middle| \begin{matrix}\alpha_1\\\alpha_2\\ b\\ \nu_b\end{matrix}\right)=
\begin{cases}
        -\alpha_2 \log \frac{\nu_r}{\nu_n} + b,& \nu_r>\nu_b,\\
        -\alpha_1 \log  \frac{\nu_r}{\nu_n}+b+(\alpha_1-\alpha_2) \log  \frac{\nu_b}{\nu_n},& \nu_r<\nu_b,
\end{cases}.\label{eq:BPL}
\end{equation}
In the broken power-law model, the break frequency, $\nu_b$, can be both a free parameter or fixed to a particular value prior to fitting.
The break frequency is derived using the Markov chain Monte Carlo (MCMC)  method \citep[implemented using the affine invariant sampler of the \emph{emcee} package][]{Foreman13} on the broken power-law fit with the break frequency as a free parameter. By inspecting the distribution of limits and detections, we find that the break frequency should lie above the highest frequency at which the last upper limit occurs. We thereby conclude that the break frequency would not be significantly affected by upper limits. 

The MCMC code samples the log-likelihood distribution of parameters, $\theta=(\alpha_1, \alpha_2, b, \nu_b),$  
 \begin{equation}
 \ln \mathcal{L}(\theta) \propto -\frac{1}{2}\sum\limits_i \left(\frac{\langle\log F\rangle_i-\log F(\nu_{i}, \theta)}{\sigma_{\log F_i}}\right)^2,\label{eq:MCMC}
 \end{equation}
where the values $\langle\log F\rangle_i$ and $\sigma_{\log F_i}$ and $\nu_i$ are  log-flux derived by using survival analysis in each log-frequency bin with the main value of log frequency in each bin being $x_i$. 
For the sake of simplicity of implementation, we considered only the errors in normalized log flux in the denominator of Eq. \eqref{eq:MCMC} when using the MCMC method. 
 
The final fitting of the broken power-law model is performed using the orthogonal distance regression method (hereafter, ODR), which accounts for the errors on both axes.
The $1\sigma$ confidence interval, $\sigma(\nu_r)$, is computed as $\sigma(\nu_r |\theta)=\sqrt{(\nabla_\theta \log F)^T \Sigma_\theta \nabla_\theta \log F}$, with $\Sigma_\theta$ being the covariance matrix of the fit parameters derived using the ODR method and  $\theta$ being the vector of free parameters of the broken power-law.

\subsection{Method verification through simulations}\label{sect:Method verification via simulations}

Our average SED was derived using survival analysis, where the percentage of upper limits below rest-frame $1\GHz$ is high. 
We tested the applicability of the survival analysis by deriving the radio spectral indices for a mock sample of galaxies. The mock samples were generated with preset spectral indices by simulating different properties of the HSFG sample and of the SED shape.

\begin{figure*}
 \includegraphics[width=\columnwidth]{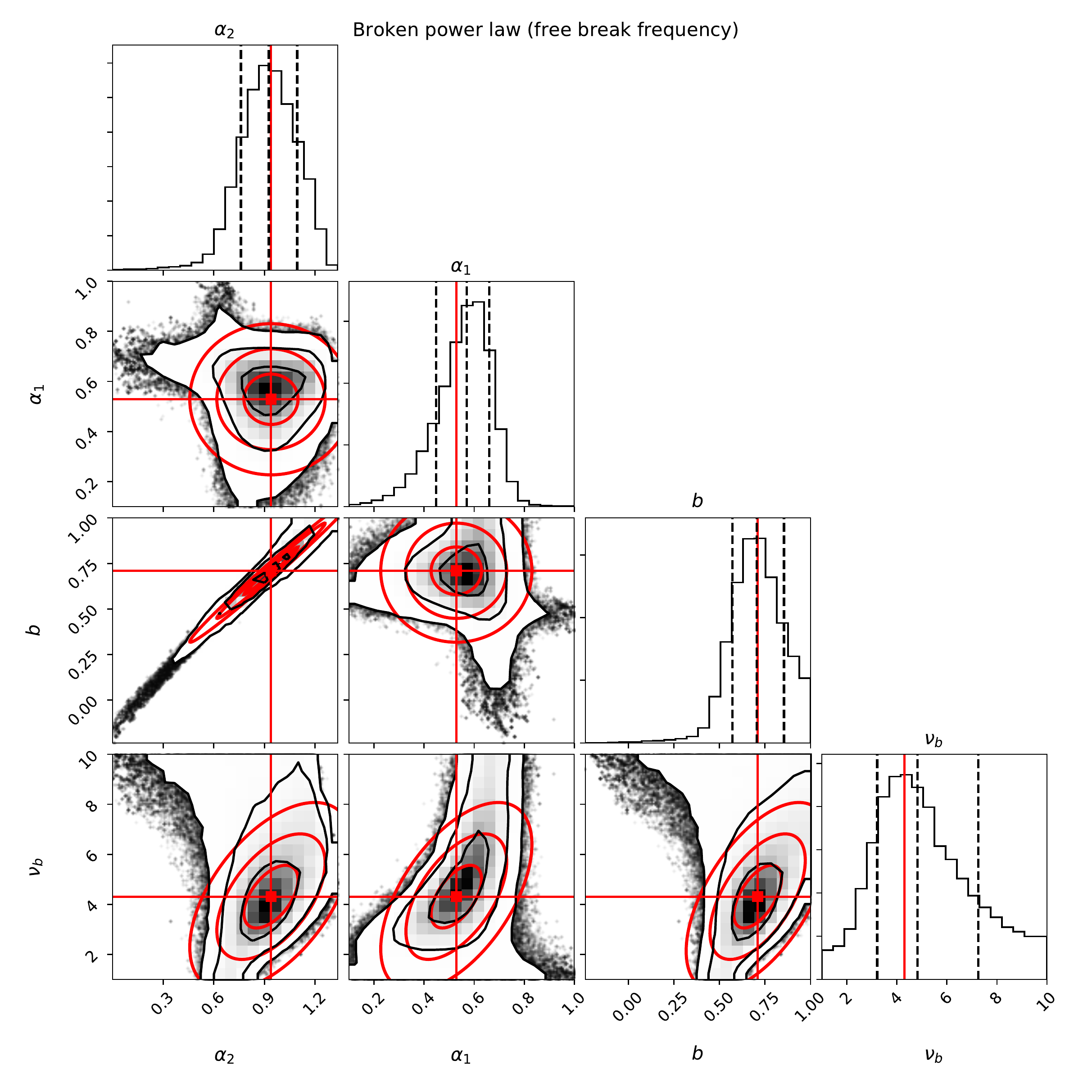}
     \hspace{1cm}
\includegraphics[width=\columnwidth]{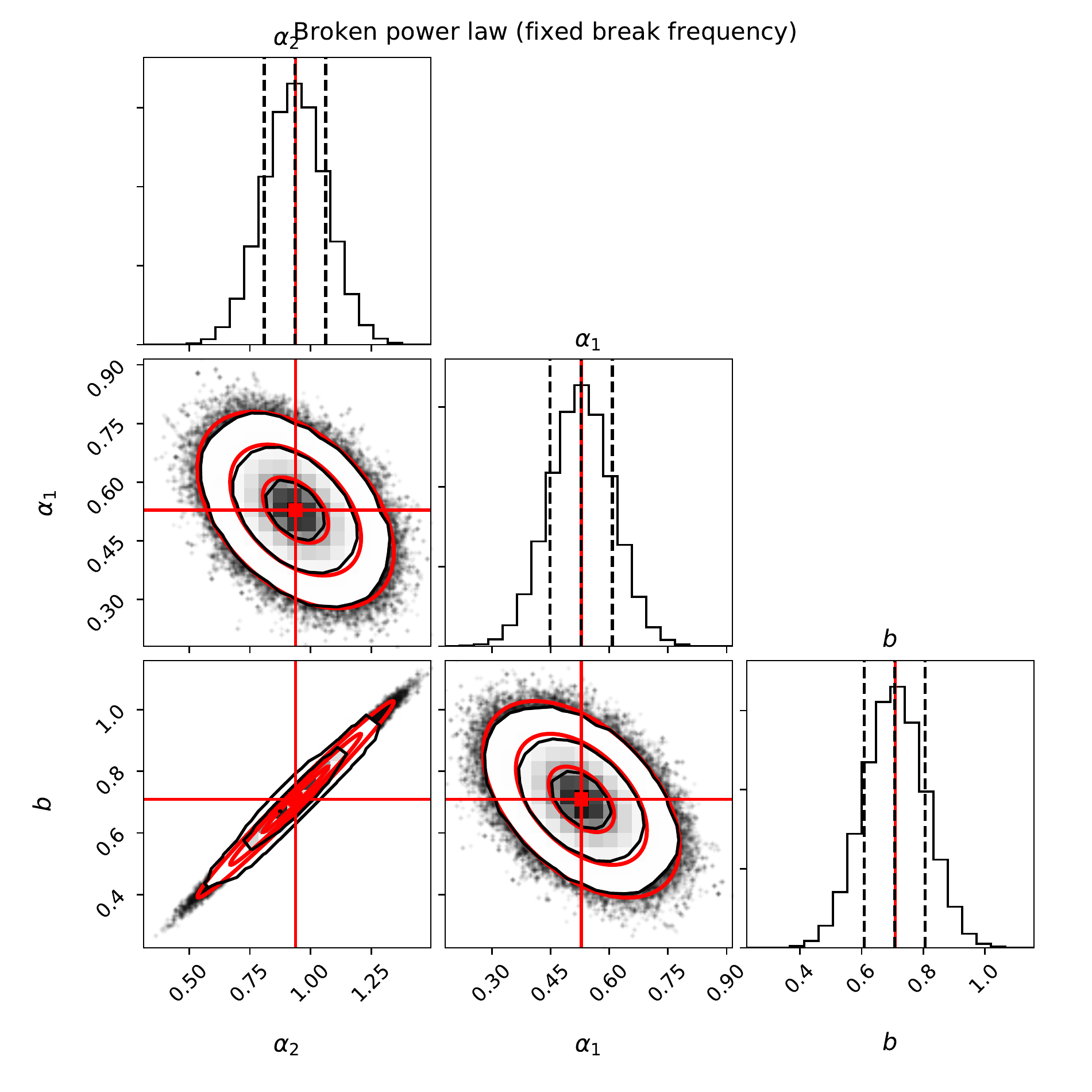}
     \hspace{1cm}
 \caption{Broken power-law parameter estimation based on the MCMC algorithm used in deriving the break frequency. The  red lines and ellipses show the results of the corresponding ODR fit, while the black contours show the 1, 2, and $3\sigma$ contours of the MCMC samples. The left panel shows the  four-parameter fit with the break frequency treated as a free parameter, while the right panel shows the three-parameter broken power-law fit derived by fixing the break frequency to the best-fit value of $\nu_b$ in the left panel. }\label{fig:MCMC3}
 
\centering \includegraphics[width=1.5\columnwidth]{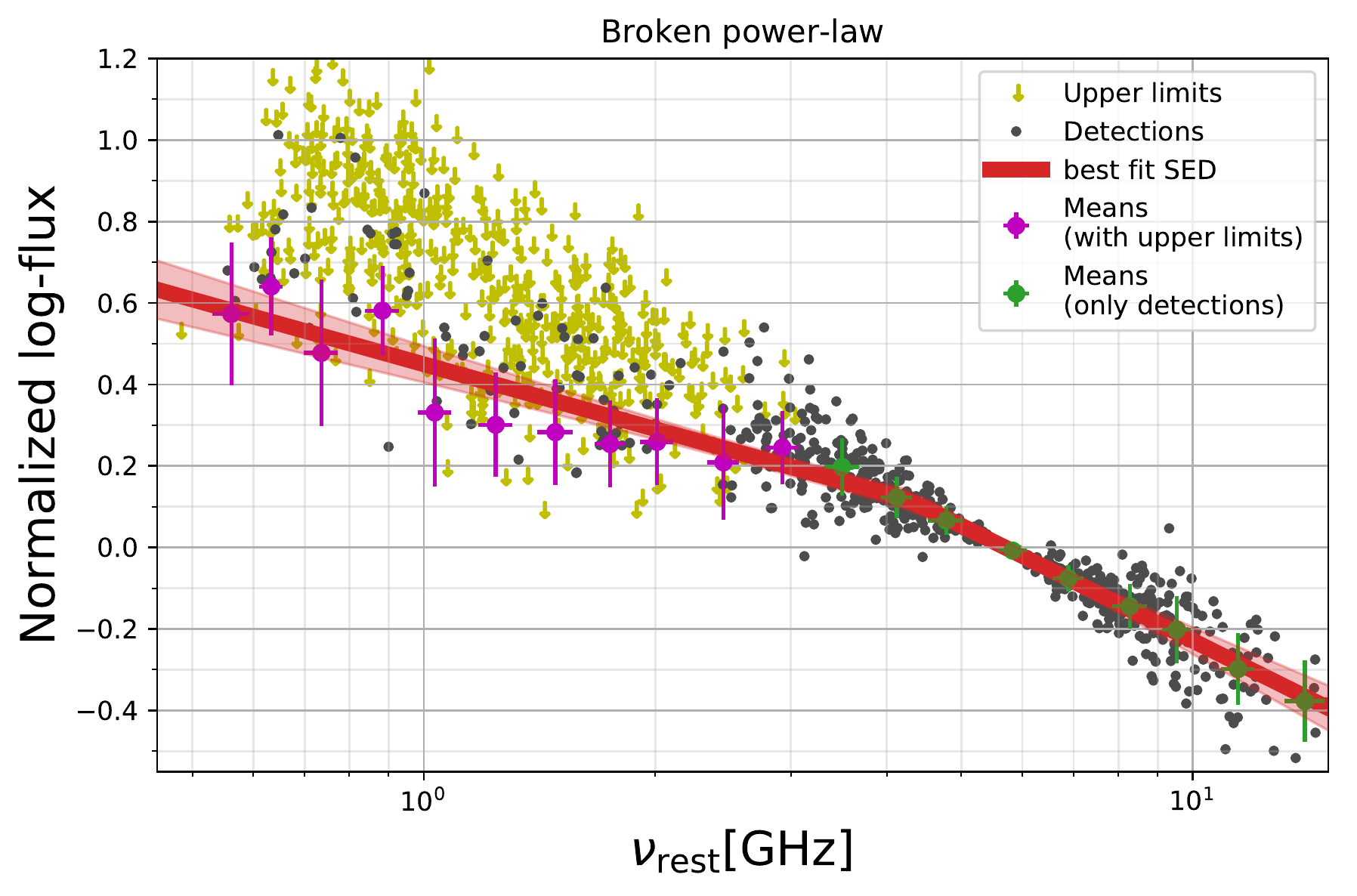}\caption{Average SED of the star-forming galaxy sample.
                        Gray data points show individual detections, yellow arrows show upper limits,
                green circles show means in bins with no upper limits, and magenta circles
                show mean values derived using survival analysis. 
                The red shaded interval shows the broken power-law fit with its $1\sigma$
                error confidence interval. {The errors derived using survival analysis are the standard deviations of the Weibull distribution that was fit on the Kaplan-Meier survival function.}
                }
                \label{fig:RealSED}
 
 \end{figure*}

\begin{figure*}
\centering \includegraphics[width=\textwidth]{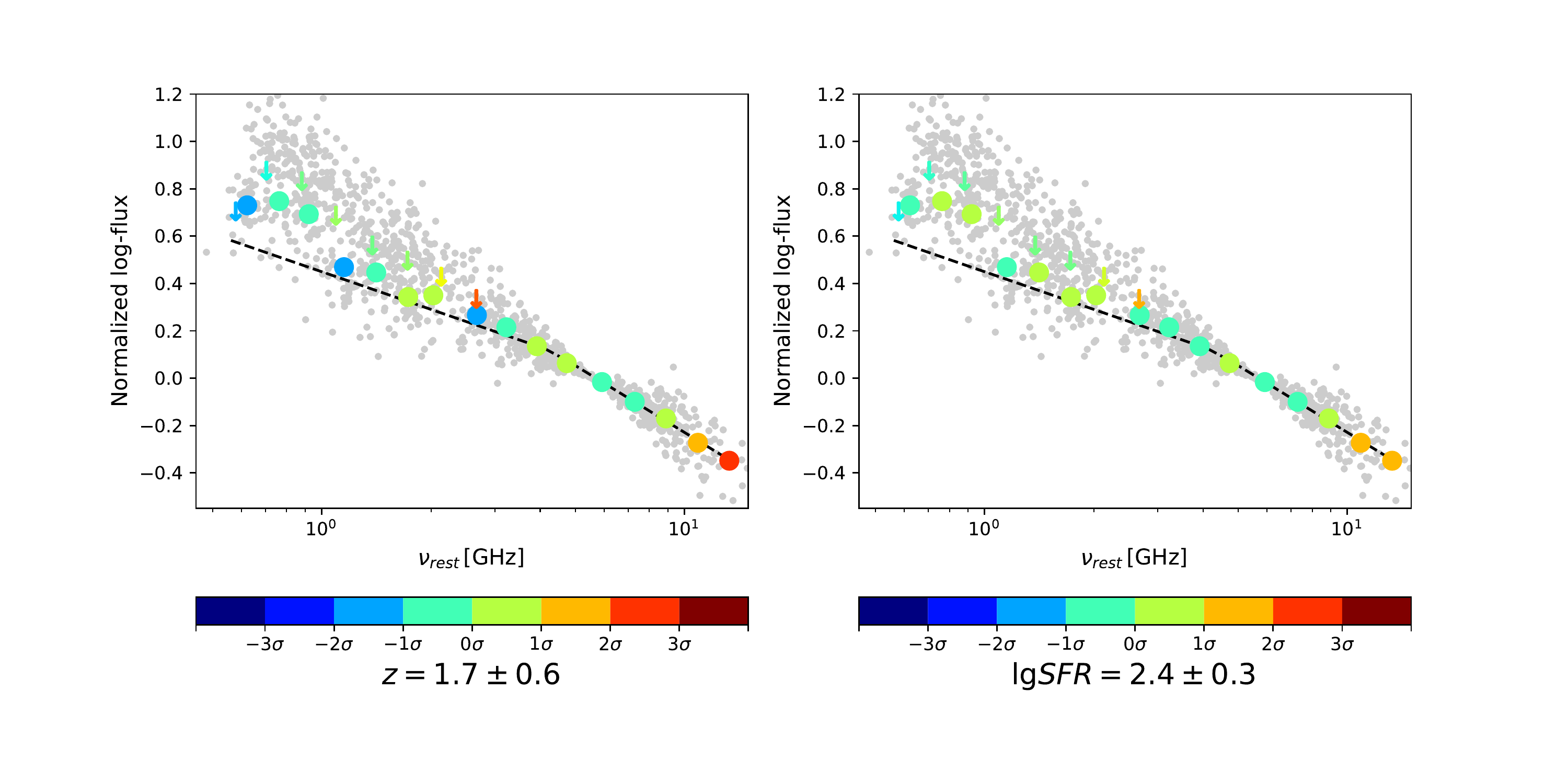}
\caption{Mean properties of various rest-frame frequency bins in the HSFG radio SED. Shown are the mean redshift (left panel) and $\log SFR$ values (right panel) for each bin. Colored circles show the means of detections, while down-pointing arrows show the mean values for non-detections in each redshift bin. The dashed line shows the best-fitting broken power-law SED.}\label{fig:properties}

\includegraphics[width=\columnwidth]{{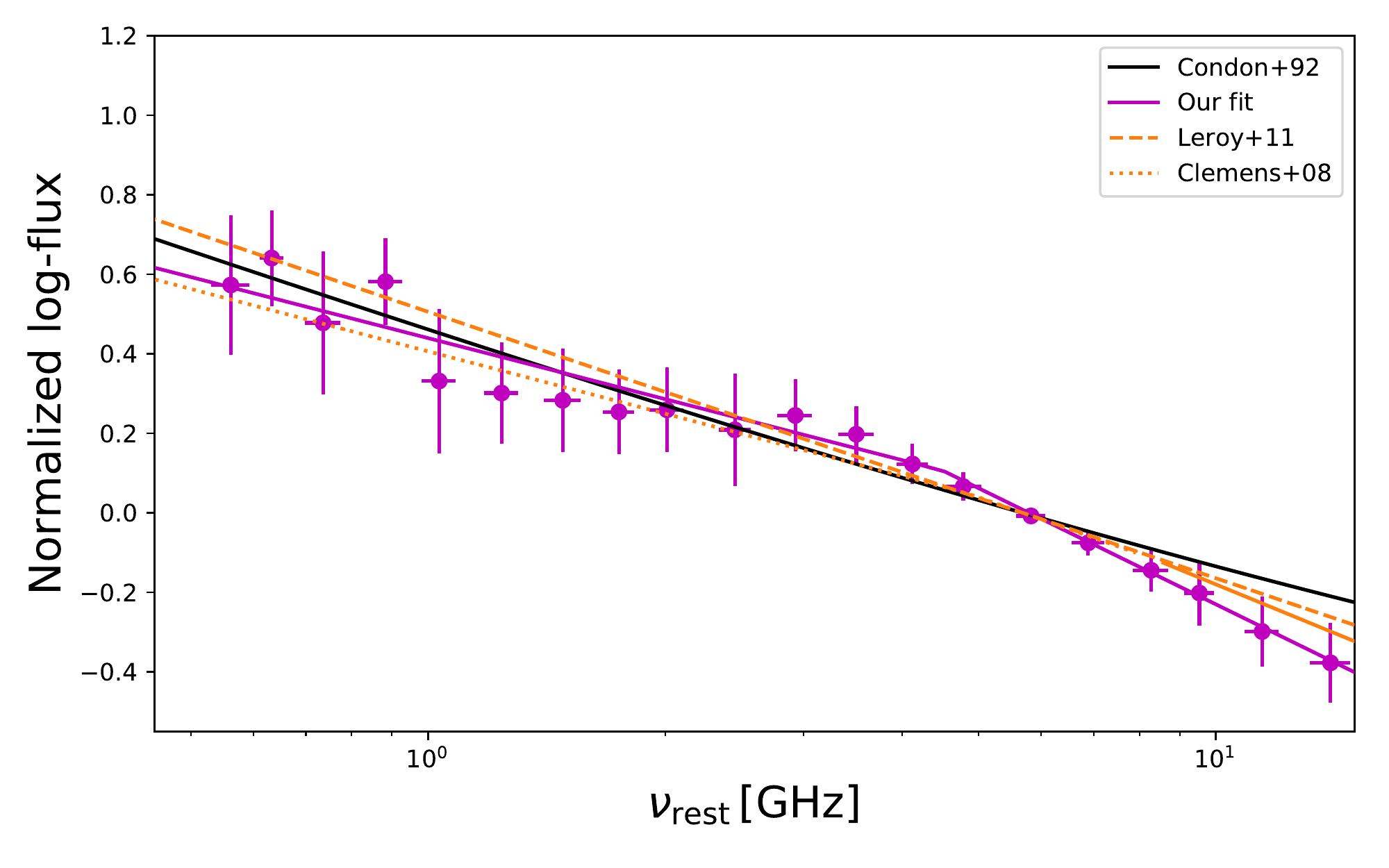}}
\includegraphics[width=\columnwidth]{{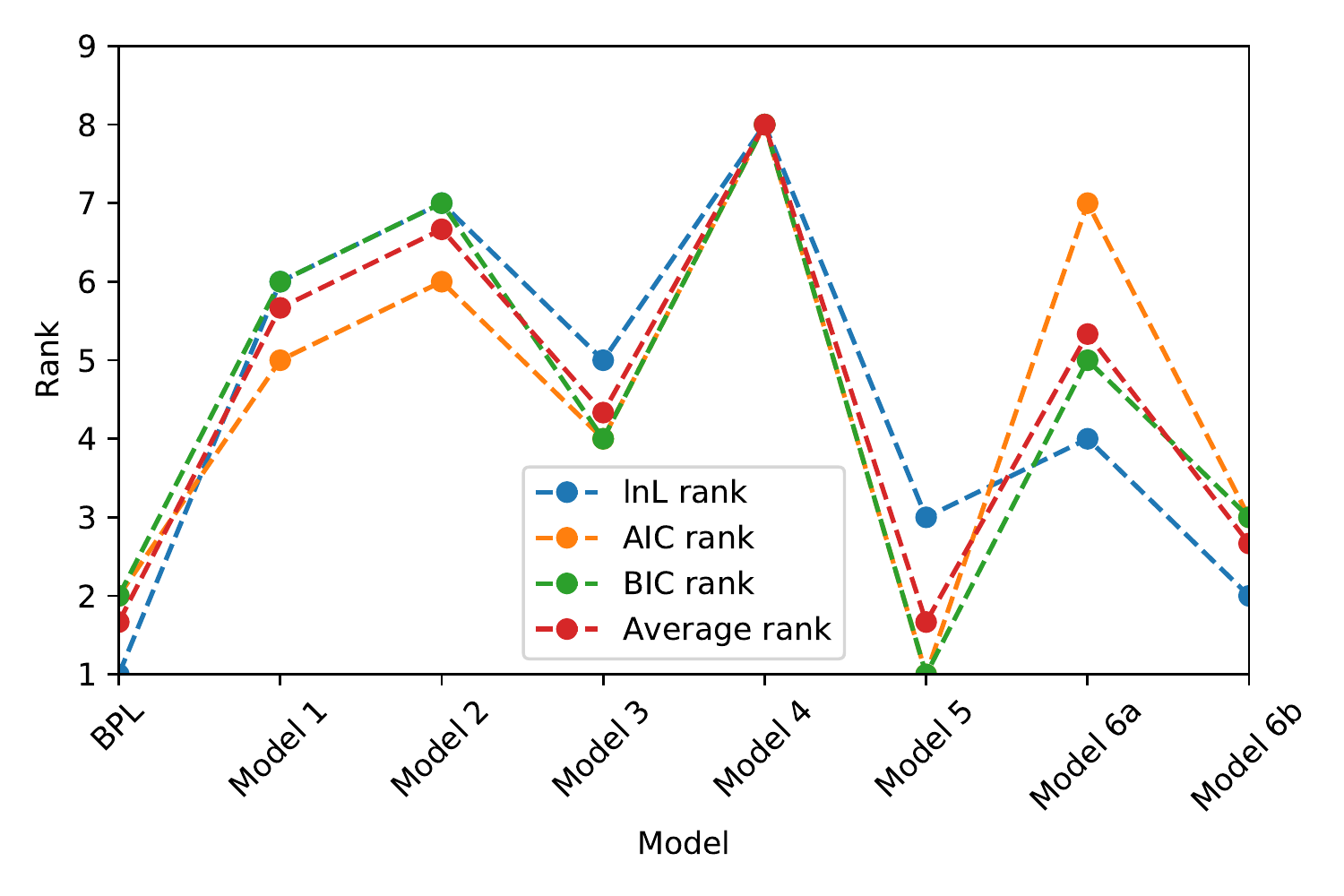}}
\caption{{Left panel:} Comparison of the average SED data (magenta points) and our broken power-law fit (magenta lines) with different literature SEDs \citep{Clemens08,Leroy11} for ULIRGs. For comparison, we show the \citet{Condon92} model for a normal SFG.
{Right panel:} Comparison of the different goodness-of-fit tests for different models. We show the log-likelihood test and the Akaike and Bayesian information criteria. We give the average of these three criteria for reference.}\label{fig:literatureandmodels}
\end{figure*}

We constructed the mock sample based on the distribution of  individual galaxy redshifts, $z$, a spectral index between $1.4$ and $3\GHz$, $\alpha_2$, and the log-flux normalization, $b$,  in the real HSFG sample. 
We have chosen to vary {the mean value of the} low-frequency spectral index, $\alpha_1$, to achieve different degrees of flattening in the simulations while keeping the number of galaxies in the simulation equal to the number of galaxies in the real sample.
In order to simulate the flattening of the spectral index below the break frequency, $\nu_b$, we assumed a broken power-law with a spectral index of $\alpha_1$ below $\nu_b$, with the following assumptions: 1) $\alpha_1$ has the same variance as $\alpha_2$ and 2) $\alpha_1$ is uncorrelated with $\alpha_2$, $z,$ and $b$.
Before creating the full mock catalog, we removed the sources in the mock $325\MHz$ and $610\MHz$ catalogs with fluxes below $5\sigma$, where $\sigma$ is the median RMS of the respective $325\MHz$ and $610\MHz$ COSMOS maps to simulate non-detections.

Overall, by analyzing the input ($\alpha_1^{in}$) and output  ($\alpha_1^{out}$) spectral index, we find that it is necessary to correct the derived spectral index by $20\percent$ (downward) for $\alpha_1^{out}\sim0.5$ and incrementally smaller correction for higher $\alpha_1^{out}$. The simulation suggests that the reason for this is the large number of upper limits occurring for lower spectral indices (i.e., flatter input slopes). The left panel of Fig. \eqref{fig:simulations} shows the dependence of the input spectral index of simulations, $\alpha_1^{in}$, on the output spectral index of simulations, $\alpha_1^{out}$, while the middle panel shows the dependence of the $\alpha_1^{in}/\alpha_1^{out}$ ratio on $\alpha_1^{out}$
Additionally, we show the behavior of the fraction of limits below $1\GHz$ as a function of $\alpha_1^{in}$ in the right panel of Fig. \eqref{fig:simulations}.
 For $\alpha_1^{out}>1$, this correction again rises to $20\percent$. Considering the results presented in the next section, this multiplicative correction is $0.8\pm0.1$ for the real dataset, as indicated by the red points in Fig. \eqref{fig:simulations}. Turning the argument around, under the assumption that $\alpha_1$ is normally distributed, the percentage of upper limits below $1\GHz$ ($88\percent$), observed in the real data,  indicates $\alpha_1\sim0.4$.

\section{Results}\label{sect:Results}
Here we present the results of the radio spectral index of the HSFG sample modeled by a broken power-law. 
In deriving the upper limits, we used the local RMS values from the GMRT maps.

The results of this procedure are shown in the left panel of Fig. \eqref{fig:MCMC3}, yielding a break frequency of $\nu_b=4.3\pm0.6\GHz$ (derived by the ODR method).
Outside the $2\sigma$ confidence interval (red ellipses  in the left panel of  Fig. \ref{fig:MCMC3}), the distribution starts to diverge from the simple form expected {by} the fit.
Furthermore, the MCMC-derived parameter medians in the left panel of  Fig. \eqref{fig:MCMC3} are not aligned with the ODR-derived means and the covariance ellipses  derived by ODR fitting diverge from the $\sigma$ contours of MCMC parameter estimates.
This divergence probably arises because the SED shape is not equally sensitive to the change of every parameter in the model, resulting in deviations in the MCMC method that are more pronounced outside the $2\sigma$ confidence interval.
Therefore, for the final result, we chose to fix the break frequency to the best-fit value provided by the ODR method and use it to fit the SED with a broken power-law with a fixed break frequency. 
The break frequency derived in this way is fixed to be  $\nu_b=4.3\GHz$ for all following considerations. We fixed the value of $\nu_b$ to the ODR-derived best-fitting value and not to the MCMC median because of the MCMC behavior outside the  $2\sigma$ confidence interval. We note that  these results using break frequency values in the range from $3-6\GHz$ are not expected to affect the presented results. This  is visible from the wide distribution of $\nu_b$ in the lower left plot in the left panel of Fig. \eqref{fig:MCMC3}. As presented in the right panel of Fig. \eqref{fig:MCMC3}, the reduced model behaves according to expectations of a normal distribution.

\begin{table*}
\centering\caption{Rankings of different fitted models with their parameters. For simplicity, we show models without their normalization parameter  and indicate the additional contribution of free-free emission with a thermal fraction at 1GHz by ‘FFE’. We show the rankings based on the Akaike and Bayesian information criteria and the log-likelihood test ($\ln L$). }\label{tab:ranking}
\scalebox{1.0}{%
\begin{tabular}{l c c c c}
\toprule
\toprule
Model& Normalized  & \multicolumn{3}{c}{Ranking}\\
parameters & (normalized) flux density & AIC & BIC &$\ln L$\\
\midrule
Broken power-law\tablefootmark{a} & \multirow{ 3}{*}{Eq. \eqref{eq:BPL}}& \multirow{3}{*}{2}&\multirow{3}{*}{2}&\multirow{3}{*}{1}\\
$ \alpha_1=0.53 \pm 0.04$ ($\alpha_1^{corr}=0.42\pm0.06$)\\
$ \alpha_2= 0.94 \pm 0.06$\\
\midrule
Model \eqref{m:MM} - mixed model\tablefootmark{b} & \multirow{ 3}{*}{$\nu^{-\alpha+2.1}\left(1-e^{-\frac{\tau_1}{\nu^{2.1}}}\right)$}& \multirow{3}{*}{6}& \multirow{3}{*}{5}& \multirow{3}{*}{6}\\
$ \alpha=0.74 \pm 0.06$ \\
$ \tau_{1}= 0.2 \pm 0.2$\\
\midrule
Model \eqref{m:FS} - foreground screen model\tablefootmark{b}& \multirow{ 3}{*}{$ \nu^{-\alpha} e^{-\frac{\tau_1}{\nu^{2.1}}}$}& \multirow{3}{*}{6}& \multirow{3}{*}{7}& \multirow{3}{*}{7}\\
$ \alpha=0.73 \pm 0.06$ \\
$ \tau_{1}=0.11 \pm 0.08$\\
\midrule
Model \eqref{m:SA} - synchrotron aging model\tablefootmark{b}&\multirow{ 3}{*}{$ \frac{\nu^{-\alpha}}{1+\left(\frac{\nu}{\nu_b}\right)^{\Delta\alpha}}$}& \multirow{3}{*}{4}& \multirow{3}{*}{4}& \multirow{3}{*}{5}\\
$ \alpha=0.24 \pm 0.06$\\
$ \Delta\alpha=0.55 \pm 0.04$\\
\midrule
Model \eqref{m:SSA} - synchrotron self-absorption model
\tablefootmark{c}& \multirow{ 3}{*}{$\nu^{5/2}\left(1-e^{-(\frac{\nu}{\nu_b})^{-\alpha-5/2}}\right)$}& \multirow{3}{*}{8}& \multirow{3}{*}{8}& \multirow{3}{*}{8}\\
$ \alpha=0.70 \pm 0.05$ \\
$ \nu_b=0.76 \pm 0.07\GHz$\\
\midrule
Model \eqref{m:CVM} - \citep{Lisenfeld04} \tablefootmark{b}&\multirow{ 2}{*}{$ \nu^{-\alpha}\left(1-\left(1-\sqrt{\frac{\nu}{\nu_b}}\right)^{1.1}\right)$}& \multirow{2}{*}{1}& \multirow{2}{*}{1}& \multirow{2}{*}{3}\\
$ \alpha=0.47 \pm 0.03$\\
\midrule
Model \eqref{m:SA+FFE} -synchrotron aging+FFE\tablefootmark{b}&\multirow{ 4}{*}{$ \frac{(1-f_{th}(\nu_n))\left(\frac{\nu}{\nu_n}\right)^{-\alpha}}{1+\left(\frac{\nu}{\nu_b}\right)^{\Delta\alpha}}+f_{th}(\nu_n)\left(\frac{\nu}{\nu_n}\right)^{-0.1}$}& \multirow{4}{*}{7}& \multirow{4}{*}{5}& \multirow{4}{*}{4}\\
$ \alpha=0.24 \pm 0.07$\\
$ \Delta\alpha=0.55 \pm 0.04$\\
$ f_{th}(1\GHz)=0.00 \pm 0.02$\\
\midrule
Model \eqref{m:CVM+FFE} - \citet{Lisenfeld04}+FFE\tablefootmark{b}&\multirow{ 3}{*}{$(1-f_{th}(\nu_n))\left(\frac{\nu}{\nu_n}\right)^{-\alpha}\left(1-\left(1-\sqrt{\frac{\nu}{\nu_b}}\right)^{1.1}\right)+f_{th}(\nu_n)\left(\frac{\nu}{\nu_n}\right)^{-0.1}$}& \multirow{3}{*}{3}& \multirow{3}{*}{3}& \multirow{3}{*}{2}\\
$ \alpha=0.50 \pm 0.06$\\
$ f_{th}(1\GHz)=0.01 \pm 0.02$\\
\bottomrule
\end{tabular}}

\tablefoottext{a}{The broken power-law model was included in the table for comparison.}
\tablefoottext{b}{See \citet{Pacholczyk80} for details.}
\tablefoottext{c}{See \citet{Condon16} for details.}\vspace{.1cm}
\end{table*}
{In Fig. \eqref{fig:RealSED} we show the average radio SED of HSFGs and the derived broken power-law SED with fixed $\nu_b$. 
The percentage of upper limits was $88\pm 2\percent$, which is consistent with $\alpha_1\sim0.5$ in Fig. \eqref{fig:simulations}.
The spectral index changes from $\alpha_1=0.53\pm 0.04$ below $\nu_b$ to $\alpha_2=0.94\pm0.06$ above $\nu_b.$ Considering the correction calculated using simulations in Sect. \ref{sect:Method verification via simulations}, the corrected spectral index is $\alpha_1^{corr}=0.42\pm0.06$. Some of the models described in Sect \ref{sect:The shape of the radio SED} can explain this difference, for example, the difference of 0.5 is expected by model \ref{m:SA} assuming continuous injection of electrons \citep{Pacholczyk80}. Although our data clearly do not follow a simple power-law, we also report the result of the simple power-law fit of $0.67\pm0.05$ to facilitate a quick comparison  to the literature.}

In the left and right panels of Fig. \eqref{fig:properties}, we compute mean values of redshift and star-formation rates for different rest-frame frequency bins of our radio SED, respectively.  
The mean redshifts and SFRs do not vary more {than one standard deviation of all the data points ($\sigma_z=0.3$, $\sigma_{\log \SFR}=0.6$)} below $10\GHz,$ and there is no clear trend between contiguous frequency bins below this frequency. Only a small fraction of the data points, corresponding to $z>2.3$, lie above $10\GHz,$ and they could not influence the SED fitting procedure significantly. We therefore conclude that our average SED represents the SED of a single population of highly star-forming galaxies (with properties $z=1.7\pm0.6$ and $\log SFR=2.4\pm0.3$) over the whole frequency range from $300\MHz$ to $10\GHz$.

\section{Discussion}\label{sect:Discussion}
 In Sect. \ref{sect:The shape of the radio SED} we discuss the shape of the radio SED of star-forming galaxies and alternative, physically motivated  fitting strategies. In Sect. \ref{sect:Infrared-radio correlation} we discuss the impact of the shape of the radio SED on the infrared-radio correlation. 
\subsection{Shape of the radio SED}\label{sect:The shape of the radio SED}
In the left panel of Fig. \eqref{fig:literatureandmodels}, we overplot the different SEDs from the literature and our broken power-law (BPL). 
Our results, covering a broad range of redshifts, are in overall agreement with literature SEDs derived for (U)LIRGS in the nearby Universe ($z\lessapprox0.1,\,\SFR>10\Msun/\yr$)  \citep[for galaxies with an $\SFR$ in the $50-150\Msun/\yr$ range ][]{Clemens08,Leroy11}. 
We find disagreement with the typical model of ``normal'' star-forming galaxies {(NSFGs)} above $\sim 10\GHz$  \citep{Klein88, Condon92,Tabatabaei17}, which may be due to the different thermal fractions in normal and highly star-forming galaxies.
In the nearby Universe ($z<0.1$), the typical SED of NSFGs is described by a simple nonthermal power-law with a spectral index of $\sim 0.8$ and a thermal fraction of $8-10\percent$  at $\sim 1 \GHz$ \citep{Klein88, Condon90}, while studies of HSFGs  find a flat spectrum around $\sim1\GHz$ and a steepening spectrum above $10\GHz$  \citep{Clemens08,Leroy11}.

For our HSFGs with $\SFR>100\Msun/\yr$, we find a relatively flat ($\alpha\sim0.5$) spectral index below $\sim 4\GHz$. 
This result is in line with previous studies of (U)LIRGs \citep{Clemens08,Leroy11,Galvin17}.
There are multiple effects that could produce a flattening in the spectrum of galaxies in the radio, but only synchrotron aging and free-free absorption play a role in the $1-10\GHz$ range.
Effects like synchrotron self-absorption and  the Razin-Tsytovi\v{c} effect \citep{Razin60,Tsytovich51}, an effect that is due to the refractive index of plasma in the radio, can be ruled out because they are usually found at frequencies below $1\GHz$  \citep{Deeg93}. 
These effects could be even more suppressed, as magnetic fields have been found to be stronger for higher SFRs \citep{Tabatabaei17}.
Furthermore, we do not expect to be able to successfully find evidence for any more complex shape of the SED such as multi-peak behavior of free-free absorption because of the different opacities of different star-forming regions within a galaxy, as any such behavior would be averaged out when constructing an average SED.
\citet{Tabatabaei17} instead pointed out that the nonthermal spectral index (which can be compared to our $\alpha_1$) flattens with increasing  SFR density, which gives a possible explanation why this effect is pronounced in our HSFG sample. 

\citet{Bressan02} modeled the radio SEDs of dusty starburst galaxies that initially have $\SFR$s in the range of $100-1000\Msun/\yr$. They concluded that if the SFR changes abruptly, the core-collapse supernova rate will lag behind as a result of the finite lifetimes of stars, leading to a suppressed synchrotron emission in young starbursts. This effect can produce flat spectra below $10\GHz$, as the thermal fraction is initially very high. 
\citet{Prouton04} analyzed the behavior of the thermal fraction with the age of the starburst for galaxies with an $\SFR$s from $30$ to $\sim 150\Msun/\yr$. They found that the thermal fraction decreases from the initial value of almost $100\percent$ to below $10\percent$ at $1.4\GHz$ within $~10^{7.5}\yr$.

For the HSFGs in the COSMOS field, we find  a steep ($\alpha\sim1$) spectrum above $4\GHz$ that does not exhibit signs of free-free emission.
A steep spectral index has also been found by \citet{Galvin17} ($\alpha=1.06$) for galaxies with an $\SFR$ in the range of $\sim 50-250\Msun/\yr$, which they associated with a steeper injection index of the CR distribution. 
\citet{Clemens08}  found a similar trend (for galaxies with an $\SFR\sim 50-150\Msun/\yr$ ) that they tried to explain  using the continuous injection model, which can explain differences in spectral indices up to  $\Delta\alpha\approx0.5$.

Since the SED of HSFGs has been found to differ from a simple power-law, we fit the following models often used in literature \citep[e.g.,][]{Calzetti94,Condon92,Clemens08, Condon16, Lisenfeld04, Klein18} that might explain this behavior:
\begin{enumerate}
\item \label{m:MM} Mixed model\begin{itemize}
\item This model assumes that the power-law synchrotron spectrum with a spectral index $\alpha$ is attenuated by free-free absorption with opacity $\tau_{\nu}=\frac{\tau_1}{\nu^{2.1}}$ arising from the same region. 
\end{itemize}
\item \label{m:FS}Foreground screen model\begin{itemize}
\item This model assumes that the power-law synchrotron spectrum with a spectral index $\alpha$ is attenuated by free-free absorption with opacity $\tau_{\nu}=\frac{\tau_1}{\nu^{2.1}}$ arising from a region that is in our line of sight, but is not cospatial with the emitting region. 
\end{itemize}
\item \label{m:SA}Synchrotron aging model\begin{itemize}
\item In contrast to the previous two models, this model does not take into account free-free absorption of synchrotron radiation initially described by a simple power-law. This model  instead explains the shape of the SED by  taking into account that the cosmic ray electron population loses energy with time. The model predicts a spectrum that was initially described by an injection index that produces a power-law  SED with a spectral index $\alpha$ that gradually steepens above a frequency $\nu_b$ to a spectral index $\alpha+\Delta\alpha$ due to aging.
\end{itemize}
\item \label{m:SSA} Synchrotron self-absorption\begin{itemize}
\item The synchrotron self-absorption model has the same form as model \ref{m:MM}, with the opacity of the form $\tau_\nu=(\nu/\nu_b)^{-(\alpha+5/2)}$, where $\alpha$ is the synchrotron spectral index and $\nu_b$ the break frequency. 
\end{itemize}
\item \label{m:CVM} \citet{Lisenfeld04} model \begin{itemize}
\item In this model, the electrons are propagated through and eventually out of the galactic halo by means of convective winds.
\end{itemize}
\item Free-free emission\begin{itemize}
\item Models \ref{m:MM}-\ref{m:CVM} can be extended by including an additional contribution of thermal free-free emission, $f_{th}$, as described by Eq. \eqref{eq:SEDfull}. We consider the following: 
\item \begin{enumerate}
\item \label{m:SA+FFE} Synchrotron aging model with free-free emission
\item \label{m:CVM+FFE} \citet{Lisenfeld04} model with free-free emission
\end{enumerate}
\end{itemize}
\end{enumerate}
Among the models listed above, free-free absorption and synchrotron aging models would reproduce flattening at lower frequencies,  while models \eqref{m:SA+FFE}-\eqref{m:CVM+FFE} can be used to estimate the contribution of free-free emission to the spectrum.
The parameterization of the fitted SEDs and their parameters can be found in Table \eqref{tab:ranking}. 
We find that models \eqref{m:MM} and \eqref{m:FS}, which  include free-free absorption,  fit the low-frequency data well. 
They overestimate the measured flux at higher frequencies (by $\sim0.1\,\mathrm{dex}$ at $10\GHz$), however.
The synchrotron aging model \eqref{m:SA} agrees with the broken power-law at both low and high frequencies. 
The low-frequency spectral index corresponds to the value typically found in supernova remnants (SNRs) and the Galactic plane in the Milky Way \citep{Green14,Klein18}.
We tried to constrain the thermal fraction with the broken power-law model by using Eq. \eqref{eq:SEDfull}, but found a strong degeneracy between parameters $f_{th}$ and $\alpha_2$ and therefore cannot draw meaningful results from this model. 
This degeneracy is expected for this class of models \citep{Transtrum10} and can only be broken by more observations at even higher frequency ($>10\GHz$).

To compare how well these models explain our dataset, we used the log-likelihood test and the Akaike and Bayesian information criteria (AIC and BIC, respectively).
The greatest log likelihood means that the broken power-law is optimal in terms of the fitting. 
 In contrast to the log-likelihood test, AIC and BIC criteria favor models with a smaller number of free parameters, penalizing the models with a higher number of free parameters by higher AIC and BIC values.
In the right panel of Fig. \eqref{fig:literatureandmodels} we show the ranking in our statistical analyses (log likelihood, AIC, and BIC) for the models in Table \eqref{tab:ranking}. The results reported in Table \eqref{tab:ranking} were computed using the ODR method restricted to a $2\sigma$ subset of the respective model's parameter space, derived using the MCMC procedure.

We found that our broken power-law model has the greatest log likelihood and  the second lowest AIC and BIC values. The  BPL model shares the same average rank of the AIC, BIC, and log-likelihood ranks with model \eqref{m:CVM} because model \eqref{m:CVM} has fewer free parameters and is therefore favored by the AIC and BIC values, while it ranks third on the log-likelihood test. As shown in Fig. \ref{fig:modelSEDs}, the broken power-law model and models \eqref{m:SA}, \eqref{m:CVM}, \eqref{m:SA+FFE}, and \eqref{m:CVM+FFE} describe the average radio SED points sufficiently well. We therefore conclude that we can favor models ranking the lowest on all three tests. In addition to the broken power-law, the highest ranking models are model \eqref{m:CVM} and its extension that includes free-free emission (model \ref{m:CVM+FFE}). Additionally, literature models of HSFGs \citep{Clemens08,Leroy11} score better in all three tests than the standard NSFG model \citep{Condon92}. 
This comparison means that HSFGs do not have the same SED shape as the NSFGs. The main difference to the NSFGs is the behavior at higher frequencies, which might be due to a smaller thermal fraction at higher SFR.
Since our empirical broken power-law describes our dataset better than simple theoretical models (models \ref{m:MM}-\ref{m:SA}), and different goodness-of-fit criteria sort their preference differently, we conclude that neither simple free-free absorption over the whole sample nor synchrotron aging are the only drivers of the radio SED of HSFGs.

\begin{figure}
\includegraphics[width=\columnwidth]{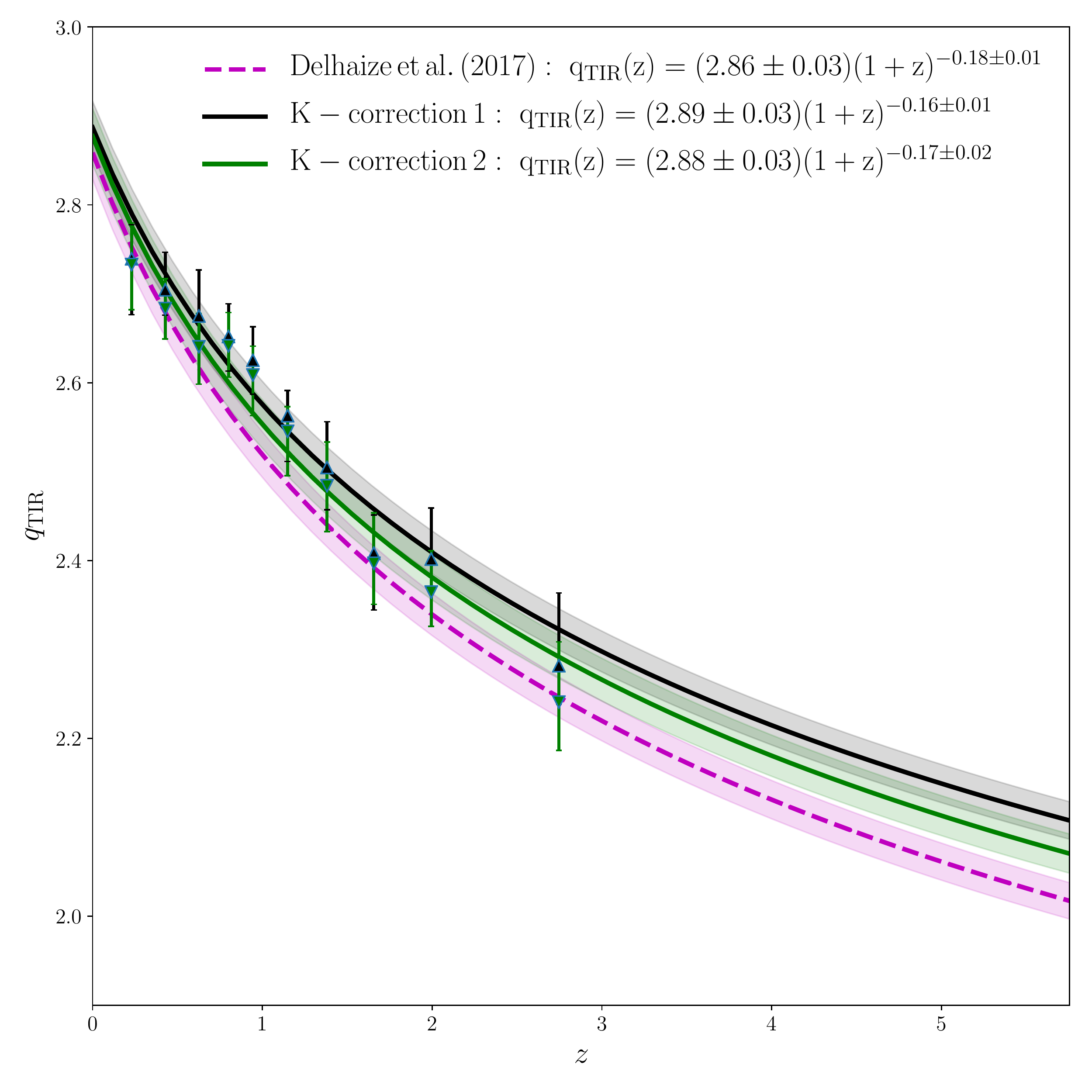}
\caption{Infrared-radio correlation, $q_{TIR}$, as a function of redshift for galaxies in the \citet{Delhaize17} joint radio and infrared sample. We used the broken power-law SED from this paper for star-forming galaxies with an $SFR>10\Msun/\yr$ and the following SED shapes  for $SFR<10\Msun/\yr$ \citep{Tabatabaei17}: 1) a simple power-law SED with a spectral index of $0.79$ (black line, K-correction 1) or  a nonthermal spectral index of $0.97$, and a $10\percent$ thermal fraction at $1\GHz$ (green line, K-correction 2). We show the \citet{Delhaize17} relation for reference (magenta line).}\label{fig:qz}
\end{figure}

\begin{figure*}
\centering\begin{minipage}{.5\textwidth}
\includegraphics[width=\columnwidth]{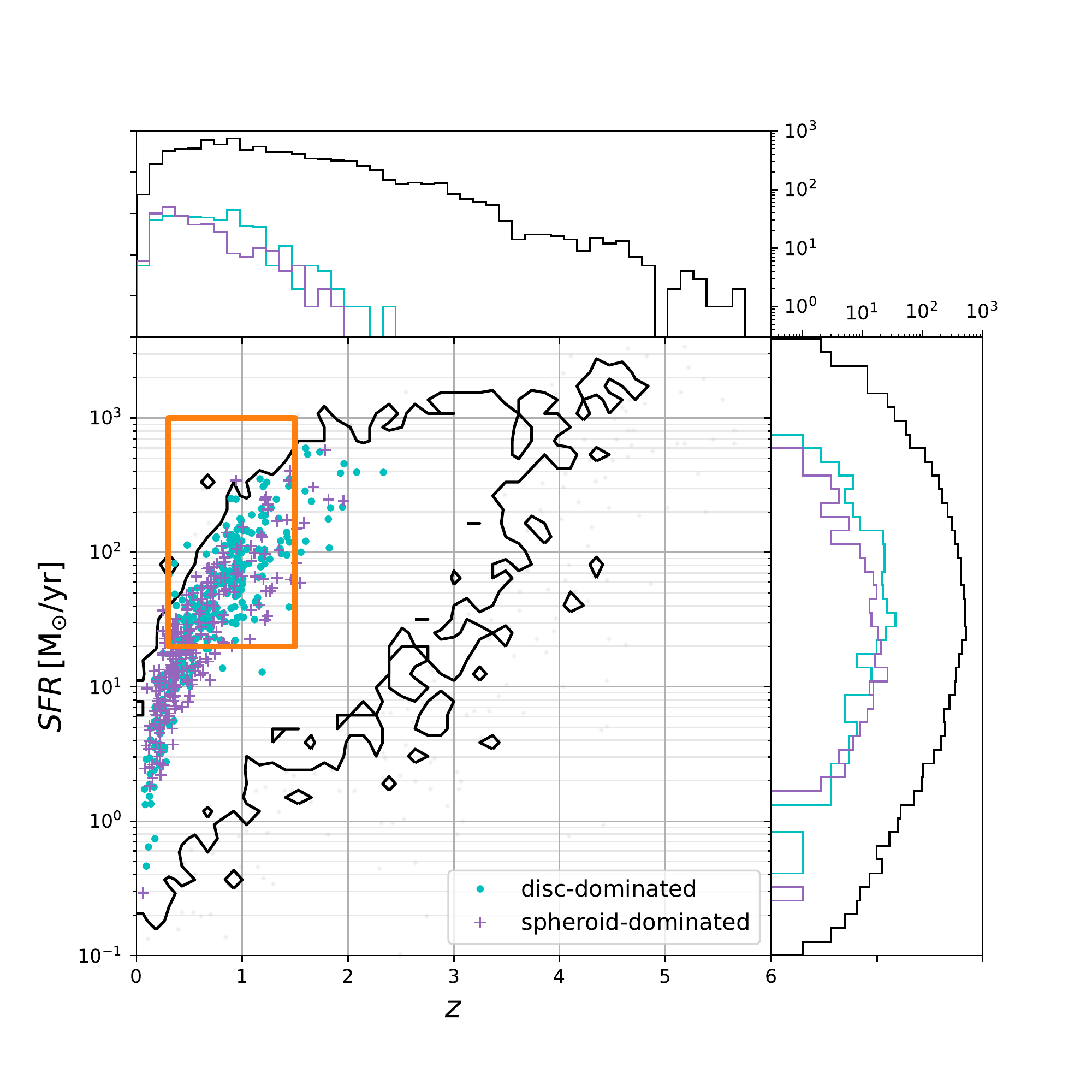}
\end{minipage}\hfill
\begin{minipage}{.5\textwidth}
\includegraphics[width=\columnwidth]{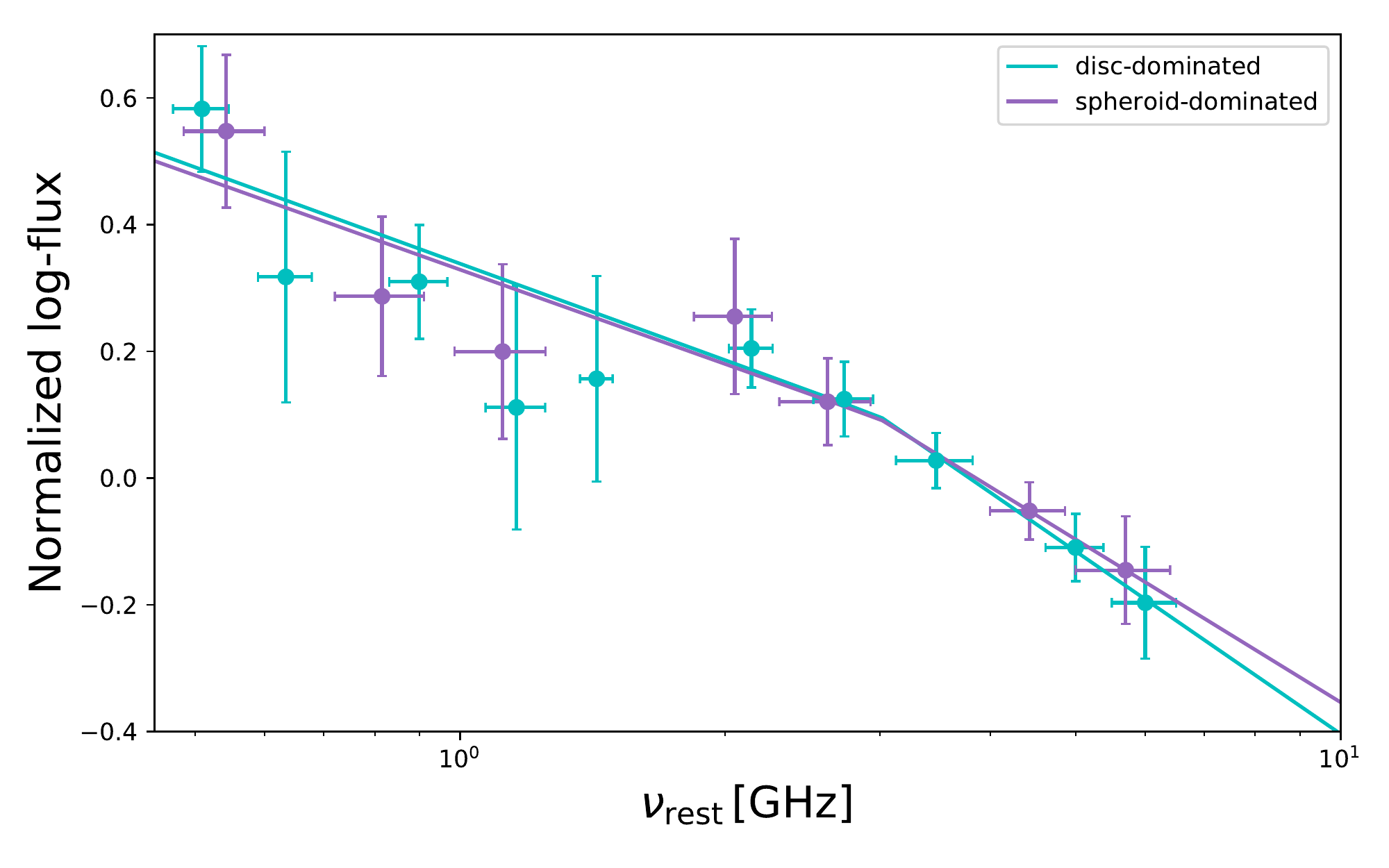}
\end{minipage}\hfill
\caption{Left panel: Distribution of sources in the $SFR-z$ plane, analogous to Fig. \eqref{fig:SFRz}. Cyan points and magenta crosses indicate the distributions of disk- and spheroid-dominated subsamples of star-forming galaxies, as defined in \citet{Molnar18}. The $3\sigma$ contours of the CSFG sample are shown as black lines, and the orange rectangle marks the edges of the chosen nearly complete disk- and spheroid-dominated samples.  The right panel shows their respective SEDs.}\label{fig:Daniel}
\end{figure*}

\subsection{Infrared-radio correlation}\label{sect:Infrared-radio correlation}
 Recent stacking and survival analysis studies \citep{Ivison10,Magnelli15, Delhaize17, CalistroRivera17} have found the infrared-radio correlation (described by the $q$ parameter) to decrease with increasing redshift.
\citet{Delhaize17}  mentioned the possibility that the computation of rest-frame radio luminosity via a K-correction using a  single power-law assumption of the star-forming galaxies' SED could produce a redshift-dependent trend.

Here we calculate the infrared-radio correlation by using a K-correction based on our best-fitting broken power-law SED  for galaxies with an $SFR>10\Msun/\yr$ and the combination of thermal and nonthermal emission, following Eq. \eqref{eq:SEDfull}, with a nonthermal spectral index of $\alpha=0.97$ and a $10\percent$ thermal fraction  \citep{Tabatabaei17}  for $SFR<10\Msun/\yr$ (described in detail in Appendix B). 
We adopted a threshold between these two types of SFGs at $10\Msun/\yr$, considering that our results are consistent with studies of (U)LIRGs \citep[$SFR>10\Msun/\yr$][]{Clemens08,Leroy11}. 
We find that the broken power-law derived luminosity is $20\percent$ lower than that derived by assuming a $0.7$ spectral index for redshifts below $z\sim 1$.  
We also find that the difference is negligible for sources above $z\gtrsim2$.
We constrained the infrared-radio correlation using the same procedure as applied by \citet{Delhaize17}. In the infrared, we used Herschel-detected $\geq 5\sigma$ objects, with the COSMOS2015 \citep{Laigle16} $24\mu m$ catalog as priors to minimize blending, while in the radio, we used the \citet{Smolcic:17b} $3\GHz$ Large Project Counterpart catalog. If an object is undetected in the radio, we estimated its upper $q$ limit as $\text{five}$ times the local RMS value in the $3\GHz$ map {convolved to a resolution at which its peak flux ceases to change significantly with increased convolution}, while we used lower $q$ limits based on upper limits to the  infrared luminosity, derived through integration of the best-fit SED. For details, see \citet{Delhaize17}. In short, we used the joint radio- and infrared-detected samples of star-forming galaxies in the COSMOS field, and then used the survival analysis method to find the mean $q$ value for different redshift bins and fit a power law of the form
\begin{equation}
 q(z)=q_0(1+z)^n,
 \end{equation} 
where $q_0$ is the value of $q$ derived for $z=0$ and $n$ is the exponent. 
Contrary to what would be expected from theory \citep{Murphy09}, but in line with previous studies \citep{Delhaize17,CalistroRivera17}, the $q$ value, shown in Fig. \eqref{fig:qz} and computed taking into account the K-corrections described above, decreases with redshift with $n=-0.17\pm0.02$, and $q_0=2.88\pm0.03$.  In Fig. \eqref{fig:qz} we show the $q-z$ trend derived under various SED assumptions, as described above. 
In all the cases, we find that $q$ decreases with increasing redshift. 
We therefore conclude that the shape of the radio SED, as assumed here, probably does not explain the $q-z$ trend. 
For comparison, in Fig. \eqref{fig:qz}, we also show the resulting $q-z$ trend based on a simple power-law fit  with $\alpha=0.79$ for galaxies with an $\SFR<10\Msun/\yr$ \citep{Tabatabaei17}. This fit is less complex, but has a lower parameter uncertainty. 
We find that this simplification does not significantly alter the resulting $q-z$ trend ($n=-0.18\pm0.01$ and $q_0=2.89\pm0.03$.).
 
\citet{Molnar18} have studied the impact of morphology on the evolution of the $q$ parameter (for star-forming galaxies covering a broad $\SFR$ range of $1-1000\Msun/\yr$ {and in the redshift range of $0.2-1.5$}) and found that spheroid-dominated {star-forming} galaxies show a steep redshift behavior ($n=-0.19\pm0.02$), while disk-dominated {galaxies} do not ($n=-0.04\pm0.01$). 
They argued that in order for the radio spectral index of spheroid-dominated galaxies to explain the $q-z$ behavior, it would have to decrease with redshift from $\alpha=0.7$ at $z=0.8$ to $\alpha=0.45$ at $z=1.5$. 
To investigate this, we have derived an average radio SED for complete samples of star-forming galaxies based on morphology. 
To match the selection performed by \citet{Molnar18}, we selected SFGs with an $SFR>20\Msun/\yr$ to be fairly complete out to a redshift of $z=1.5$, with the requirement that we included only galaxies marked as either spheroid- or disk-dominated, based on the classification of \citet{Molnar18}.
We show the position {in the $\SFR-z$ plane } of their respective complete subsets in the left panel of Fig. \eqref{fig:Daniel} and their respective broken power-law fits in the right panel. Overall, we find that the median MCMC-derived break frequency is $3\GHz$  and that the differences in the parameters of the broken power-law fit between spheroid  ($\alpha_2=0.9\pm0.3$, $\alpha_1=0.5\pm0.1$) and disk-dominated  ($\alpha_2=1\pm0.2$, $\alpha_1=0.51\pm0.09$) galaxies are smaller than $1\sigma$. 
{This suggests that a difference in radio-SEDs of these morphologically different galaxies, and thus a potentially different K-correction, is not the cause of the differing trend found by \citet{Molnar18}.}

\section{Summary}\label{sect:Summary}
We constructed the average radio spectral energy distribution for a $1.4\GHz$-selected sample of highly star-forming galaxies with an $\SFR>100\Msun/\yr$ in the COSMOS field. To achieve a broad rest-frame frequency range, we combined previously published VLA observations at $1.4\GHz$ and $3\GHz$ with unpublished GMRT observations at $325\MHz$ and $610\MHz$ by employing survival analysis to account for non-detections in the GMRT maps caused by their higher RMS values.
By fitting a broken power-law to the SED, we find the spectral index to change from $\alpha_1=0.42\pm0.06$ below $4.3\GHz$ to  $\alpha_2=0.94\pm0.06$ above $4.3\GHz$. 
Our results are in line with previous low-redshift studies of star-forming galaxies with an $\SFR>10\Msun/\yr$ that show that their SED differs from the one found in normal star-forming galaxies by {having} a steeper spectral index around $10\GHz$, which could imply a smaller thermal fraction than in normal star-forming galaxies.

We further constructed the IR-radio correlation by using our broken power-law SED for galaxies with an $\SFR>10\Msun/\yr$ and  an SED based on a steep nonthermal synchrotron spectral index ($\alpha=0.97$) and a $10\percent$ thermal fraction at $1.4\GHz$ \citep{Tabatabaei17} for galaxies with $\SFR<10\Msun/\yr$. We find that the shape of the radio-SED is probably not the root cause of the redshift trend found by previous studies, as we see a clear trend of the IR-radio correlation decreasing with increasing redshift {also} when a more sophisticated K-correction is applied to the data.
\section*{Acknowledgements}
This research was funded by the European Union's Seventh Frame-work program under grant agreement 337595 (ERC Starting Grant, `CoSMass').

EV  acknowledges  funding  from  the  DFG  grant  BE1837/13-1.   EV  also acknowledges  support  of  the  Collaborative  Research  Center  956,  subproject  A1  and  C4, funded  by  the  Deutsche  Forschungsgemeinschaft  (DFG).  JD acknowledges financial assistance from the South African SKA Project (SKA SA; \url{www.ska.ac.za}).
\bibliography{lit}
\begin{table*}
\centering
\caption{Summary of data reduction results of the GMRT maps. Columns  $b_{maj}\,\mathrm{[arcsec]},$ $ b_{min}\,\mathrm{[arcsec]},\text{and}$  $BPA\,\mathrm{[^\circ]}$  show the beam major and minor axes and the beam position angle, respectively. Column \emph{Sources} shows the number of sources that are detected toward the COSMOS field, column \emph{Resolved} shows the number of resolved sources, column \emph{$FDR$} shows the false-detection rate, column \emph{Cross-matched} shows the number of sources that are present in the respective GMRT catalog and in the VLA $3\GHz$ catalog, and column \emph{HSFG Sample} shows the number of sources in the subset of the GMRT catalog that is present in the highly star-forming galaxy sample described in Sect. \ref{sect:Sample}. 
Percentages in the table are derived relative to the column \emph{Sources}.}\label{tab:GMRTInfo}
\scalebox{1.0}{%
\begin{tabular}{c c c c c c c c c}
\hline
\hline
Map & $b_{maj}\,\mathrm{[arcsec]}$ &$ b_{min}\,\mathrm{[arcsec]}$ & $BPA\,\mathrm{[^\circ]}$ & Sources & Resolved  & $FDR$   & Cross-matched & HSFG Sample  \\
\hline
\rule{0pt}{0.25cm}    
$325\MHz$ & $10.8$ & $9.5$ & $-62.8$& $634$ & $177$ ($28.0\percent$) & $3.0 \percent$ & $567$ ($89.6 \percent$) &  29\\
$610\MHz$ &  $5.6$ & $3.9$ & $65.5$ & $999$ &  $196$ ($ 19.6 \percent$) & $1.3 \percent$  &  $ 870$ ($87.1 \percent$)  & 52\\
\hline
\end{tabular}}
\end{table*}

\appendix

\section{GMRT data reduction}\label{sect:GMRT data reduction}
Here we describe the calibration, imaging, source extraction, and validation (including all of the corrections applied) of the GMRT 325 and 610 MHz data.
\begin{figure*}[ht]
\centering 
\includegraphics[width=.3\textwidth]{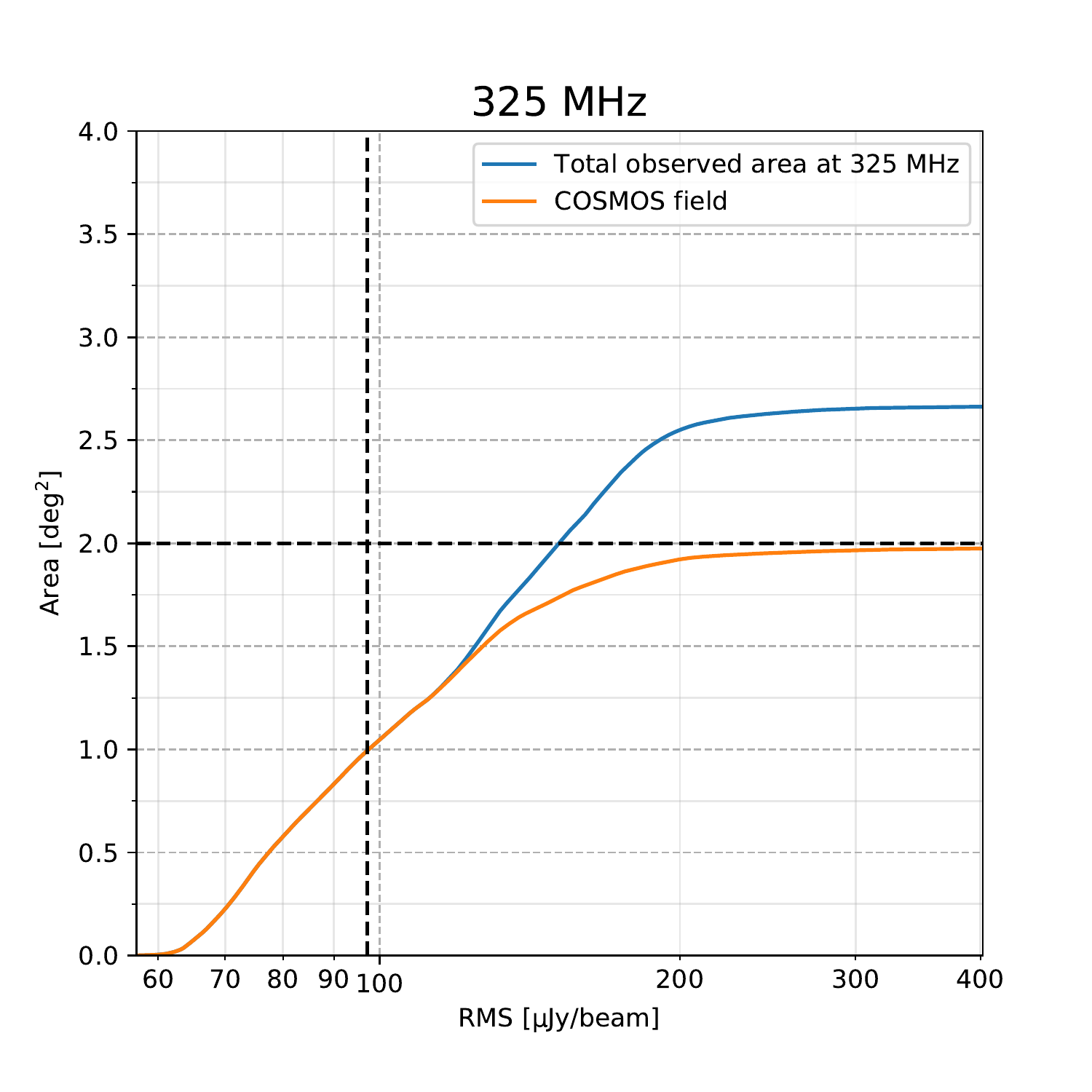}
\includegraphics[width=.3\textwidth]{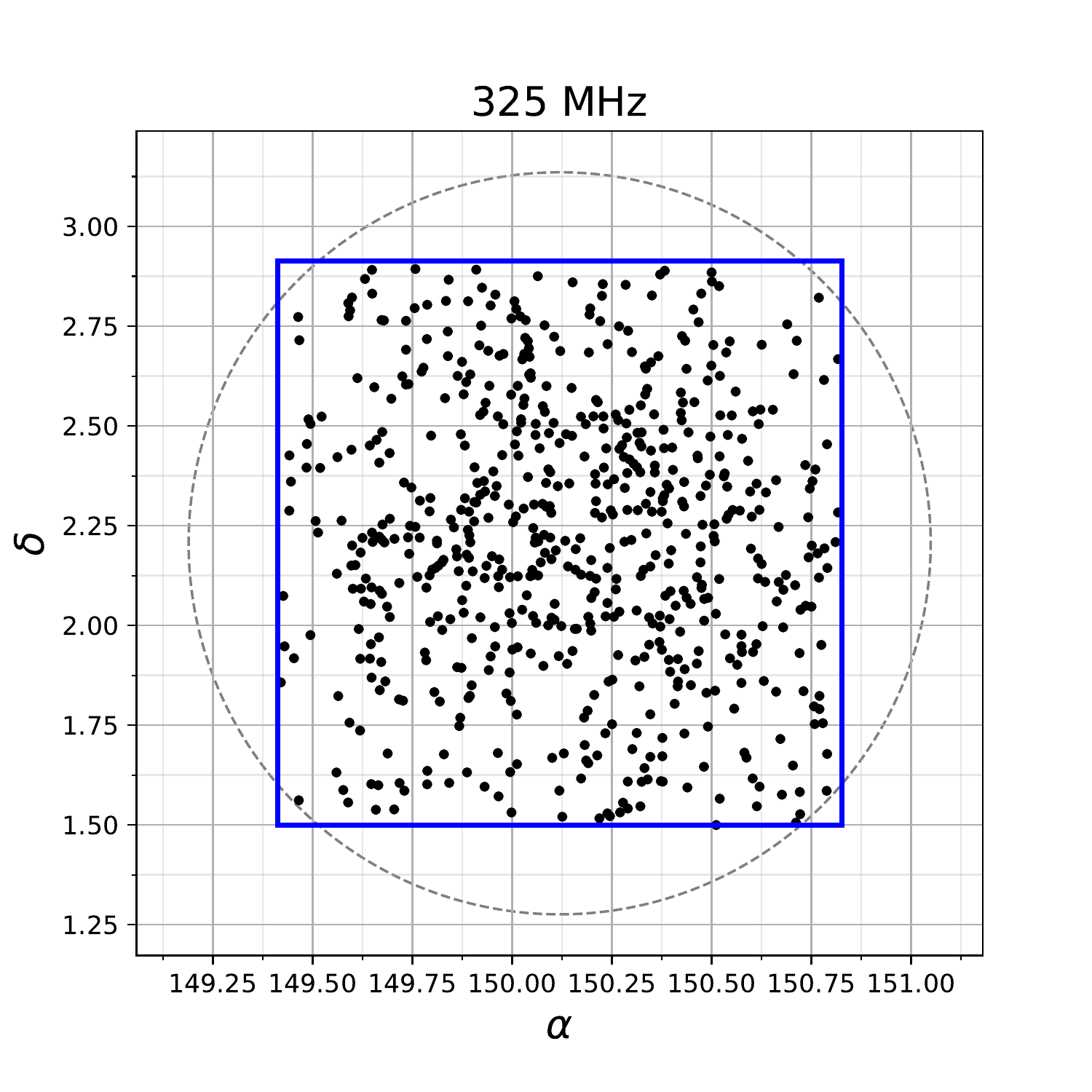}
\includegraphics[width=.3\textwidth]
            {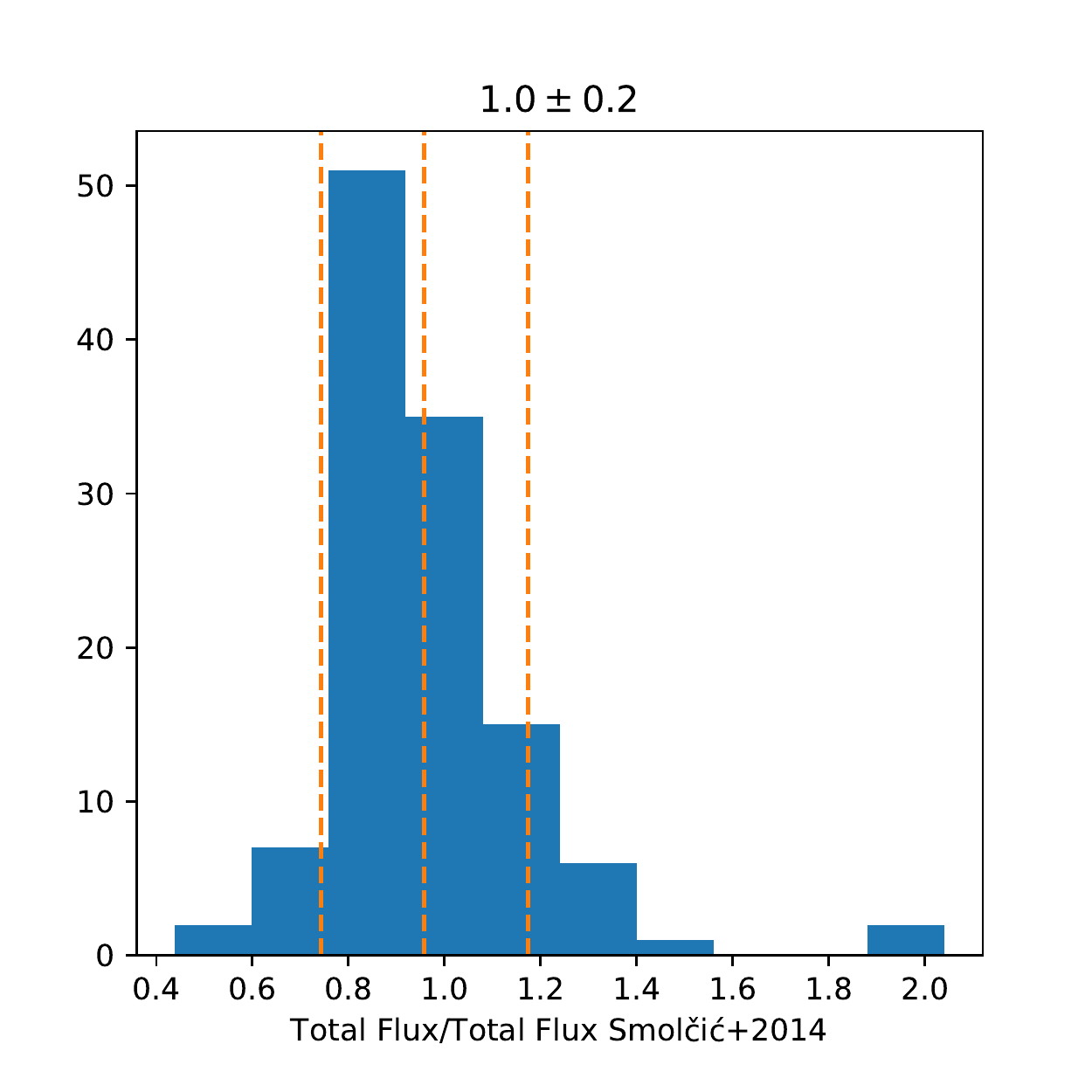}

        \includegraphics[width=.3\textwidth]
            {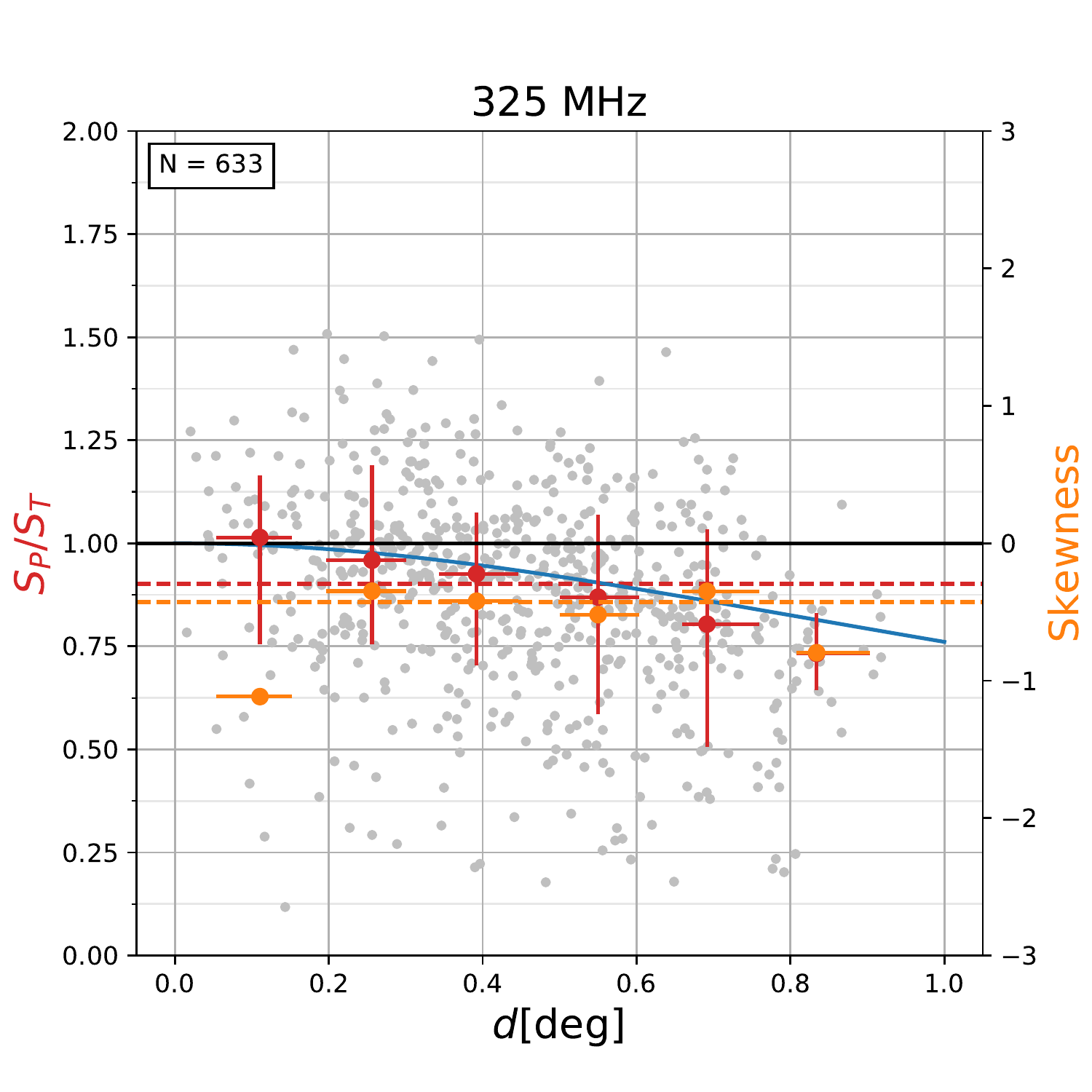}
        \includegraphics[width=.3\textwidth]
            {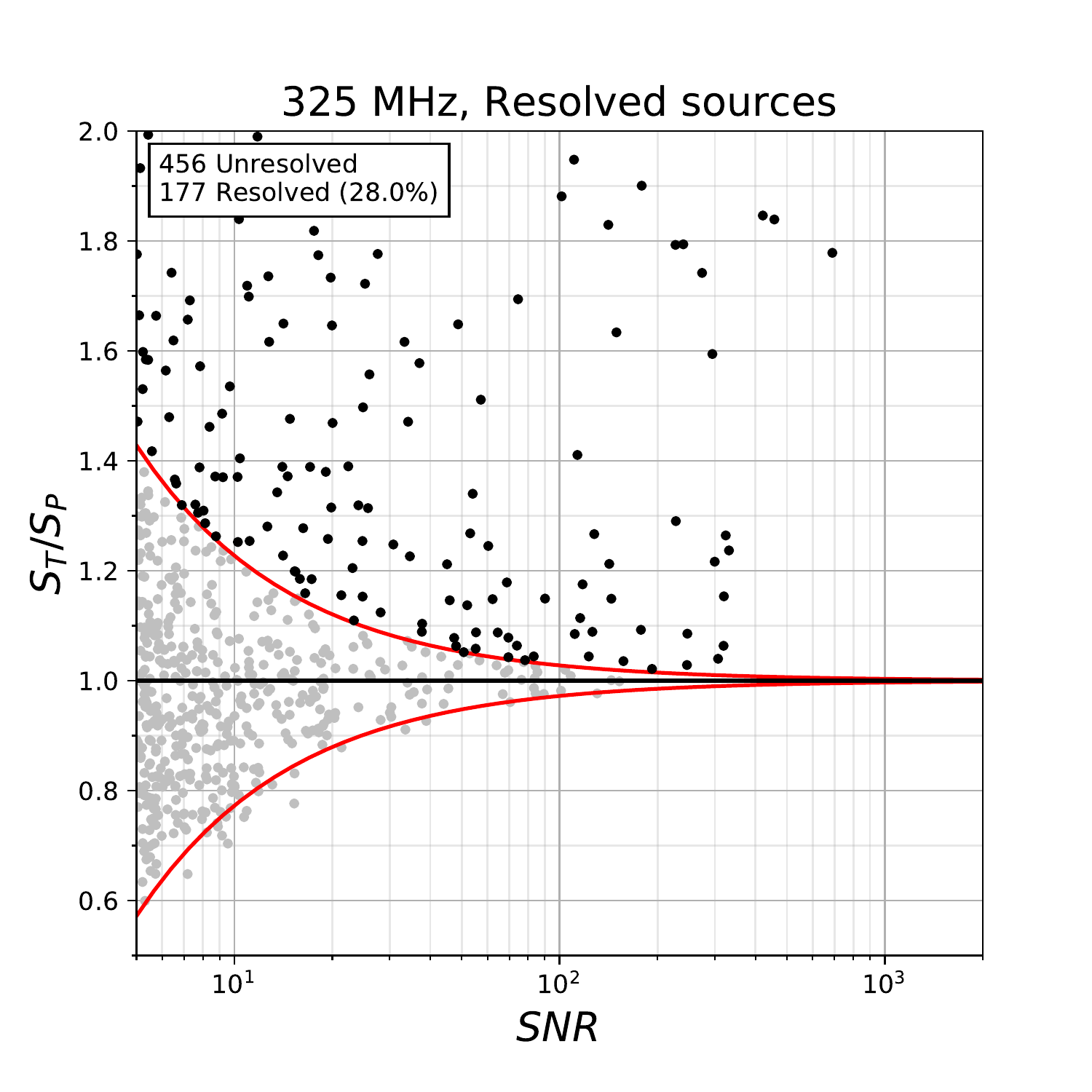}

\includegraphics[width=.3\textwidth]
            {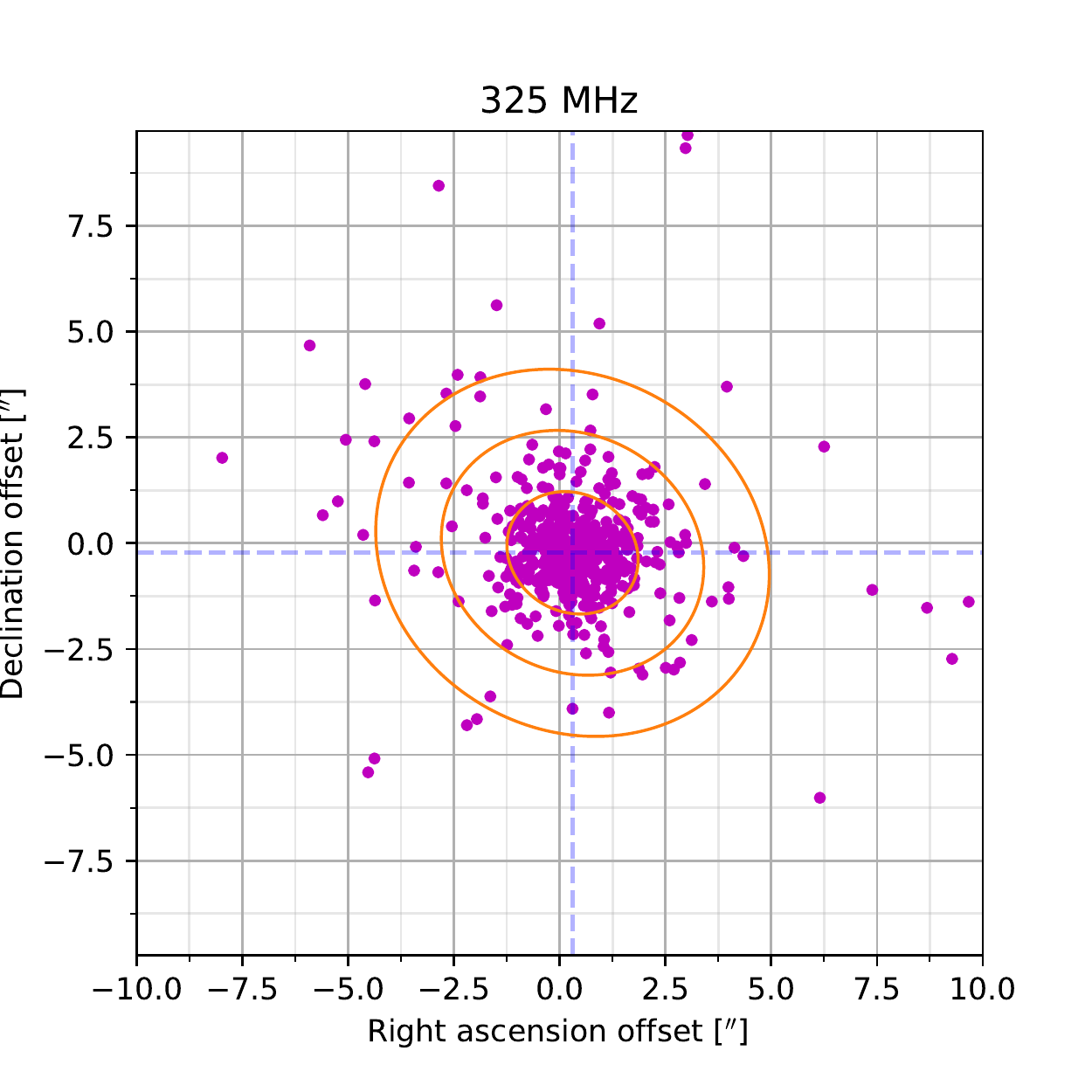}
        \includegraphics[width=.6\textwidth]
            {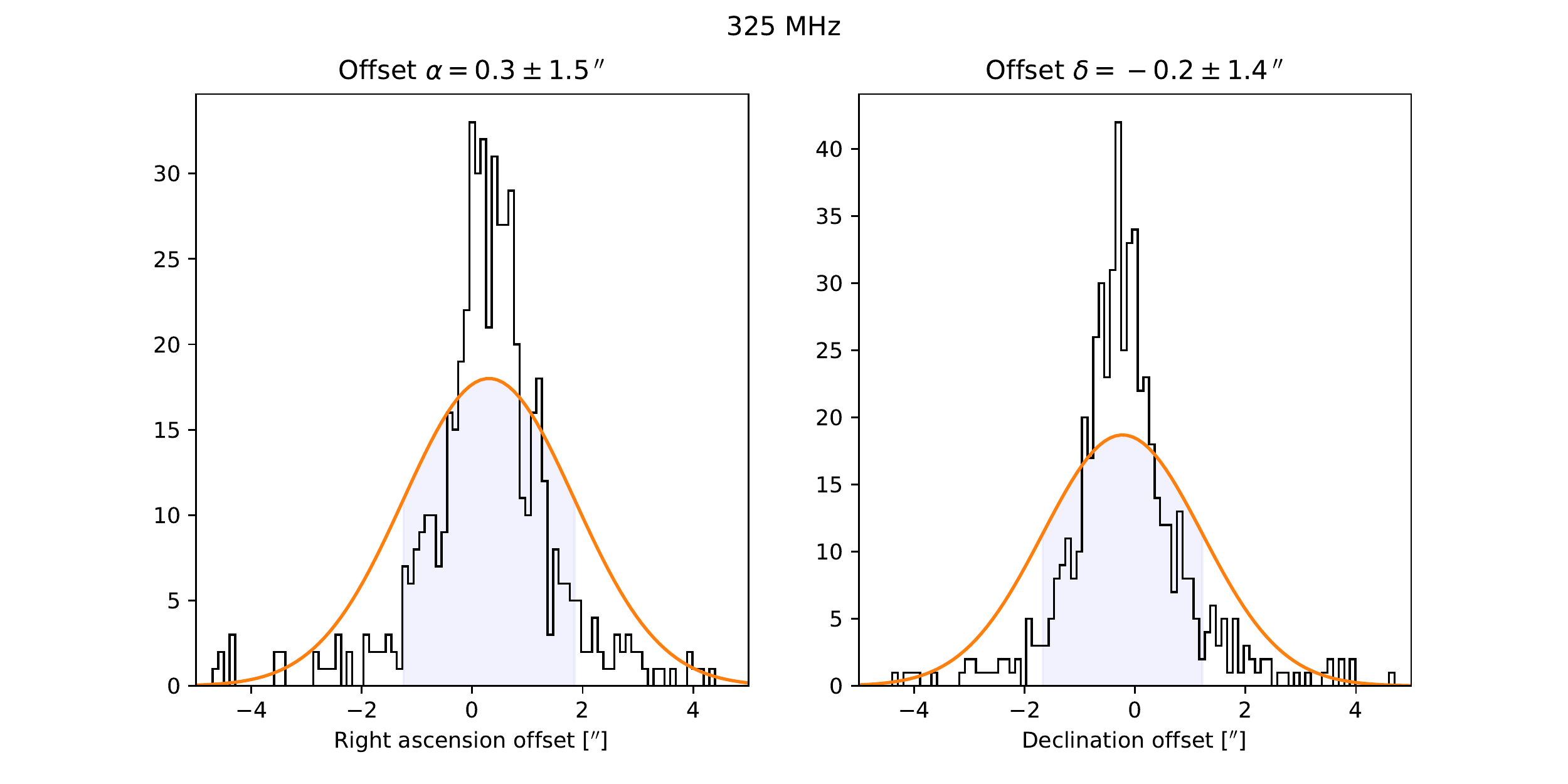}
     \caption{
    \textbf{Top row:} The left panel shows the visibility function for the entire $325\MHz$ map (blue line) and the $2\deg^2$ COSMOS field (orange line). The median RMS within the COSMOS field is denoted by the black dashed line. The positions of sources detected using blobcat in the $325\MHz$ map are shown in the middle panel. The blue rectangle denotes the $2\deg^2$ COSMOS field, while the dashed circle shows the pointing half-power radius. The right panel shows the ratio of total fluxes detected in the $325\MHz$ map and in the \citet{Smolcic14} map. The mean and the $1\sigma$ interval of the ratio are shown as vertical orange lines.
    \textbf{Middle row:} The left panel shows the peak-over-total flux ratio, $S_P/S_T$, as a function of the distance to the pointing center ($d$) for the $325\MHz$ map.  Red dots show the medians of $S_P/S_T$ for different $d$ bins,  with $1\sigma$ percentiles as error bars. The fitted \citet{Bridle99} is shown by a blue line. The orange points show the skewness for different $d$ bins, while the orange dashed line shows the overall skewness in this dataset. The right panel shows the total-over-peak flux ratio, $S_T/S_P$, as a function of the signal-to-noise ratio ($S/N$) for the $325\MHz$ map. The red line shows the fit we used to discern resolved sources. Sources above the red envelope were considered resolved.
    \textbf{Bottom row:} The panels show the astrometric offsets of the $325\MHz$ pointing positions to the $3\GHz$ catalog. The left panel shows the two-dimensional distribution of offsets. Dashed blue lines indicate the median offset in right ascension and declination, while the blue shaded region shows the $1$, $2,$ and $3\sigma$ covariance ellipses. The middle and right panels show their respective right ascension and declination offset histograms with fitted Gaussians.
  }\label{fig:325AstroBWS}
\end{figure*}

\begin{figure*}[ht]

\centering \includegraphics[width=.3\textwidth]{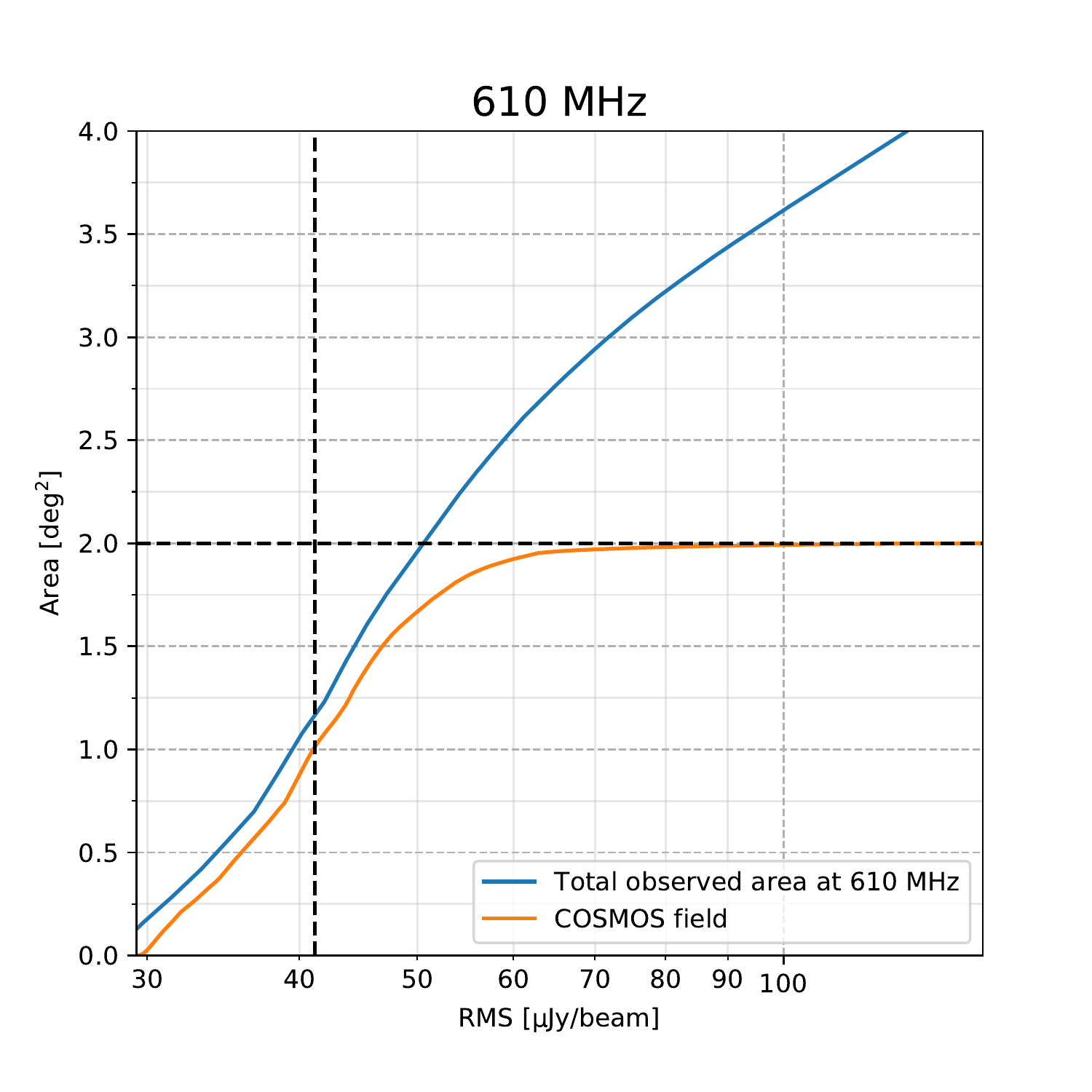}
\includegraphics[width=.3\textwidth]{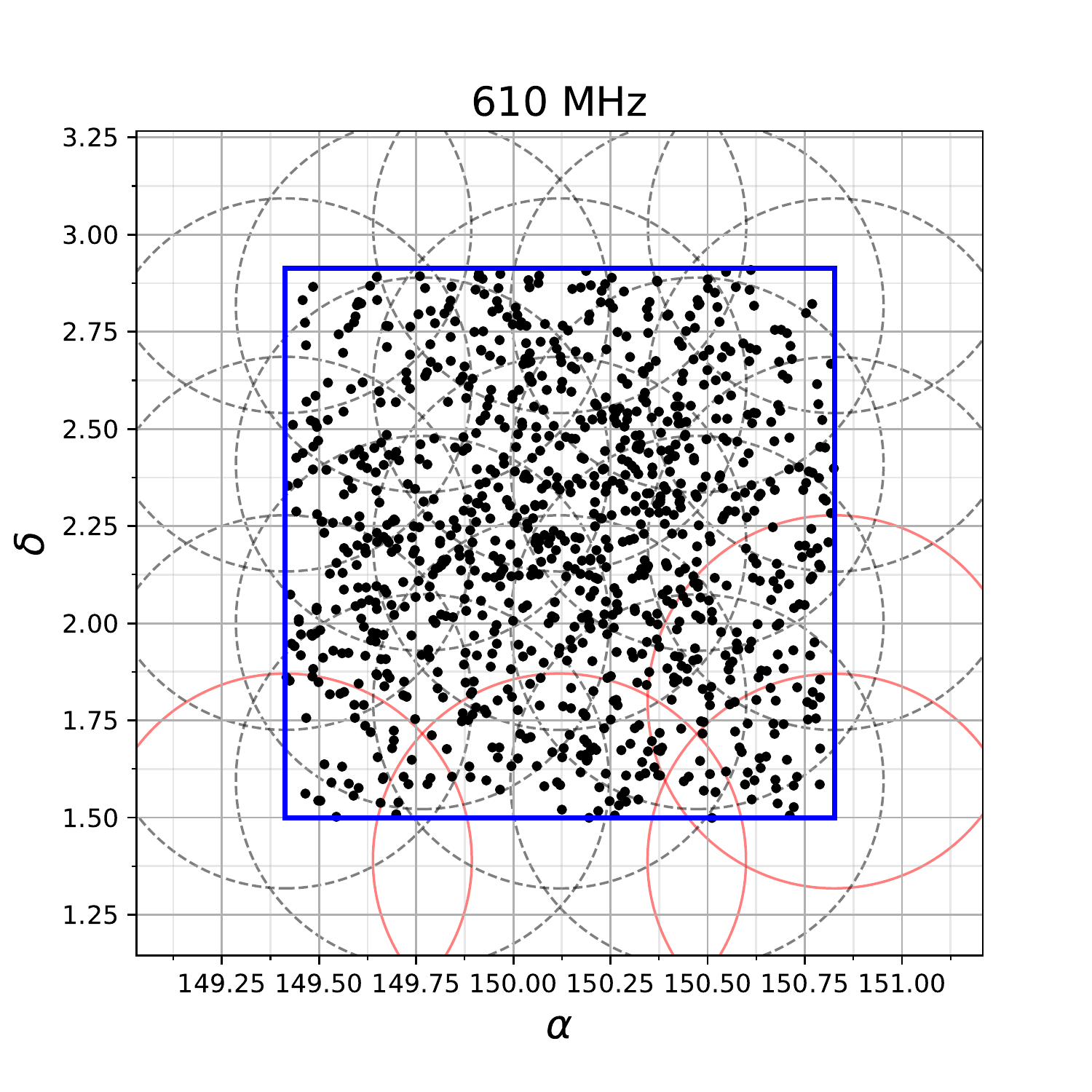}
 \includegraphics[width=.3\textwidth]
            {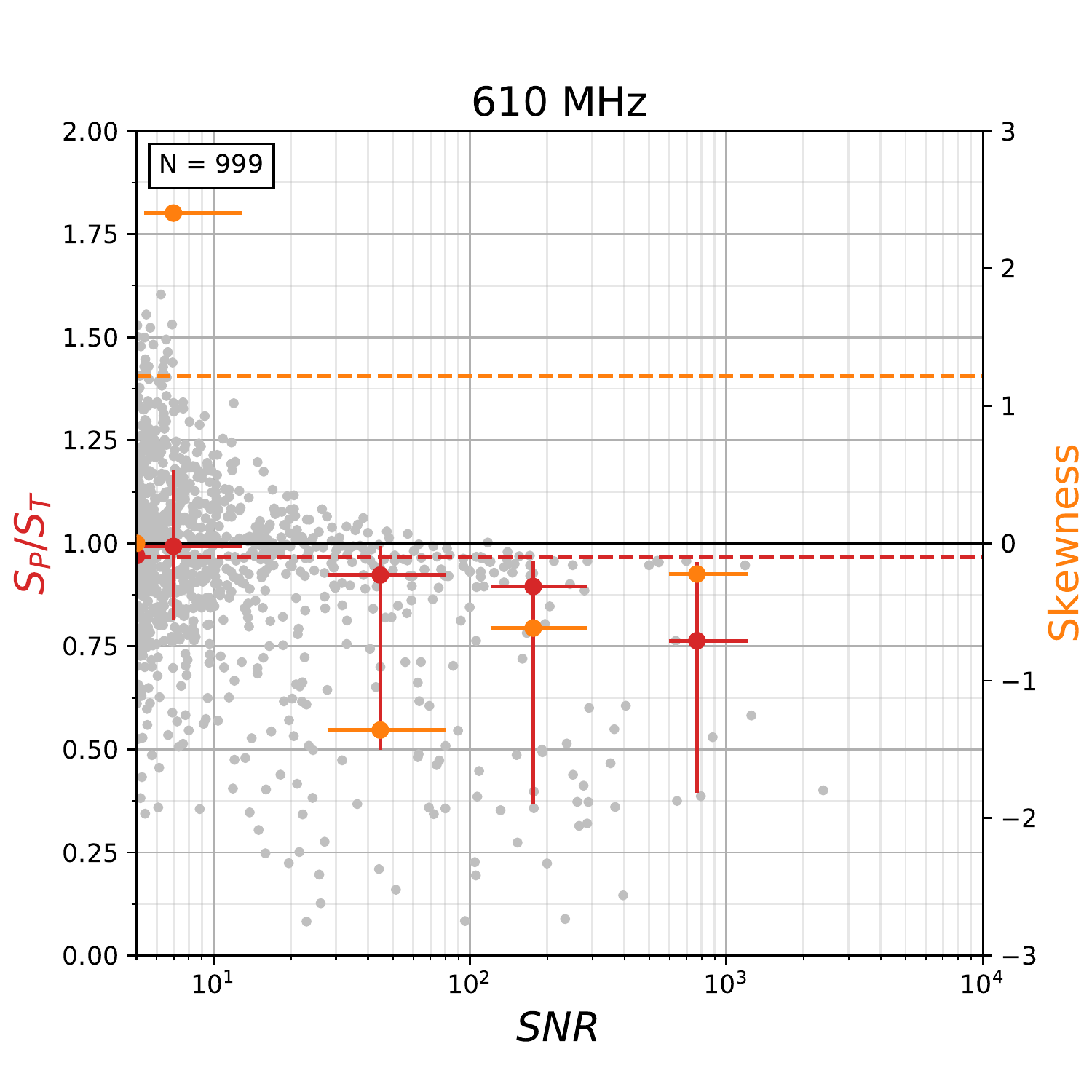}

\includegraphics[width=.3\textwidth]
            {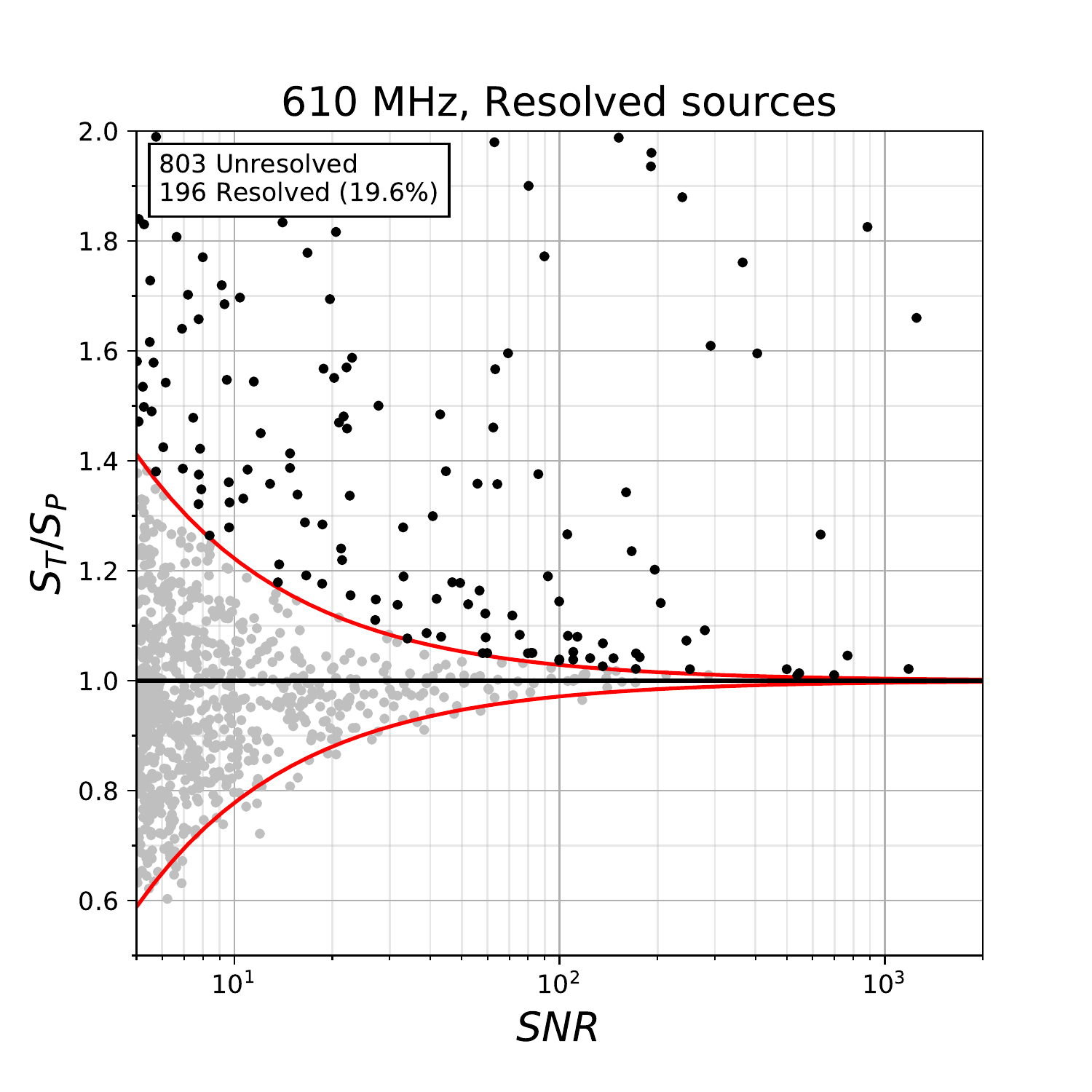}
\includegraphics[width=.6\textwidth]
            {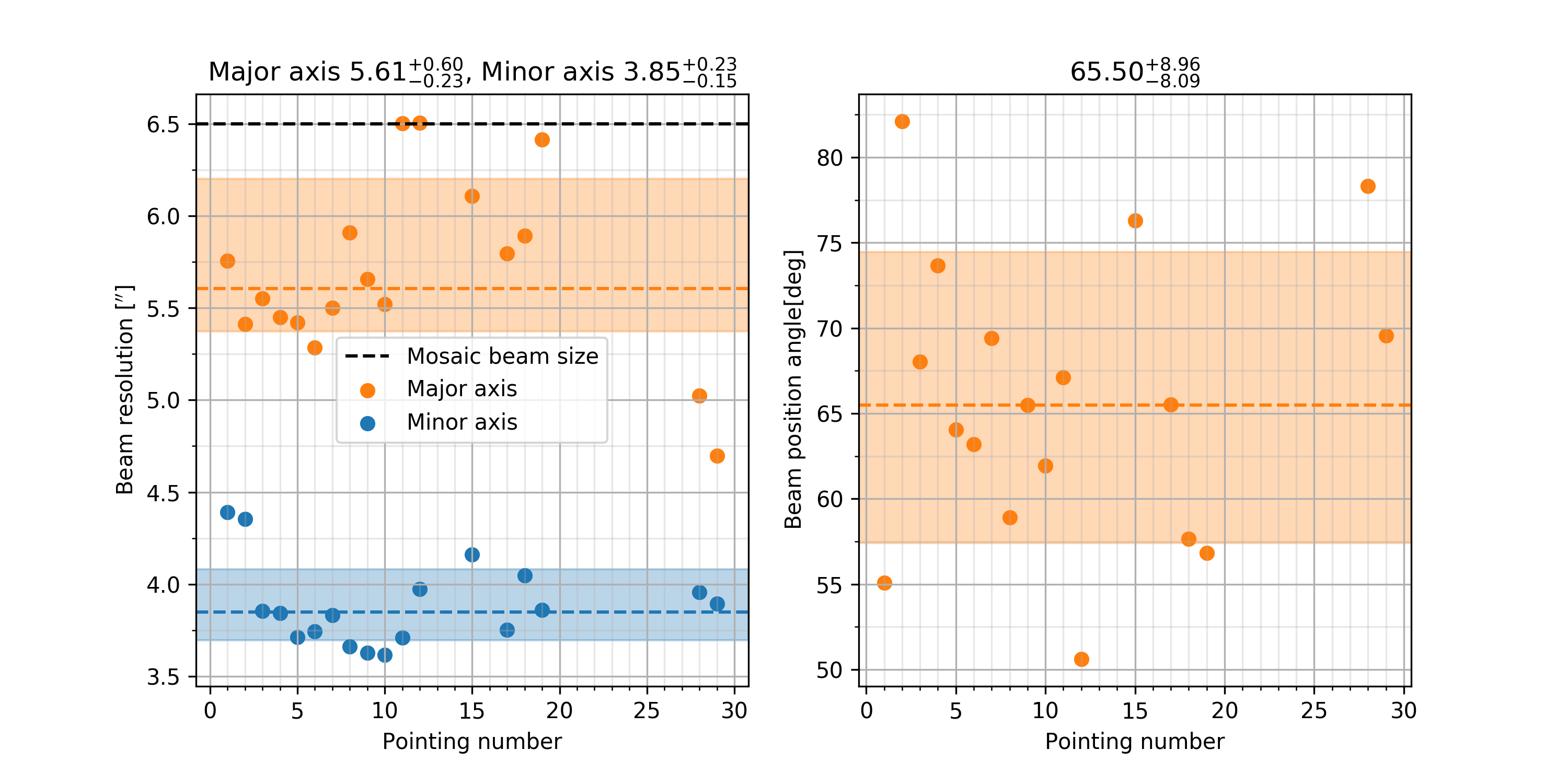}
        
\includegraphics[width=.3\textwidth]
            {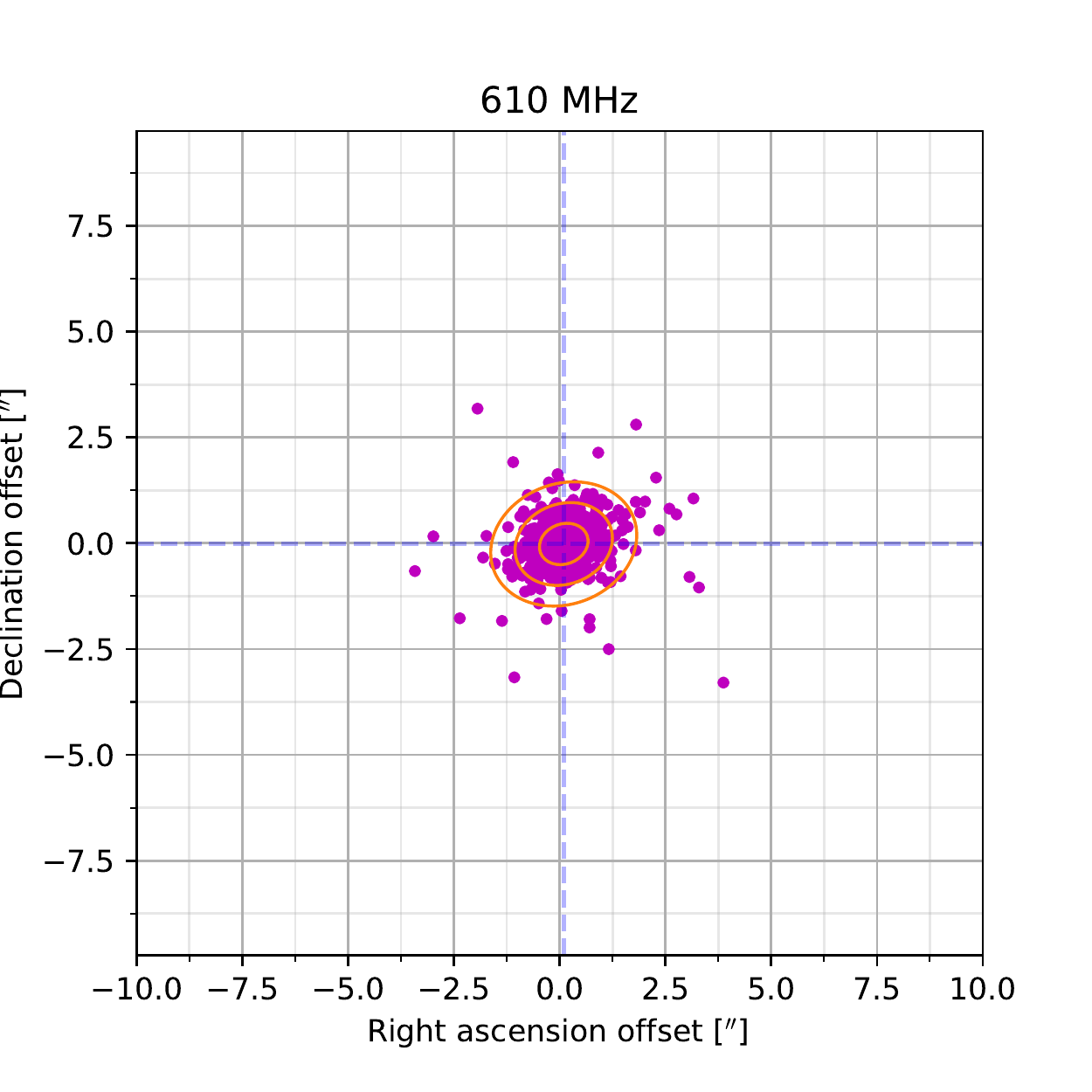}
        \includegraphics[width=.6\textwidth]
            {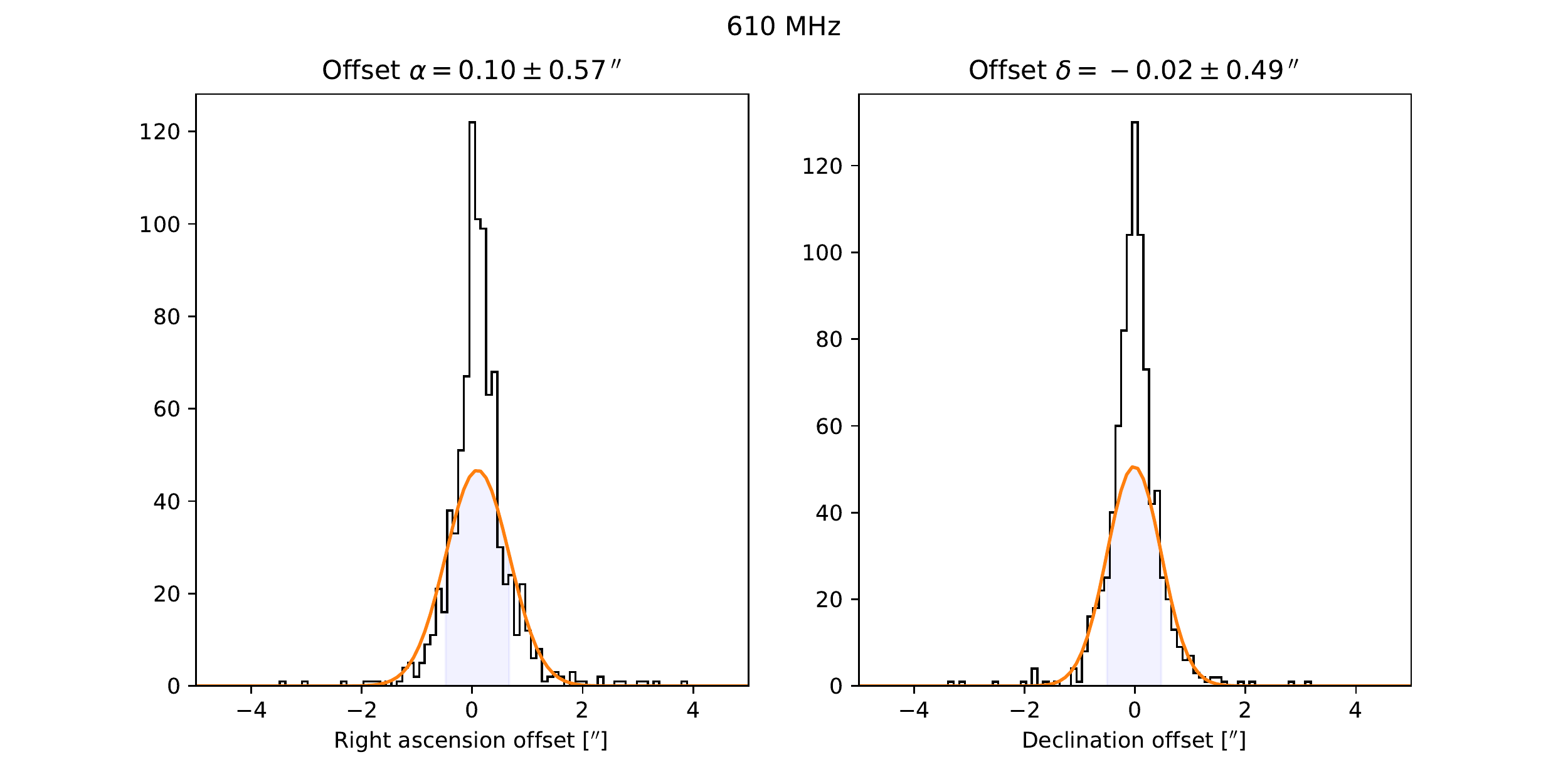}
        \caption{
        \textbf{Top row:} The left panel shows the visibility function for the entire $610\MHz$ map (blue line) and the $2\deg^2$ COSMOS field (orange line). The median RMS within the COSMOS field is denoted by the black dashed line. The  middle panel shows the positions of sources detected using blobcat in the $610\MHz$ mosaic. The blue rectangle shows the COSMOS field. Black dashed lines show the areas covered by individual pointings, while the red dashed lines show pointings that were excluded from the analysis. The right panel shows the peak-over-total flux ratio, $S_P/S_T$, as a function of the $S/N$ for the $610\MHz$ map with sources extracted using blobcat. Red dots show the medians of $S_P/S_T$ for different $S/N$ bins,  with $1\sigma$ percentiles as error bars. The red dashed line shows the median $S_P/S_T$ of all the data points,  while the orange dashed lines shows the overall skewness in this dataset.
          \textbf{Middle row:} The left panel shows the total-over-peak flux ratio, $S_T/S_P$, as a function of the $S/N$ for the $610\MHz$ map after bandwidth smearing correction. The red line shows the fit used to discern resolved sources. The middle and right panels show the beam major and minor axes and the beam position angle for the GMRT $610\MHz$ pointings. The $1\sigma$ intervals for the respective beam parameters are shown as colored regions, with dashed lines showing their medians.
             \textbf{Bottom row:} The panels show the astrometric offsets of the $610\MHz$ mosaic positions to the $3\GHz$ catalog. The left panel shows two-dimensional distribution of offsets. Dashed blue lines indicate the median offset in right ascension and declination, and the blue shaded region shows the $1$, $2,$ and $3\sigma$ covariance ellipses.  The middle and right panels show their respective right ascension and declination offset histograms with fitted Gaussians.
      }\label{fig:610AstroBWS}
\end{figure*}

\subsection*{Observations and imaging}
The GMRT observations of the $2\,\mathrm{deg^2}$ COSMOS field  were conducted using  30  antennas, their longest baseline being  $25\km$. 
The channel width of observations was $125\kHz,$ with a total bandwidth of  $32\MHz$.
The observations were  reduced with the source peeling and atmospheric modeling pipeline \citep[SPAM, described in detail in][]{Intema17}. 
It accounts for radio interference at low radio frequencies and for direction-dependent effects, such as ionospheric dispersive delay.

The $325\MHz$ observations  were carried out in a single pointing (see the top middle panel of Fig. \ref{fig:325AstroBWS}) under the project 07SCB01 (PI: S. Croft).
The observations were allocated $45\hr$  in total and comprised four observations with an average time per observation of  $\sim120\min$  and a total time on target of  $\sim40\hr $. 
Data reduction and imaging were performed using the \textsc{spam} pipeline, with the final map being imaged at a resolution of $10.8\times 9.5\,\mathrm{arcsec}^2$.
A primary beam correction was then performed using the parameters from the GMRT manual. 
The RMS as a function of fractional area is shown in the top left panel of  Fig. \eqref{fig:325AstroBWS}. We find a median RMS of $97\muJy/\mathrm{beam}$ over the $\sim 2 \deg^2$ COSMOS field, uniformly rising toward the edges of the map. 

The $610\MHz$ observations  were carried out over 19 pointings (see the top middle panel of Fig. \ref{fig:610AstroBWS}) under the project 11HRK01 (PI: H. R. Kl\"ockner).
The observations were allocated $86\hr$ in total and were spread over eight observations with an average  time on source of  $\sim4\hr$ per pointing.
We found that four pointings (labeled 14, 16, 22, and 23) deviated by more than $2\sigma$ from the mean beam axes and beam position angles, and these were therefore excluded.
The beam properties of the remaining pointings are shown in the middle row of Fig. \eqref{fig:610AstroBWS}.
After excluding the four pointings, we used one common restoring beam for the remaining pointings.
The pointing images were imaged at a common resolution of $5.6\times 3.9\,\mathrm{arcsec}^2$.
A primary beam correction was then performed on each pointing image before we  combined them into one mosaic using the parameters from the GMRT manual, following the procedure outlined in \citet{Smolcic18}.
The RMS as a function of fractional area is shown in the top left panel of  Fig. \eqref{fig:610AstroBWS}, yielding a median RMS of $39\muJy/\mathrm{beam}$ over the $ 2 \deg^2$ COSMOS field.

A summary of the GMRT data reduction statistics can be found in Table \eqref{tab:GMRTInfo}\footnote{The final catalogs will be publicly available on IRSA.}.
Below we describe the source extraction method, reliability tests, and  corrections we applied to the catalogs.

\subsection*{Source extraction}
\textsc{Blobcat} \citep{Hales12} was used to extract sources down to five times the local RMS value ($5\sigma$) in each map.
The local RMS values were taken from the RMS map produced by the \textsc{aips} task \textsc{rmsd}.
We used default \textsc{blobcat} input parameters except for the minimum blob size, which was set to $3$ pixels in right ascension and declination, as described in \citet{Smolcic:17a}. 
Even though the imaged pointing extends beyond the COSMOS field, we restricted the catalog and further analysis to the $2\,\mathrm{deg^2}$ COSMOS field.

We find 633 source components in the $325\MHz$ map and 999 source components in the $610\MHz$ map.
After accounting for multicomponent sources and blending, the number of sources in the COSMOS field is 633 (see below for details)  in the $325\MHz$ map, while the number of sources remaining in the $610\MHz$ map is 986.
All of the sources detected by \textsc{blobcat} within the $2\,\mathrm{deg^2}$ COSMOS field are shown in the top middle panels of Figs. \eqref{fig:325AstroBWS} and \eqref{fig:610AstroBWS}.
Below we describe the various tests and corrections performed to generate the final source catalogs.

\subsection*{Bandwidth smearing}
Given the finite bandwidth of the map, a certain amount of bandwidth smearing is expected.
To assess the effect of bandwidth smearing on the extracted fluxes, we computed the peak-over-total flux ratio, $S_P/S_T$.

For the $325\MHz$ map, which consists of only one observed pointing, in the middle left panel of Fig. \eqref{fig:325AstroBWS} we show the $S_P/S_T$ ratio as a function of the distance to the pointing center. 
Bandwidth smearing reduces the value of peak fluxes compared to the total fluxes with increasing distance to the pointing center \citep{Bridle99}.
We binned the data, calculated the median $S_P/S_T$ , and used the expected functional dependence of the flux ratio and distance to the pointing center in the presence of bandwidth smearing \citep[as explained in detail in][]{Bridle99} to account for bandwidth smearing of the peak fluxes. 
We used this relation to correct the peak flux for each source, and we verified that the $S_P/S_T$ ratio does not depend on distance to the pointing center after the correction.
The final total fluxes are consistent with the shallower \citet{Smolcic14} $324\MHz$ Catalog (top right panel of Fig. \ref{fig:325AstroBWS}). 

In the top right panel of Fig. \eqref{fig:610AstroBWS}, we show the peak-over-total flux ratio, $S_P/S_T$, as a function of the S/N of sources extracted from the $610\MHz$ mosaic, combined from 19 pointings.
We corrected for bandwidth smearing in the $610\MHz$ mosaic by dividing the peak flux of each source with the median value of the  peak-over-total flux distribution.

\subsection*{Resolved and unresolved sources}
In radio maps, {ideal unresolved sources with infinitely high S/Ns have peak fluxes in $\mathrm{Jy/beam}$ equal to their total fluxes because their flux is contained within the area that is covered by the beam.}
However, because of the noise properties of the map, we expect the $S_T/S_P$ ratio for unresolved sources to increasingly scatter around unity with decreasing S/N. 
We thus discerned resolved sources in the $325\MHz$ map and the $610\MHz$ mosaic through the following procedure.
{We first fit the fifth percentile in the total-over-peak ratios that are lower than 1 as a function of the S/N, as the total flux can only be lower than 1 as a result of statistical effects.}
We did this by fitting all detections with a total flux lower than the peak flux with a relation of the form 
$S_T/S_P = 1-a (S/N)^{b}$,
as is commonly used in the literature \citep[see, e.g.,][]{Smolcic:17a,Bondi03}. 
Second, we assumed that the noise is symmetric around unity. We therefore mirrored this envelope  around $S_T/S_P=1$ and considered all sources above it resolved.

The best-fit parameters  are $a=3.58$ and $b=-1.13$  and $a=1.26$ and $b=-0.76$ for the $325\MHz$and $610\MHz$ catalogs, respectively.
Therefore, we consider 177 (196) sources with $S_T/S_P>1+3.58(S/N)^{-1.13}$ ($S_T/S_P>1+1.26 (S/N)^{-0.76}$) to be resolved at $325\MHz$ ($610\MHz$), as shown in the middle right (left)  panel of Fig. \eqref{fig:325AstroBWS}  (Fig. \ref{fig:610AstroBWS}).
We set their total fluxes of unresolved sources to the values of their respective peak fluxes.
{We note that the fraction of resolved sources (see Table \ref{tab:GMRTInfo}) is higher in the $610\MHz$ catalog because  the  $610\MHz$ map is $1.3-1.8$ times more sensitive than the  $325\MHz$ map (derived by assuming spectral indices in the range from 1 to 0.5).}

\subsection*{Multicomponent sources and deblending}
If the resolution of the map is good enough, source extraction algorithms start to separate jets, lobes, and star-forming regions of radio sources into separate sources. 
We therefore need to associate these components with a single source, termed multicomponent.
To quantify this effect in our GMRT maps and find multicomponent sources, we made use of the $3\GHz$ source catalog. 
This catalog was used because it reaches a much higher depth ({it is $4-13$ times more sensitive than the $325\MHz$ map and $3-7$ times more sensitive than the $610\MHz$ map for spectral indices of 1-0.5})  and a higher resolution ($0.75\arcsec$ at $3\GHz$ vs $\sim 10\arcsec$ at $325\MHz$ and $\sim 5.6\arcsec$ at $610\MHz$).
If the source has multiple components, {that is, components that are not connected by detectable emission}, in this higher resolution $3\GHz$ map \citep[see][for details]{Smolcic:17a}, it is a good candidate for a multicomponent source in the map of lower resolution. 
Conversely, we do not expect single-component sources at $3\GHz$ to appear as multicomponent at lower resolution.

By visually inspecting cutouts of the maps extracted at positions of multicomponent sources in the $3\GHz$ source catalog, we found components that needed to be combined into one multicomponent source.
The method used to detect multicomponent sources was independently verified using the algorithm described in detail in \citet{Vardoulaki18:in-review}. 
This algorithm generated $3\GHz$ cutouts that were convolved to a $10\arcsec$ resolution, and we then used them to create masks for the component-finding algorithm.
The algorithm searches for positions within the mask and within a radius that is defined by the linear projected size of the source \citep[linear projected sizes are measured by a semi-automatic machine-learning technique and have been verified by eye,][]{Vardoulaki18:in-prep}.
{To be consistent with \citet{Smolcic:17a}, the total flux  was measured after the multicomponent source was identified using the \textsc{AIPS} task \textsc{TVSTAT} after clipping all the pixels below $2\sigma$.  } 
This is consistent with the method used for all other  radio source catalogs in the COSMOS field \citep{Schinnerer07,Schinnerer10, Smolcic14, Smolcic:17a}.
In the $325\MHz$ map, we find only 2 components  that needed to be combined into 1 multicomponent source, and in the $610\MHz$ mosaic we find 25 components  that needed to be combined into 12 multicomponent sources.
As our analysis is based on the $3\GHz$ source catalog, in the final catalog, we set the position of the multicomponent source in the GMRT catalogs to its respective position in the $3\GHz$ catalog, its peak flux to -99, and we replaced its total flux by the value derived above and set the multicomponent flag to 1. 

We further identified two sources that were blended into one source at $325\MHz$. 
This number of blended sources was expected, based on the analysis of the $3\GHz$ source catalog (middle right panel of Fig. \ref{fig:325AstroBWS}).
We deblend the identified blended source in the $325\MHz$ map by using \textsc{blobcat} to be consistent with other flux estimates of the catalog. 
To ensure that \textsc{blobcat} separated the blended source into two sources, we set the $fS/N$ parameter to 5.78.
We tested these measured fluxes using the CASA task IMFIT with visually identified masks that separated the two sources.
The sum of total fluxes of the two sources derived using \textsc{blobcat} equals the total flux of the blended source.
In the final catalog, we replaced the blended source with two entries in the catalog and the \textsc{blobcat} -derived properties of the respective sources.

To summarize, we associated two source components with one source and deblended one source into two separate sources in the $325\MHz$ catalog. We also associated 25 source components with 12 multicomponent sources in the $610\MHz$ catalog.
The total number of sources in the GMRT catalogs of the COSMOS field is therefore 633 and 986, respectively.

\subsection*{False-detection rate}
{As noise in radio-interferometric images is non-Gaussian and source detection is based on a $5\sigma$-base peak detection,} we statistically expect that a small percentage of detections might  be random noise fluctuations. 
 To assess the false-detection rate, we {assumed that the background noise is symmetric around zero} and ran \textsc{blobcat} on the inverted maps (i.e., map multiplied by -1), with the same input parameters as in the original maps. 
 Given that sources with negative fluxes are not expected in the radio map, any source extracted from the inverted map can be reliably identified as a false detection.
 We find 19 ``detections'' in the inverted $325\MHz$ map, yielding a false-detection rate of $3\percent$ and 13 ``detections'' in the inverted $610\MHz$ map, yielding a false-detection rate of $1.3\percent$.
 
\subsection*{Astrometry}
To test the astrometric accuracy of the data, we show in the bottom row panels of Figs. \eqref{fig:325AstroBWS} and  \eqref{fig:610AstroBWS}   the offsets of the source positions in the $325\MHz$ and $610\MHz$ catalogs, respectively, from those present in the $3\GHz$ catalog.
The mean offsets in right ascension are within one tenth of the beam size of the GMRT maps, which means that the astrometric accuracy is good.

\subsection*{Final catalogs }
In the final catalog we report the positions of sources as well as their peak (in $\mathrm{Jy/beam}$), and total fluxes (in $\mathrm{Jy}$).
For the multicomponent sources, the total flux of both components is indicated in one catalog entry. 
The 'Multicomponent' column indicates the presence of more than one component to the source.
For resolved sources, the column 'Resolved' is set to 1.

\section{K-corrections}\label{sect:K-Corrections}

Generally, the luminosity at a rest-frame frequency $\nu_{e_2}$, $L(\nu_{e_2})$, is connected to the flux measured at the observed frequency $\nu_{o_2}$, $S(\nu_{o_2})$, as
\begin{equation}
L(\nu_{e_2})=\frac{4\pi D_L^2}{1+z} S(\nu_{o_2}),
\end{equation}
where $z$ is the object redshift and the observed frequency is connected to the emitted frequency by $\nu_e=\nu_o(1+z).$
This relation does not depend explicitly on the shape of the SED, so we have the relation connecting ratios of luminosities at two different frequencies
\begin{equation}
\frac{L(\nu_{e_2})}{L_(\nu_{e_1})}=\frac{S(\nu_{e_2})}{S_(\nu_{e_1})}=\frac{S(\nu_{o_2})}{S_(\nu_{o_1})}.\label{eq:ratiosS}
\end{equation}
We can model $\frac{S(\nu_{e_2})}{S_(\nu_{e_1})}$ by an SED. 
Inserting Eq. \eqref{eq:ratiosS} into the first equation, we obtain
\begin{equation}
L(\nu_{e_2})=\frac{4\pi D_L^2}{1+z} \frac{S(\nu_{e_2})}{S\left(\nu_{o_1}(1+z)\right)} S(\nu_{o_1}).
\end{equation}
We can now proceed to find the K-correction for a broken power-law.

\begin{figure*}
\includegraphics[width=\textwidth]{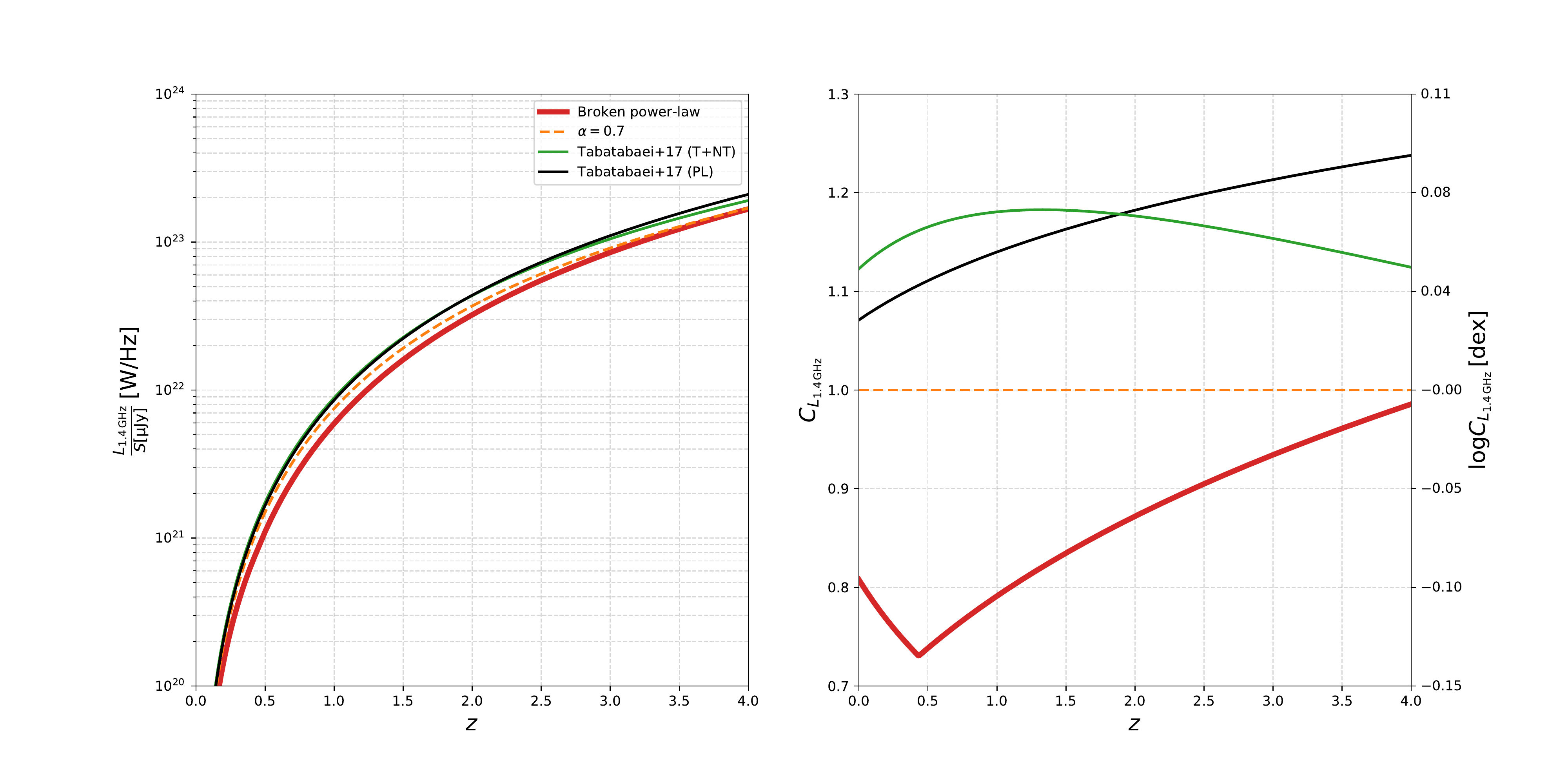}
\caption{{Left panel: Luminosity at a rest-frame frequency of $1.4\GHz$ for different redshifts computed using the fiducial value of $1\,\muJy$ for (observer frame) $3\GHz$ flux densities, while the right panel shows the correction of $1.4\GHz$ radio luminosity, $C_{L_{1.4\GHz}}$ computed using the spectral index of $\alpha=0.7$ as reference. Solid lines are corrections based on either a broken power-law (red line), derived in Sect. \ref{sect:Results} for our HSFG sample, and an SED (black line) of normal star-forming galaxies based on a $0.97$ nonthermal spectral index and a 10\percent\  thermal fraction \citep{Tabatabaei17}. The green line shows the SED of NSFGs based on a simple power law \citep[$\alpha=0.79$][]{Tabatabaei17}. } }\label{fig:Kcorr}
\end{figure*}
\subsection{Broken power-law}
We assumed that the spectrum is that of a broken power-law with spectral indices $\alpha_0$, $\alpha_1$, and so on, with each new spectral index, $\alpha_k$, beginning above a fixed frequency $\nu_k$.
In this case, the flux for each frequency range $k$ can be expressed as
\begin{equation}
\mathrm{lg} S_k = -\alpha_k \mathrm{lg}\nu +b_k,
\end{equation}
where $b_k$ is the $k$-th bin normalization constant. Using the fact that the flux needs to be a continuous function of $\nu$, we can roll back $b_k$ to the first bin, $b_0$ as 
\begin{equation}
b_k = b_0 +\sum\limits_{j=0}^{k-1}(\alpha_{k-j}-\alpha_{k-(j+1)})\mathrm{lg}\nu_{k-j}=b_0+\sum\limits_{j=0}^{k-1}\Delta\alpha_{k-j}\mathrm{lg}\nu_{k-j}.
\end{equation}
The bin number, $k$, is easily determined from $\nu$ and the list of break frequencies $\nu_i$, giving functions $k(\nu)$ and $\alpha(\nu)$.
We can repeat this process for each frequency $\nu_{e_1}$ and $\nu_{e_2}$, yielding
\begin{equation}
\frac{S(\nu_{e_2})}{S\left(\nu_{o_1}(1+z)\right)}=\frac{1}{(1+z)^{-\alpha(\nu_{o_1}(1+z))}}\left(\frac{\nu_{e_2}^{-\alpha(\nu_{e_2})}}{\nu_{o_1}^{-\alpha(\nu_{o_1}(1+z))}}\right)10^{b_k(\nu_{e_2})-b_k(\nu_{o_1}(1+z))}.
\end{equation}
Finally, we obtain the luminosity as
\begin{equation}
L(\nu_{e_2})=\frac{4\pi D_L^2 S(\nu_{o_1})}{(1+z)^{1-\alpha(\nu_{o_1}(1+z))}}\left(\frac{\nu_{e_2}^{-\alpha(\nu_{e_2})}}{\nu_{o_1}^{-\alpha(\nu_{o_1}(1+z))}}\right)10^{b_{k(\nu_{e_2})}-b_{k(\nu_{o_1}(1+z))}}.\label{eq:KBPL}
\end{equation}
For the broken power-law with a single break frequency, $\nu_{e_2}=1.4\,\mathrm{GHz}$ and $\nu_{o_1}=3\,\mathrm{GHz}$, we can express Eq. \eqref{eq:KBPL} as
\begin{equation}
\tiny
L_{1.4\GHz}=4\pi D_L^2 S_{3\GHz}\begin{cases}
\frac{1}{(1+z)^{1-\alpha_1}}\left(\frac{1.4^{-\alpha_1}}{3^{-\alpha_1}}\right),\, \,3\GHz(1+z)<\nu_b\\
\frac{1}{(1+z)^{1-\alpha_2}}\left(\frac{1.4^{-\alpha_1}}{3^{-\alpha_2}}\right)10^{-(\alpha_2-\alpha_1)\log\nu_b},\, \,3\GHz(1+z)>\nu_b
\end{cases}.
\end{equation}

The left panel of Fig. \eqref{fig:Kcorr} shows the luminosity at $1.4\GHz$ rest-frame derived using various K-corrections.
From the right panel of the same figure, we see that luminosities are by up to $26\percent$ overestimated compared to the broken power-law for $z\leq 1$ if we assume a single spectral index of $0.7$.
\subsection{K-corrections and thermal fraction}
We used the results of \citet{Tabatabaei17} to derive luminosities of normal star-forming galaxies. They have found that the SED of normal star-forming galaxies has a steep nonthermal spectral index ($\alpha_2=0.97$) and a $\sim10\percent$ thermal fraction at $1.4\GHz$. The K-correction in this case is
\begin{equation}
L_{1.4\GHz}=\frac{4\pi D_L^2}{1+z} \frac{S_{3\GHz}}{(1-f_{th})\left(\frac{3 (1+z)}{1.4}\right)^{-\alpha_2}+f_{th}\left(\frac{3 (1+z)}{1.4}\right)^{-0.1}} .
\end{equation}
As discussed in Sect. \ref{sect:Infrared-radio correlation}, we compared the $q-z$ trends derived using this K-correction and also the K-correction based on the \citet{Tabatabaei17}  simple power-law spectral index of $0.79$. 
In the next section, we derive the average correction of radio luminosity needed for a sample of galaxies consisting of both NSFGs and HSFGs. 
\begin{table*}
\caption{{Average corrections of luminosity at $1.4\GHz$ needed if the K-correction was performed by assuming a power-law SED with a spectral index of $0.7$. The corrections were derived for the average luminosity at $1.4\GHz$, as described in Sect. \ref{sect:K-Corrections}, and are based on the assumption that galaxies with an $\SFR>10\Msun/\yr$ follow the broken power-law (BPL) SED and that galaxies with an $\SFR<10\Msun/\yr$ follow the SED based on Eq. \eqref{eq:SEDfull}, with the nonthermal spectral index and thermal fraction taken from \citet{Tabatabaei17}. The percentage $p_{<10}$ is the percentage of $3\GHz$ flux that is due to galaxies with an $\SFR<10\Msun/\yr$. Columns of $p_{<10}=0\percent$ and $p_{<10}=100\percent$ are also exact corrections for single objects derived by assuming the BPL or the \citet{Tabatabaei17} SEDs, respectively. The values in parentheses are correction $C_{L_{1.4\GHz}}$ expressed in $\mathrm{dex}$.}}\label{tab:Kcorr}
\centering\begin{tabular}{c c c c c c}
\toprule
\toprule
$z$     & \multicolumn{5}{c}{ $C_{L_{1.4\GHz}}$ ($\log C_{L_{1.4\GHz}}\,\mathrm{[dex]}$)} \\
\midrule
& \multicolumn{5}{c}{$p_{<10}$} \\
\cmidrule(r){2-6}
   & 0\percent          & 25\percent       & 50\percent      &   75\percent    & 100\percent \\
\midrule
0.0      &       0.81 (-0.09)    &       0.89 (-0.05)    &       0.97 (-0.02)    &        1.04 ( 0.02)    &       1.12 ( 0.05)\\
0.5      &       0.74 (-0.13)    &       0.85 (-0.07)    &       0.95 (-0.02)    &        1.06 ( 0.02)    &       1.17 ( 0.07)\\
1.0      &       0.79 (-0.10)    &       0.89 (-0.05)    &       0.99 (-0.01)    &        1.08 ( 0.03)    &       1.18 ( 0.07)\\
1.5      &       0.83 (-0.08)    &       0.92 (-0.04)    &       1.01 ( 0.00)    &        1.10 ( 0.04)    &       1.18 ( 0.07)\\
2.0      &       0.87 (-0.06)    &       0.95 (-0.02)    &       1.02 ( 0.01)    &        1.10 ( 0.04)    &       1.18 ( 0.07)\\
2.5      &       0.90 (-0.04)    &       0.97 (-0.01)    &       1.04 ( 0.02)    &        1.10 ( 0.04)    &       1.17 ( 0.07)\\
3.0      &       0.93 (-0.03)    &       0.99 (-0.00)    &       1.04 ( 0.02)    &        1.10 ( 0.04)    &       1.15 ( 0.06)\\
3.5      &       0.96 (-0.02)    &       1.01 ( 0.00)    &       1.05 ( 0.02)    &        1.09 ( 0.04)    &       1.14 ( 0.06)\\
4.0      &       0.99 (-0.01)    &       1.02 ( 0.01)    &       1.06 ( 0.02)    &        1.09 ( 0.04)    &       1.12 ( 0.05)\\
\bottomrule
\end{tabular}
\end{table*}
\subsection{Average K-corrections}
\begin{figure}[hb]
\includegraphics[width=\columnwidth]{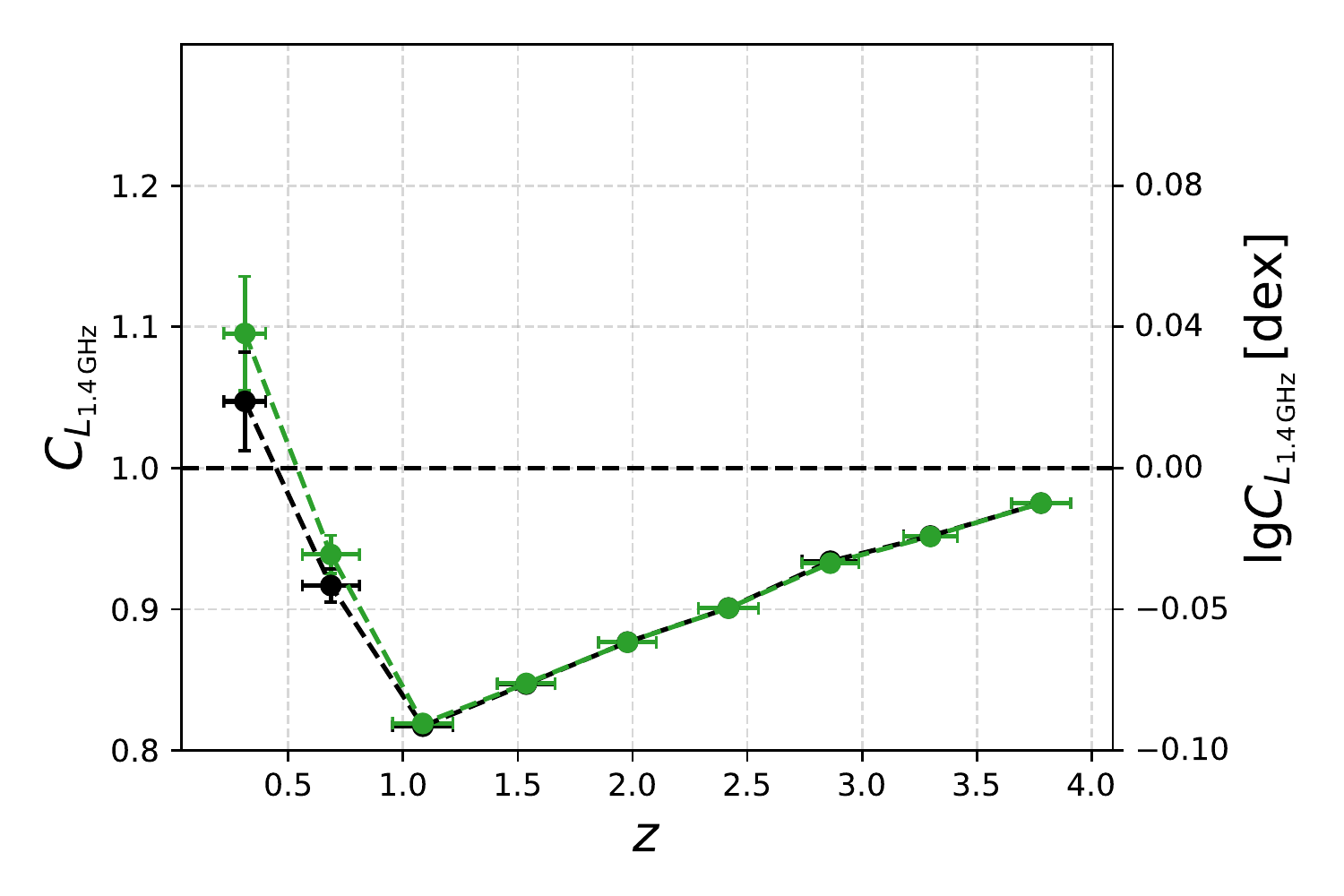}
\caption{Average correction of $L_{1.4\GHz}$ for galaxies detected in the radio from the \citet{Delhaize17} sample. The dashed black line connects corrections in different redshift bins derived by using our broken power-law SED for galaxies with an $\SFR>10\Msun/\yr$, and an SED based on a $0.97$ nonthermal spectral index and a 10\percent\  thermal fraction \citep{Tabatabaei17} for NSFGs ($\SFR<10\Msun/\yr$). The dashed green line similarly connects points based on the broken power-law for an $SFR>10\Msun/\yr$ and a simple power-law SED for NSFGs \citep[$\alpha=0.79$][]{Tabatabaei17}. }\label{fig:LcorrDel}
\end{figure}
To better understand the impact of the different K-corrections on the $q-z$ trend, we also derived corrections of $L_{1.4\GHz}$ for an average $1.4\GHz$ luminosity by assuming a distribution of fluxes, $f(\SFR, z)$, that is dependent on both $\SFR$ and $z$.
We started by rewriting Eq. \eqref{eq:KBPL} for SFGs as $L_{1.4\GHz}=K(\SFR, z) S_{3\GHz}$, where $S_{3\GHz}$ is the observed flux density. As discussed in Sect. \ref{sect:Infrared-radio correlation}, we used a K-correction, $K_{BPL}$, based on the broken power-law SED for galaxies with an $\SFR>10\Msun/\yr$ and the \citet{Tabatabaei17} SED-based K-correction, $K_{Tab}$, for $\SFR<10\Msun/\yr$.
The average $1.4\GHz$ luminosity at redshift $z$ is
\begin{equation}
\langle L_{1.4\GHz}\rangle (z) = \int f(\SFR,z) K(\SFR, z) S(\SFR, z) \,\mathrm{d} \SFR.
\end{equation}
We divided this average luminosity by the average luminosity, $\langle L_{1.4\GHz}\rangle (z)_{\alpha}$, derived by assuming a K-correction, $K_\alpha$, based on a simple power-law SED with a single spectral index $\alpha=0.7$, yielding the correction of $L_{1.4\GHz}$, 
\begin{equation}
C_{L_{1.4\GHz}}(p_{<10}, z)= (1-p_{<10}(z)) \frac{K_{BPL}(z)}{K_{\alpha=0.7}(z)}+p_{<10}(z) \frac{K_{Tab}(z)}{K_{\alpha=0.7}(z)},\label{eq:Lave}
\end{equation}
where we have introduced the fraction of sources with an SFR below $10\Msun/\yr$, $p_{<10}$, defined as
\begin{equation}
p_{<10}=\frac{\int\limits_{<10\Msun/\yr} f(\SFR, z) S(\SFR, z)\,\mathrm{d} \SFR}{\int f(\SFR, z) S(\SFR, z)\,\mathrm{d} \SFR}.\label{eq:p10def}
\end{equation}
Given the finite number of sources in our sample, this fraction is estimated as the sum of fluxes of all sources within a redshift bin that have an $SFR<10\Msun/\yr$ divided by the sum of fluxes of all sources within the redshift bin.
We see that $C_{L_{1.4\GHz}}(p_{<10}, z)$  is a linear interpolation between K-corrections derived by assuming a broken power-law SED and the \citet{Tabatabaei17} SED and can be useful for quickly estimating the impact of using a simple  K-correction based on an $\alpha=0.7$ power-law SED instead of a more elaborate SED. Average corrections for selected redshifts and $p_{<10}$ values can be found in Table \eqref{tab:Kcorr}. In Fig. \eqref{fig:LcorrDel} we calculated average corrections for a subset of galaxies from the \citep{Delhaize17} sample that were detected in the VLA-COSMOS $3\GHz$ Counterpart catalog \citep{Smolcic:17b}. The corrections are significant only below $z\sim1$. The fact that the corrections are not needed for higher redshift bins indicates that the shape of the radio SED cannot explain the $q-z$ relationship.   
\section{Radio SED of HSFGs with models from Table \ref{tab:ranking}}

\begin{figure*}[ht]
\centering \includegraphics[width=.4\textwidth]{SEDs/{SED_data1.pdf}}
\centering \includegraphics[width=.4\textwidth]{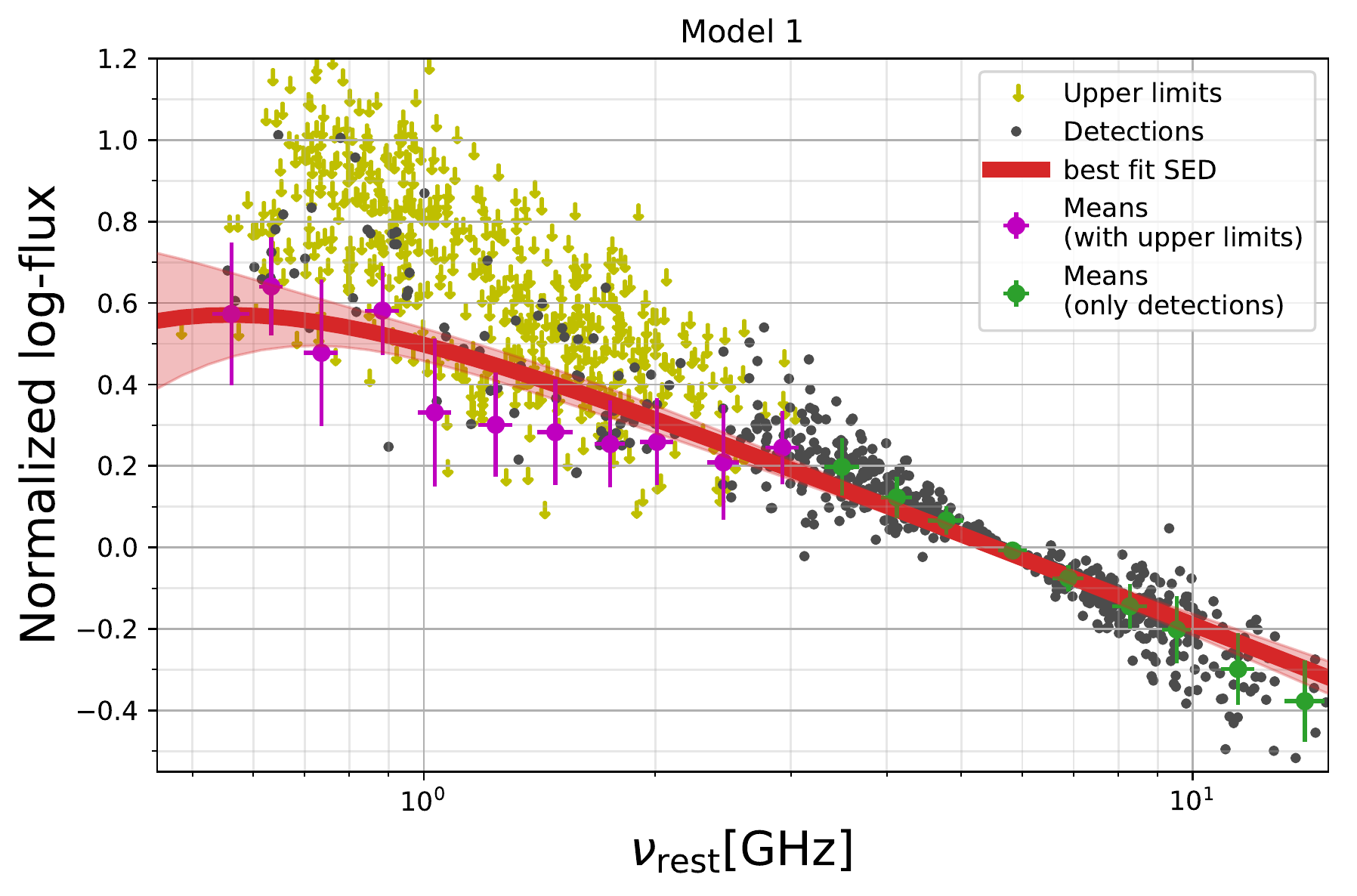}

\centering \includegraphics[width=.4\textwidth]{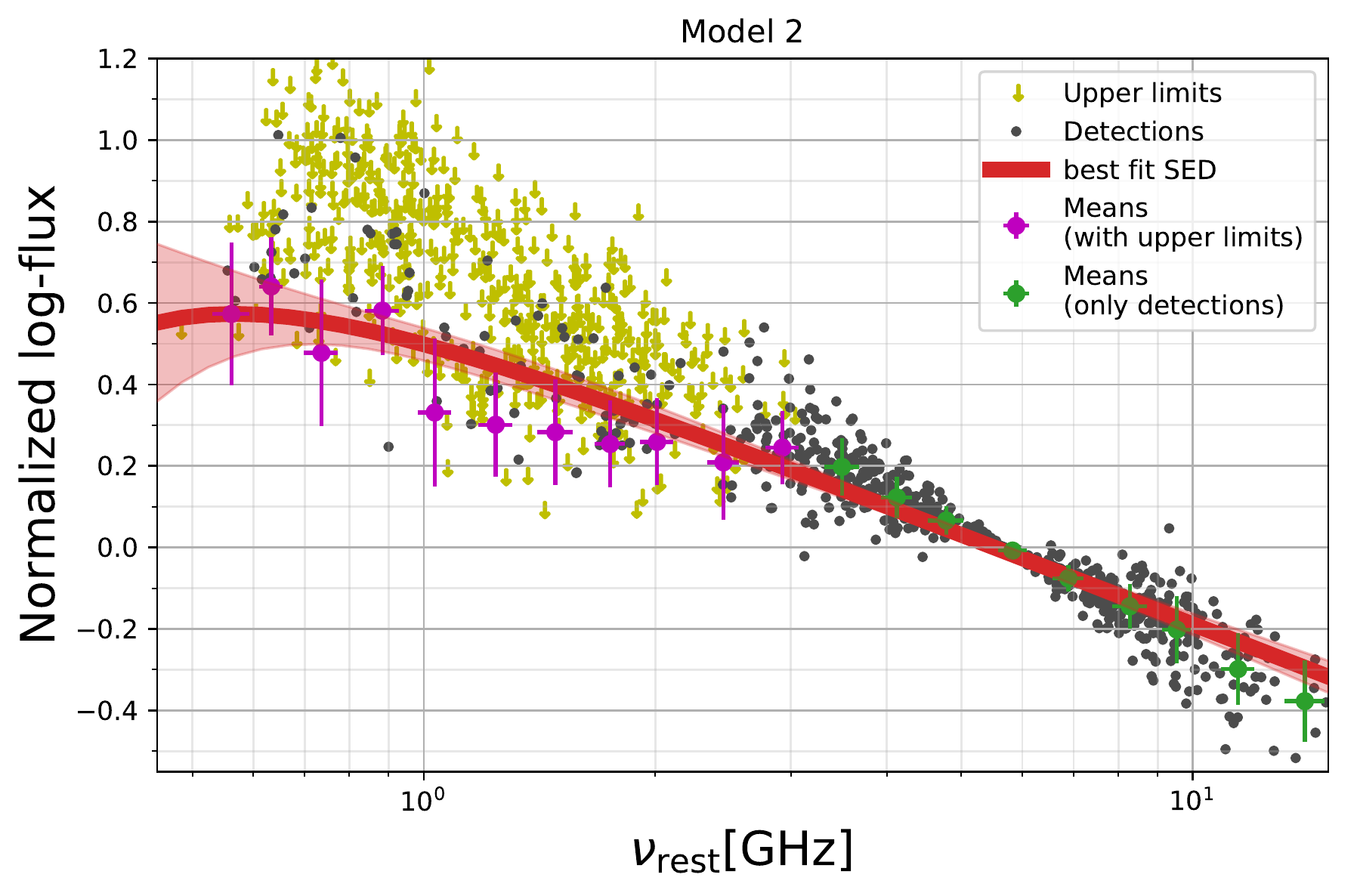}
\centering \includegraphics[width=.4\textwidth]{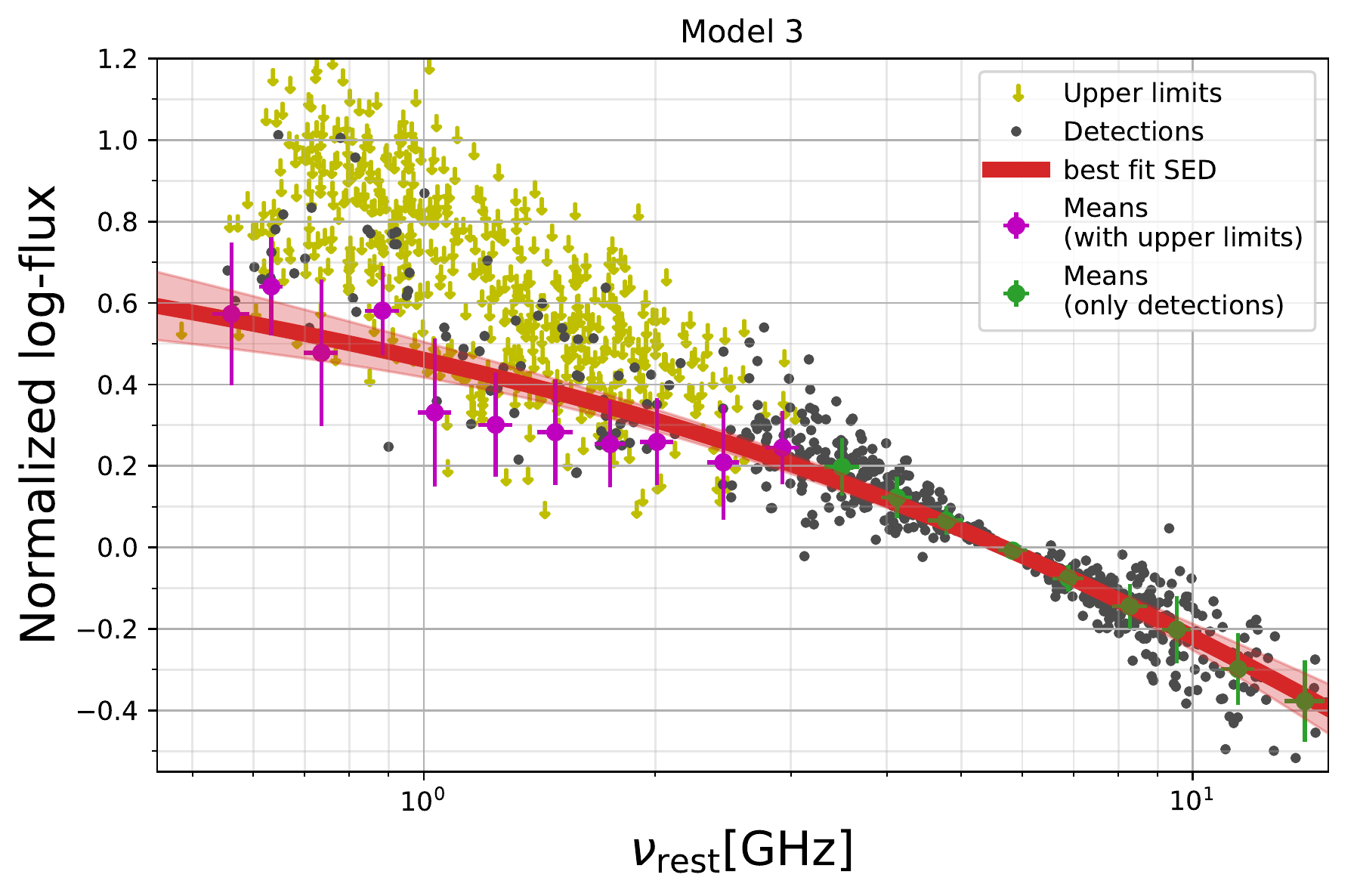}

\centering \includegraphics[width=.4\textwidth]{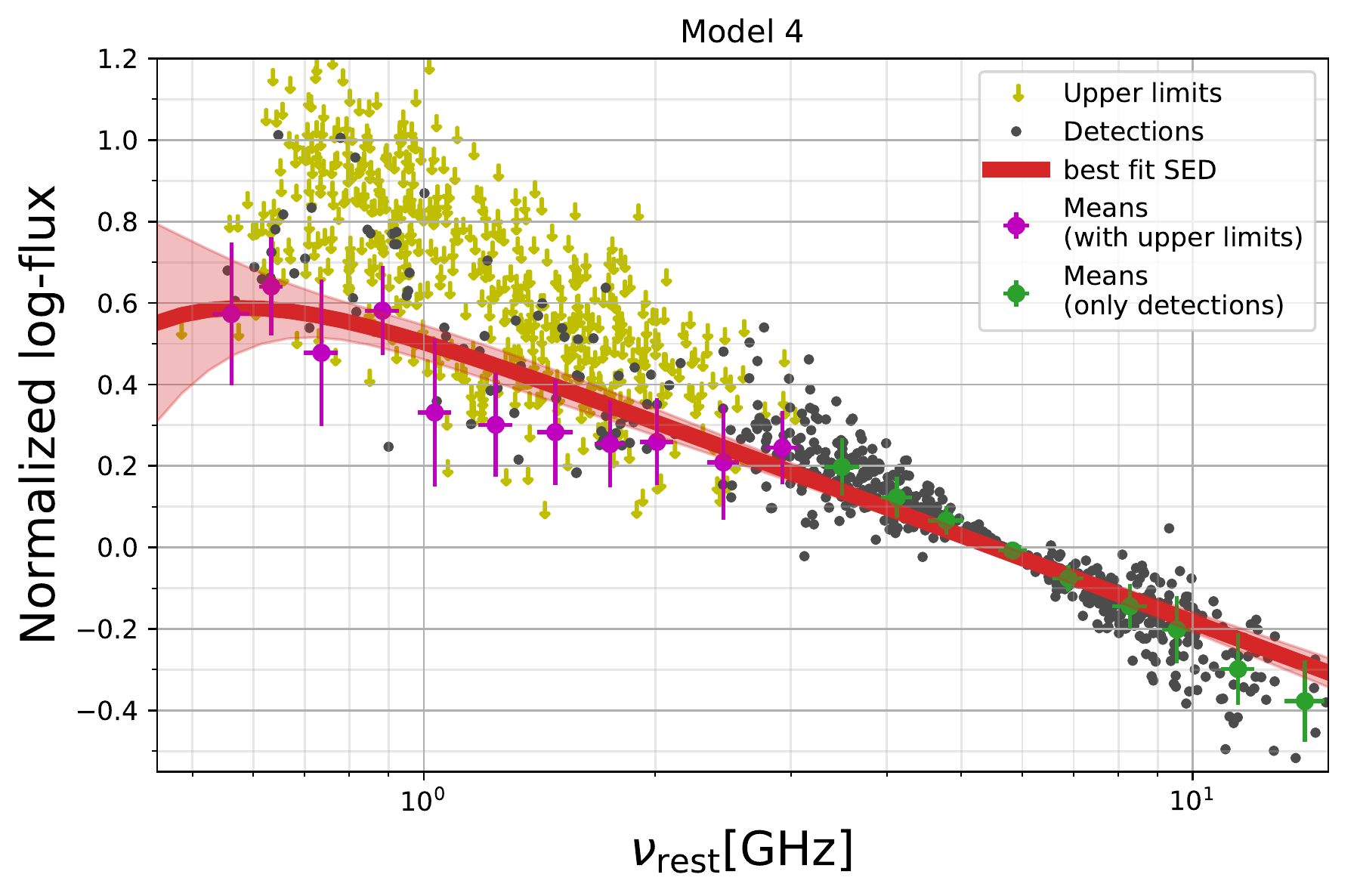}
\centering \includegraphics[width=.4\textwidth]{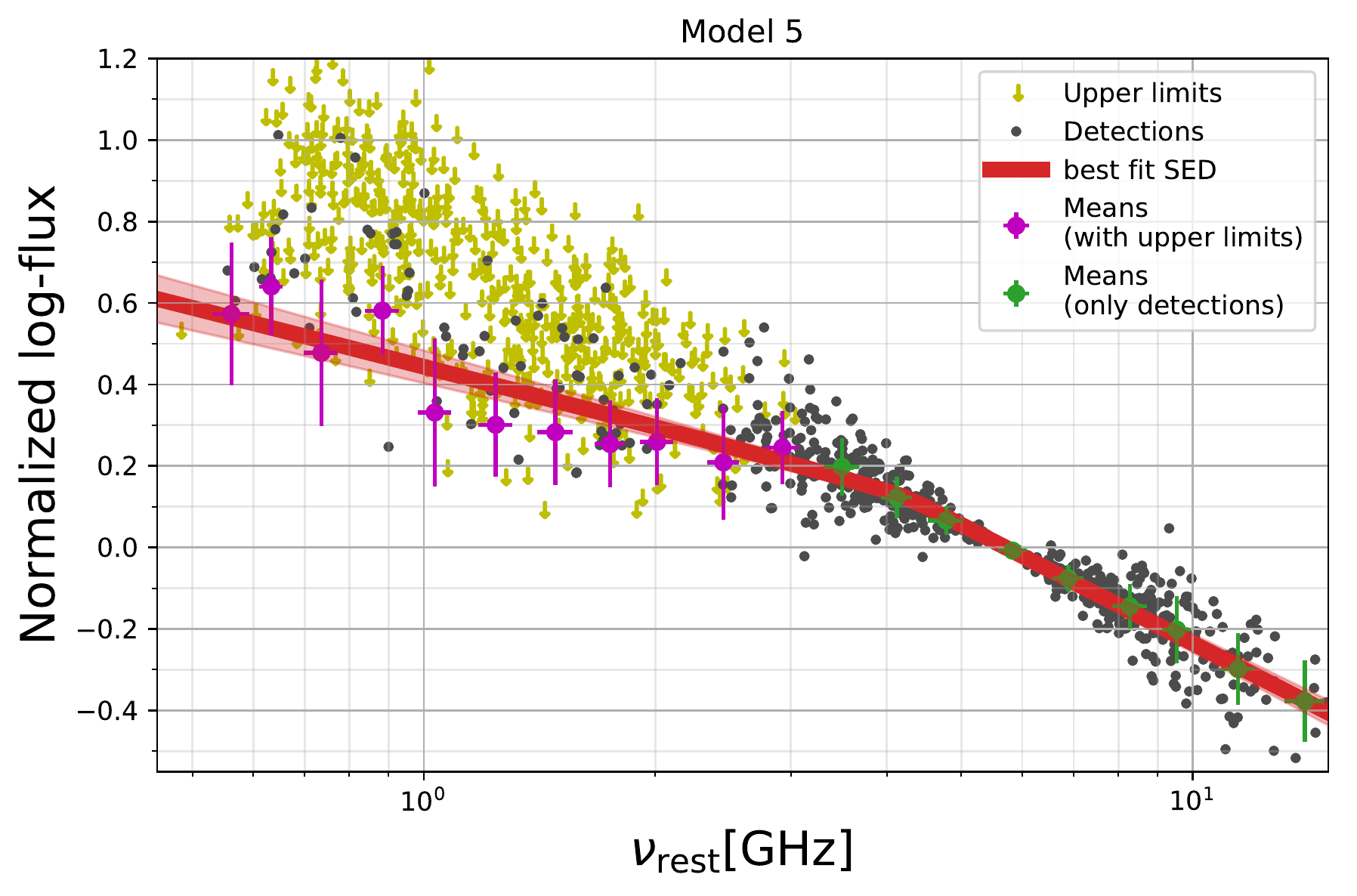}

\centering \includegraphics[width=.4\textwidth]{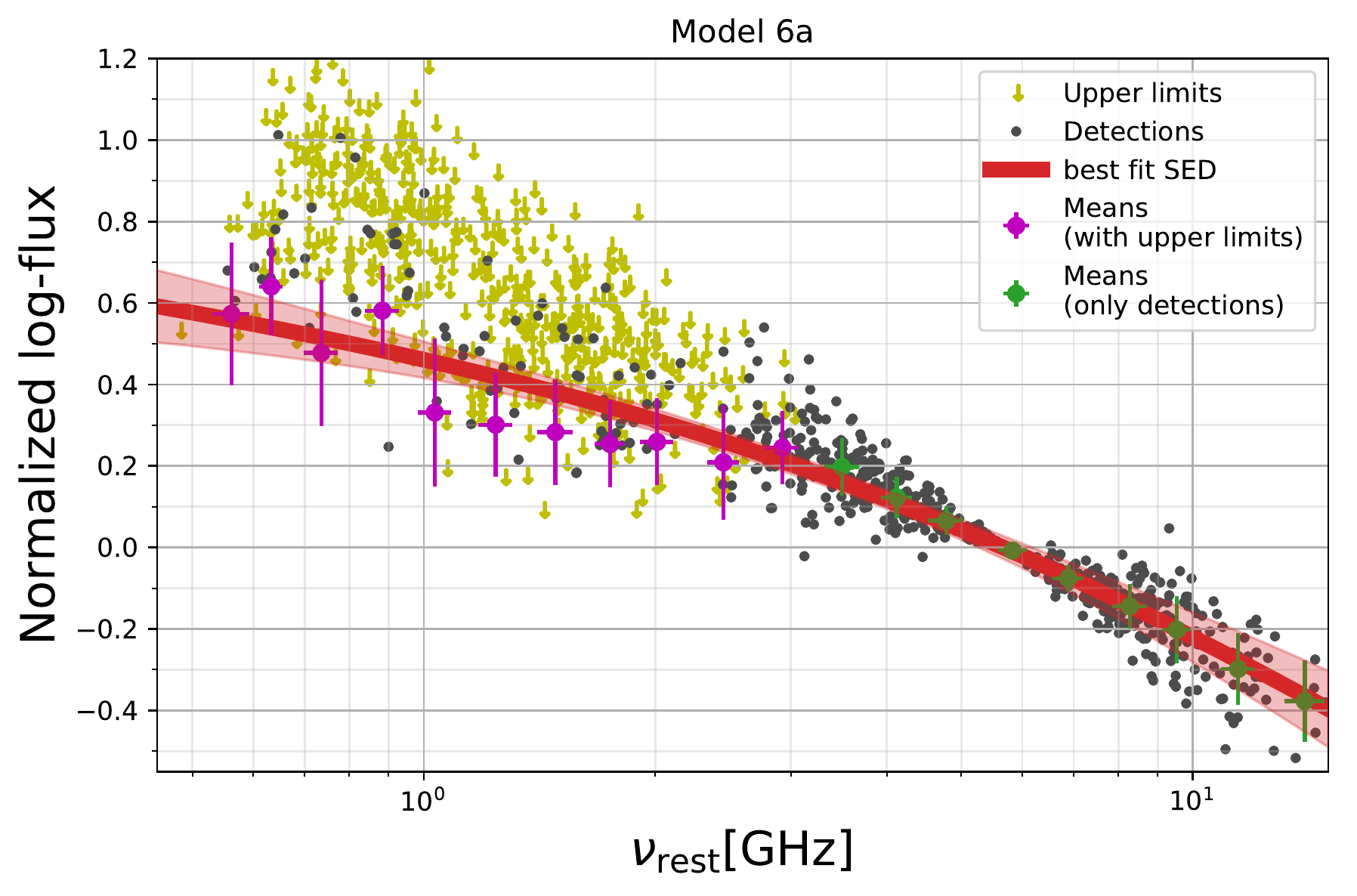}
\centering \includegraphics[width=.4\textwidth]{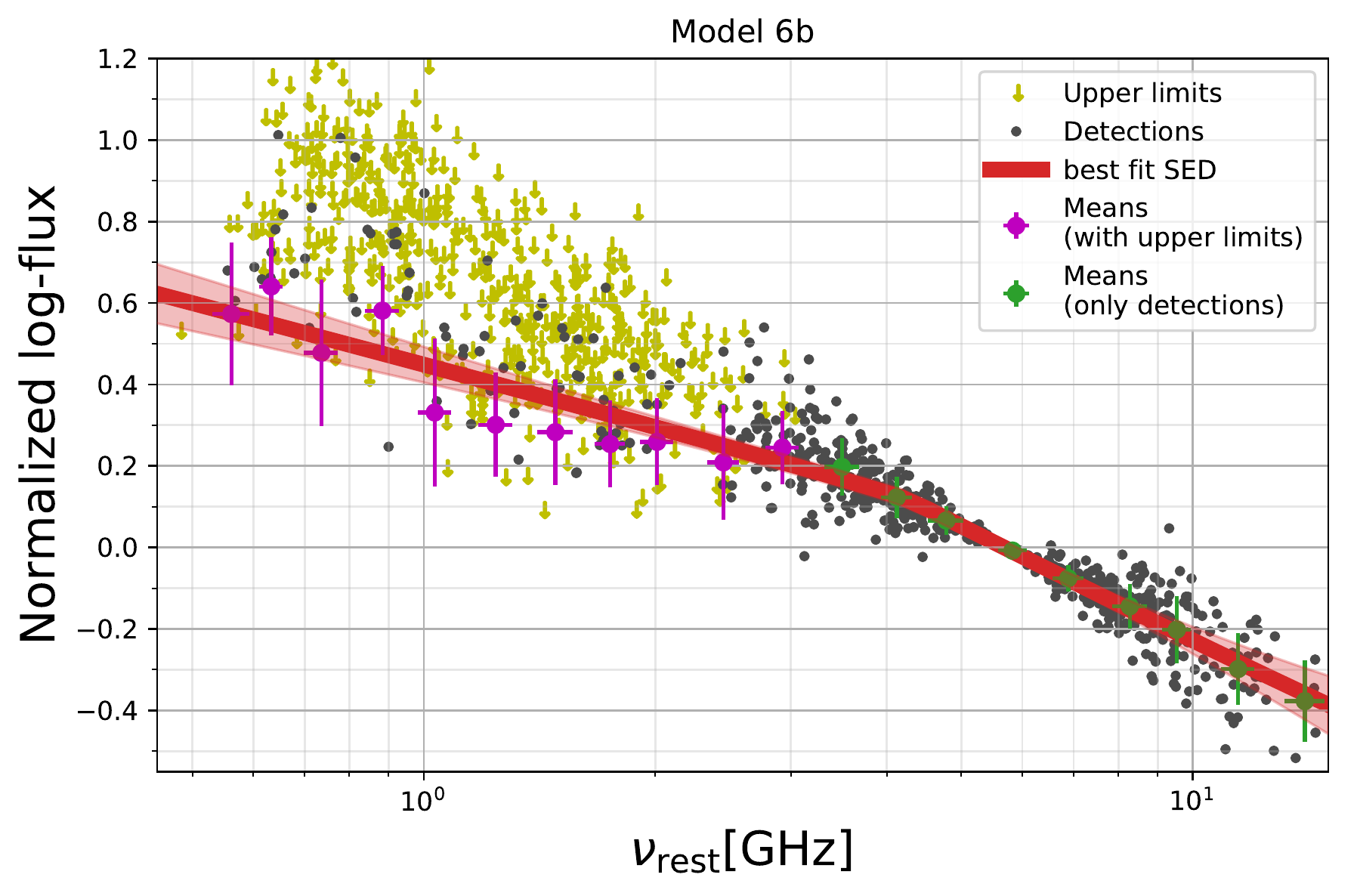}

\caption{Average radio SED of HSFGs for the models from Sect. \ref{sect:The shape of the radio SED} plotted as in Fig. \ref{fig:RealSED}. The red solid lines and shaded intervals represent the best-fitting parameters of Table \ref{tab:ranking} and confidence intervals, respectively.}\label{fig:modelSEDs}
\end{figure*}
\end{document}